\documentclass{elsarticle}
\usepackage[width=430pt,height=650pt,centering]{geometry}
\usepackage{latexsym}
\usepackage{pstricks}
\usepackage{epsf}
\usepackage{graphicx}
\usepackage[latin1]{inputenc}
\usepackage{amssymb}
\usepackage{amsmath}
\usepackage{verbatim}
\usepackage{color}
\usepackage{array}
\usepackage{ifthen}
\usepackage{theorem}
\usepackage{dsfont}

\newboolean{showFigures}
\newboolean{finalManuscript}

\setboolean{showFigures}{false}
\setboolean{finalManuscript}{true}

\newtheorem{Theorem}{Theorem} 
\newtheorem{Proposition}[Theorem]{Proposition}

\newtheorem{Definition}[Theorem]{Definition}

{\theoremstyle{break}\theorembodyfont{\upshape}\newtheorem{Example}{Example}}
{\theoremstyle{plain}\theorembodyfont{\upshape}\newtheorem{Remark}[Theorem]{Remark}}

\newcommand{\CombClasses}{\mathbf{C}}

\def\Proof{{\noindent \bf Proof.~}}

\def\qed{\hfill $\Box$}

\def\ADcomment#1{}

\newcommand{\YannCom}[1]{}
\newcommand{\AlainComm}[1]{}
\newcommand{\Fig}[1]{\ifthenelse{\boolean{showFigures}}{#1}{$\emptyset$}}
\newcommand{\OnlyFinal}[1]{\ifthenelse{\boolean{finalManuscript}}{#1}{}}
\newcommand{\CCS}[2]{{\mathcal{#1}_{#2}}}

\newcommand{\bigO}[1]{\mathcal{O}(#1)}
\renewcommand{\emptyset}[0]{\ensuremath{\varnothing}}

\newcommand{\Diff}[2]{\frac{\partial #1}{\partial #2}}

\newcommand{\unrule}[0]{ \;|\; }

\newcommand{\url}[2]{ \emph{#1} }
\DeclareMathAlphabet{\stmt}{T1}{cmsy}{m}{sc}
\newcommand{\WeightsToFreqs}{\ensuremath{\Phi}}
\newcommand{\ObsFreq}{\ensuremath{f^*}}
\newcommand{\TargFreq}{\ensuremath{\mu}}
\newcommand{\TargFreqs}{\ensuremath{\boldsymbol{\mu}}}
\newcommand{\ObjFun}{\ensuremath{F}}
\newcommand{\GRGFreqs}{\ensuremath{\mbox{\tt GRGFreqs}}}

\newcommand{\Atoms}{\boldsymbol{\mathcal{Z}}}
\newcommand{\At}{\mathcal{Z}}
\newcommand{\CombClass}{{C}}

\newcommand{\SG}{S}
\newcommand{\CombSpec}{\Psi}
\newcommand{\CombSpecs}{{\bf \Psi}}
\newcommand{\WF}{\pi}
\newcommand{\Weights}{{\boldsymbol\pi}}
\newcommand{\Def}[1]{{\bf #1}}

\newcommand{\WPlus}{{\W_{+}}}
\newcommand{\WOne}{{\W_{1}}}

\newcommand{\TPlus}{{\boldsymbol{+}}}
\newcommand{\TMinus}{{\boldsymbol{-}}}
\newcommand{\TDot}{\;\bullet\;}
\newcommand{\TOpenPar}{\mbox{ {\bf (} }}
\newcommand{\TClosePar}{\mbox{ {\bf )} }}
\newcommand{\TZero}{\mathbf{0}}
\newcommand{\TOne}{\mathbf{1}}
\newcommand{\Gram}{\ensuremath{\mathcal{G}}}
\newcommand{\IdM}{\mathbb{I}}
\newcommand{\ProdOp}{\times}

\newcommand{\W}{\pi}

\renewcommand{\Pr}{\mathbb{P}}

\newcommand{\ZoomSquares}{0.10}
\newcommand{\ZoomTrees}{0.038}

\newcommand{\ShowTree}[2]{\Fig{\includegraphics[scale=\ZoomTrees]{QuadTrees/Results/#1-#2-viz}}}
\newcommand{\ShowQuad}[2]{\Fig{\includegraphics[scale=\ZoomSquares]{QuadTrees/Results/#1-#2}}}

\newcommand{\FigRatio}{0.02}

\def\k#1{\kern#1em}
\def\Ib#1{{I\kern-.25em#1}}
\def\Ibb#1{{I\kern-.23em#1}}

\begin{document}
\title{Controlled non uniform random generation of decomposable structures}

\author[lri,igm]{A.~Denise\corref{cor}}
\ead{Alain.Denise@lri.fr}
\author[lri,lix]{Y.~Ponty}
\ead{Yann.Ponty@lri.fr}
\author[igm]{M.~Termier}
\ead{termier@igmors.u-psud.fr}
\cortext[cor]{To whom correspondance should be addressed}

\address[lri]{LRI, Universit\'e Paris-Sud, CNRS, INRIA. Bat 490, 91405 Orsay cedex, France}
\address[igm]{IGM, Universit\'e Paris-Sud CNRS. Bat. 400, 91405 Orsay cedex, France}
\address[lix]{LIX, Ecole Polytechnique, CNRS, INRIA. 91128 Palaiseau cedex, France}

\date{\today}

\begin{abstract}

Consider a class of decomposable combinatorial structures, using
different types of atoms $\Atoms = \{\At_1,\ldots ,\At_{|{\Atoms}|}\}$.
We address the random generation of such structures with
respect to a size $n$ and a targeted distribution in $k$ of its
\emph{distinguished} atoms.
We consider two variations on this problem.

In the first alternative, the targeted distribution is given by $k$
real numbers $\TargFreq_1, \ldots, \TargFreq_k$ such that $0 <
\TargFreq_i < 1$ for all $i$ and $\TargFreq_1+\cdots+\TargFreq_k \leq 1$. We aim to generate
random structures among the whole set of structures of a given size
$n$, in such a way that the {\em expected} frequency of any
distinguished atom $\At_i$ equals $\TargFreq_i$. We address this problem by
weighting the atoms with a $k$-tuple $\Weights$ of real-valued
weights, inducing a weighted distribution over the set of structures
of size $n$.  We first adapt the classical recursive random generation
scheme into an algorithm taking $\bigO{n^{1+o(1)}+mn\log{n}}$
arithmetic operations to draw $m$ structures from the
$\Weights$-weighted distribution.  Secondly, we address the analytical
computation of weights such that the targeted frequencies
are achieved asymptotically, i. e. for large values of $n$.
We derive systems of functional equations
whose resolution
gives an explicit relationship between $\Weights$ and $\TargFreq_1, \ldots, \TargFreq_k$.  Lastly,
we give an algorithm in $\bigO{k n^4}$ for the inverse problem,
{\it i.e.} computing the frequencies associated with a given $k$-tuple
$\Weights$ of weights, and an optimized version in $\bigO{k n^2}$ in
the case of context-free languages. This allows for a heuristic
resolution of the weights/frequencies relationship suitable for
complex specifications.

In the second alternative, the targeted distribution is given by a
$k$ natural numbers $n_1, \ldots, n_k$ such that
$n_1+\cdots+n_k+r=n$ where $r \geq 0$ is the number of undistinguished
atoms. The structures must be generated uniformly among the set of
structures of size $n$ that contain {\em exactly} $n_i$ atoms $\At_i$
($1 \leq i \leq k$). We give a $\bigO{r^2\prod_{i=1}^k n_i^2 +m n k
\log n}$ algorithm for generating $m$ structures, which simplifies
into a $\bigO{r\prod_{i=1}^k n_i +m n}$ for regular specifications.
\end{abstract}

\maketitle

\section{Introduction}

The problem of {\it uniform} random generation of combinatorial
structures has been extensively studied in the past few years.
Notably, the wide class of {\em decomposable} structures, that is
combinatorial structures that can be constructed recursively in an
unambiguous way, has been subject to great attention. Two general
methods have been developed for the uniform generation of these
structures: the recursive method~\cite{Flajolet94} and, more recently,
the so-called~{\em Boltzmann} method~\cite{Duchon04,FFP07}.
%
In the present paper, we generalize this
problem to the problem of generating combinatorial
structures according to a given (non uniform) distribution. The distribution is
defined by the desired frequencies of some given {\it atoms} in the
structures that are generated.

According to~\cite{Flajolet94}, decomposable structures are defined by
{\em combinatorial specifications}. Briefly, a combinatorial specification
of a given class $C$ of combinatorial structures is a
tuple $\CombClasses$ of combinatorial classes which are interrelated by means of
productions made from basic objects of size zero (empty structures) or
size one (atoms), and from {\em constructions\/} ($+$ for disjoint
union, $\ProdOp$ for products, {\tt sequence} for sequences, {\tt set}
for multisets and {\tt cycle} for directed cycles).

We are interested in the following problem.  Let $C$ be a
combinatorial class, whose set of atoms is ${\Atoms}=\{\At_1,
\ldots, \At_{|{\Atoms}|}\}$. Let us distinguish $k \leq |{\Atoms}|$
atoms in $\Atoms$, say $\At_1, \ldots \At_k$.  Now let $n$ be an
integer, and let us denote $\CCS{C}{n}$ the set of structures of $C$ of
length $n$. The problem consists in generating random structures in
$\CCS{C}{n}$ while respecting a distribution of the $k$ distinguished
atoms. We consider two variations of the problem:
\begin{enumerate}
\item {\em Generation according to expected frequencies}.
The targeted distribution is given by $k$
real numbers $\TargFreq_1, \ldots, \TargFreq_k$ such that $0 <
\TargFreq_i < 1$ for all $i$ and $\TargFreq_1+\cdots+\TargFreq_k \leq 1$. The structures
must respect {\em on the average} the given frequency k-tuple. More precisely,
we generate structures at random in such a way that
\begin{enumerate}
\item any structure of $\CCS{C}{n}$ has a positive probability to be
generated;
\item for any $i \in \{1,\ldots,k\}$, the expected frequence of
occurrences of $\At_i$ in the structures is equal
to $\TargFreq_i$: if $\Pr(s)$ is the probability of
the structure $s$ to be generated by the algorithm, we must have
$\sum_{s \in \CCS{C}{n}} |s|_{\At_i} \Pr(s) = n \TargFreq_i$\,;
\item two structures having the same distribution of the $k$ distinguished
atoms have the same probability of being generated.
\end{enumerate}
\item { \em Generation according to exact frequencies}. Here the
distribution is given by $k$ natural numbers $n_1,
\ldots, n_k$ such that $n_1 + n_2 + \cdots n_k \leq n$.  The
distribution of the number of distinguished atoms of any structure
must respect the given k-tuple exactly. In other words, we generate
structures uniformly at random in a subset of $\CCS{C}{n}$ constituted
of all the structures $s \in {C}$ such that $|s|_{\At_i}=n_i$ for all
$i \in \{1,\ldots,k\}$, where $|s|_{\At_i}$ stands for the number of
atoms $\At_i$ in $s$.
\end{enumerate}

The above two problems arise when one tries to model \emph{naturally occurring objects}
or to circumvent some limitations of generative descriptions, therefore both were addressed
under fairly specific settings. For instance, a \emph{non-uniform} scheme was used by
Brlek~\emph{et al}~\cite{Brlek2006} to perform a generation of generalized Motzkin paths
according to their area.
The generation according to exact frequencies was implicitly used in~\cite{Dutour98},
where the problem of randomly generating structures while fixing more than one parameter was
addressed. One also needs to mention a very elegant $\Theta(n)$ algorithm for generating words
from regular languages with two types of atoms~\cite{Bertoni2003}. Finally, the original
presentation of the recent Boltzmann method~\cite{Duchon04} features the generation
of adsorbing staircase walks according to both the size and number of contacts to the origin.

Our approach is based on the recursive method, which
was initiated by Nijenhuis and Wilf~\cite{Nijenhuis79}, and then
generalized and formalized by Flajolet, Zimmermann and Van
Cutsem~\cite{Flajolet94}.  Section~\ref{secUniform} is devoted to a
short presentation of this methodology in the classical context of
{\em uniform} generation. In Section~\ref{secMoyenne}, we focus on generating
structures according to expected frequencies, with an emphasis on the computation
of suitable weights.
Finally, we present in Section~\ref{secExact} another algorithm which allows to generate structures according to exact
frequencies.

\section{Combinatorial specifications and uniform generation}
\label{secUniform}

As seen above, a combinatorial specification of a
given class $C$ of combinatorial structures is a
tuple of classes which are interrelated by means of
productions made from basic objects (empty structures denoted $\varepsilon$ and atoms, of size $0$
and $1$ respectively) and from {\em constructions\/} ($+$ for disjoint
union, $\ProdOp$ for products, {\tt sequence} for sequences, {\tt set}
for multisets and {\tt cycle} for directed cycles).

The algorithm works as follows: First translate the specification into
a {\em standard\/} one, where all products are binary, and the {\tt
sequence}, {\tt set}, {\tt cycle} constructions have been replaced
with the marking and unmarking constructions $\Theta$ and
$\Theta^{-1}$ (see \cite{Flajolet94}).
Then the standard specification translates directly into procedures for counting the
number of structures of a given size generated from a given non-terminal
(see Table~\ref{tabSS}),
\begin{table}[t]

\begin{minipage}{0.525\textwidth}
\begin{eqnarray}
  C = 1
  \!&\Rightarrow &\!
  c_0 = 1\; (\mbox{\rm $\varepsilon$ struct.}) \label{unif1}
\\
  C = A + B
  \!&\Rightarrow &\!
  c_n = a_n + b_n\label{unif3}
\\
  \Theta C = A \ProdOp B
  \!&\Rightarrow &\!
  c_n = {\frac 1 n} \sum_{k=0}^n a_k b_{n-k} \label{unif4bis}
\end{eqnarray}
\end{minipage}
\begin{minipage}{0.475\textwidth}
\begin{eqnarray}
  C = \At_i
  \!&\Rightarrow&\!
  c_1 = 1\ \label{unif2} \ (\mbox{\rm atom})
\\
  C = A \ProdOp B
  \!&\Rightarrow&\!
  c_n = \sum_{k=0}^n a_k b_{n-k} \label{unif4}
\\
  C =  \Theta A
  \!&\Rightarrow&\!
  c_n = n a_n . \label{unif4ter}
\end{eqnarray}
\end{minipage}
\caption{Counting procedures for standard specifications.}
\label{tabSS}
\end{table}
or for generating one such object uniformly at random
(see Table~\ref{tabRG}).
\begin{table}
\parbox{6.5cm}{
\hspace{0cm} Case: $C = 1$.

\hspace{0.5cm} gC := procedure($n$: integer);

\hspace{1cm} if $n=0$ then Return($1$)

\hspace{0.5cm} end.

\medskip
\hspace{0cm} Case: $C = \At$.

\hspace{0.5cm} gC := procedure($n$: integer);

\hspace{1cm} if $n=1$ then Return($\At$)

\hspace{0.5cm} end.

\medskip
\hspace{0cm} Case: $C = A+B$.

\hspace{.5cm} gC := procedure($n$: integer);

\hspace{1cm} $U$:=Uniform($[0,1]$);

\hspace{1cm} if $U< a_n/c_n$

\hspace{1.5cm}     then Return(gA($n$))

\hspace{1.5cm}     else Return(gB($n$))

\hspace{.5cm} end.
}
\parbox{6cm}{
\hspace{0cm} Case: $C = A \ProdOp B$.

\hspace{.5cm} gC := procedure($n$: integer);

\hspace{1cm} $U$:=Uniform($[0,1]$);

\hspace{1cm} $k:=0$;

\hspace{1cm} $S:= a_0 b_n / c_n$;

\hspace{1cm} while $U > S$ do

\hspace{1.5cm} 	$k:=k+1$;

\hspace{1.5cm} 	$S := S + a_k b_{n-k} / c_n$;

\hspace{1cm} Return($\left\langle \mbox{gA}(k),\mbox{gB}(n-k)\right\rangle$)

\hspace{.5cm} end.
}
\caption{Uniform random generation procedures for standard
  specifications. The straightforward pointing and
  unpointing cases are omitted.}
\label{tabRG}
\end{table}
The computation of all tables up to size $n$ requires $\bigO{n^2}$
operations on coefficients, which can be lowered to $\bigO{n (\log
n)^2 \log \log n}$ by using Joris van der Hoeven's technique for
computing the coefficients~\cite{Hoeven02}. Then one random
generation needs $\bigO{n \log n}$ operations in the worst case using
the boustrophedonic method. These complexities can be lowered for some
particular classes of combinatorial structures, notably those that
give rise to holonomic generating functions, so that the counting
sequences satisfy linear
recurrences~\cite{Lipshitz1989,BostanChyzakLecerfSalvySchost2007},
leading to $\bigO{n}$ operations only for computing the tables. This
is the case for context-free specifications for
example~\cite{Goldwurm95}.

The integer coefficients used in the algorithm usually have an
exponential growth with respect to the size $n$: $\bigO{n \log n}$ in
the labelled case and $\bigO{n}$ in the unlabelled
case~\cite{Flajolet94}. Therefore, with Schönhage's multiplication
algorithm~\cite{fft71} for integer arithmetic or Fürer's recent
improvement~\cite{Furer2007}, the precomputation and the generation
have bit-complexity $\bigO{n^{2 + o(1)}}$. Meanwhile, using adaptative
floating point computations, the bit-complexity of the generation step
can be lowered to $\bigO{n^{1 + o(1)}}$~\cite{Denise99}. Furthermore,
combining~\cite{Denise99} and the later work in \cite{Hoeven02} leads to a
precomputation step in $\bigO{n^{1 + o(1)}}$ bit-complexity too.

Another work extends this approach to unlabeled objects
\cite{Flajolet97b}.
From now on, we suppose we are given an unlabeled
standard specification, with union, product, marking and unmarking
constructions. Tables~\ref{tabSS} and~\ref{tabRG}, respectively,
summarize the counting and generating procedures.  The labeled case
is very similar, with additional binomial coefficients.


\section{Generation according to expected frequencies}
\label{secMoyenne}

\subsection{Weighted combinatorial structures and random generation}

In this section, we consider the problem of generating structures of
$\CCS{C}{n}$
at random
in such a way that each structure $s$ is generated with positive
probability $\Pr(s)$, and the k-tuple of expected frequencies of the
atoms $\At_1,  \ldots, \At_k$ equals the given k-tuple $(\TargFreq_1,
\ldots, \TargFreq_k)$.
Formally:
\begin{equation} \label{condGen1}
  \Pr(s)>0\ \ \forall s  \in \CCS{C}{n}
\end{equation}
and
\begin{equation} \label{condGen2}
  \sum_{s \in \CCS{C}{n}} |s|_{\At_i} \Pr(s) \ = n \TargFreq_i
        \ \ \forall i \in \{1,2,\ldots,k\}.
\end{equation}
Moreover, any two structures $(s,s')\in \CCS{C}{n}\times \CCS{C}{n}$ having the same distribution in atoms $\At_1, \ldots, \At_k$ must be equally generated:
\begin{equation} \label{condGen3}
  (|s|_{\At_i} = |s'|_{\At_i} \ \forall i \in \{1,\ldots,k\})
        \ \Rightarrow\ \Pr(s) = \Pr(s').
\end{equation}
Our method consists in adjoining a $k$-tuple of weights  $\Weights = (\W_1,\ldots,\W_k)$ to the specification, assigning a real-valued {\it weight}
$\W_i\in\mathbb{R}_+^*$ to each distinguished atom $\At_i\in\Atoms$.
The weight of any combinatorial structure is then defined to be the product of the
weights of its distinguished atoms:
$$
  \WF(s) = \prod_{1 \leq i \leq k} \W_i^{|s|_{\At_i}},
$$
and the weight of a finite combinatorial class is the sum of the weights of
its members. In particular, for $\CCS{C}{n}$ we have:
$$
  \WF(\CCS{C}{n}) = \sum_{s \in \CCS{C}{n}} \WF(s).
$$
If the algorithm is such that
\begin{equation} \label{condGen4}
  \Pr(s)= {\frac {\WF(s)} {\WF(\CCS{C}{n} )}}, \ \ \ \forall s \in \CCS{C}{n},
\end{equation}
then the larger the weight of any given atom is (with regard to the
weights of the other ones), the more this atom occurs in a
random sample. On the other hand, formula (\ref{condGen4}) implies
conditions (\ref{condGen1}) and (\ref{condGen3}).

\medskip
Now we have to solve two problems:
\begin{enumerate}
\item Find a $k$-tuple $\Weights$ that satisfies (\ref{condGen2}) assuming
that (\ref{condGen4}) holds;
\item Design a generation algorithm which satisfies (\ref{condGen4}).
\end{enumerate}
Let us first solve the latter, for which we adapt the recursive scheme.

\medskip
\begin{Proposition}
Suppose that $\Weights$ is given.
  Then an adaptation of the recursive approach gives an
  algorithm which takes $\bigO{n^{1+o(1)} + mn \log n}$ arithmetic operations
  for generating $m$ structures of size $n$ such that each structure $s$
  is generated with probability $\Pr(s)$.
\end{Proposition}

In order to generate words with the required distribution
(\ref{condGen4}), we use the methodology presented in Section~\ref{secUniform}, with just a slight change: Now the rule
\begin{eqnarray*}
  C = \At_i
  &\Rightarrow&
  c_1 = \WF(\At_i) \equiv \W_i. 
\end{eqnarray*}
replaces rule (\ref{unif2}) in Table~\ref{tabSS}.
The generation process then works exactly like the uniform one described in Section
\ref{secUniform}. It can be easily shown that the probability of
generating a structure $s$ occurs will be proportional to its weight $\WF(s)$.

The $\bigO{n \log n}$ behavior of a Boustrophedon search follows from the
facts that: i) The {\it worst-case complexity} of the \emph{uniform} generation
is in $\bigO{n\log(n)}$, as was shown in~\cite{Flajolet94}; ii) For any sampled structure $s$,
the costs of generating $s$ in the weighted and uniform distribution are strictly identical.
Since the generation cost of \emph{any structure} is in $\bigO{n\log(n)}$, then so is
the expected cost of a generation, regardless of the distribution.

\medskip
From now on, given $C$, $\Weights$ and $n$, let us write $f_\Weights(\At_i,\CombClass,n)$
for the average number of atoms $\At_i$ in the structures of $\CombClass_n$
generated by the above scheme.
Our problem is then the following: given the k-tuple $(\TargFreq_1, \ldots,
\TargFreq_k)$, find the $k$-tuple $\Weights$ of weights that achieves targeted
frequencies, that is such that
$$
  f_\Weights(\At_i,\CombClass,n) = n \cdot \TargFreq_i \ \ \
          \mbox{\rm  for all}\ i\ \mbox{\rm such that}\ 1 \leq i \leq k.
$$

We give two different approaches to tackle this problem.  The
first one, detailed in Subsections~\ref{sec:asympt},
is analytic and gives, if some conditions on $C$ hold, asymptotic
formulas for $f_\Weights(\At_i,\CombClass,n)$ when $n$ is large, assuming we are
able to solve some system of functional equations. By contrast, our
second programme, described in Subsection~\ref{sec:opt}, leads to an
heuristic for approximating $\Weights$ in the general case.

\subsection{Computing weights suitable for asymptotical frequencies}
\label{sec:asympt}

\subsubsection{The (non-rational) context-free case}
\label{sec:drmota}

A combinatorial class is said to be context-free if it can be specified
without using set and cycle operations.
A result of Drmota~\cite{Drmota97}, applied by Denise \emph{et al}~\cite{Denise00} to the case of weighted
context-free grammars allows us to foresee a symbolic approach to the
computation of weights compatible with expected frequencies. More specifically, it defines
sufficient conditions such that the number $c_n$ of structures of size $n$ asymptotically follows
the ubiquitous behavior
	$$c_{n} \sim \kappa_\Weights\cdot\frac{\rho_\Weights^n}{n\sqrt{n}}(1+\bigO{1/\sqrt{n}})$$
and such that the coefficients $c_{n}^{i}$ that count the total number of symbols $\At_i$ in all words
of size $n$ follow asymptotic expansions of the form
	$$c_{n}^i \sim \kappa_{\Weights,i}\cdot\frac{\rho_\Weights^n}{\sqrt{n}}(1+\bigO{1/\sqrt{n}})$$
for $\kappa_\Weights$ and $\kappa_{\Weights,i}$ some explicit constants of $n$.
It follows that a relationship exists between the weights
and the asymptotical frequencies of occurrence for each atom $\At_i$.
This relationship is in most cases quite simple, and allows to
derive suitable weights $\Weights$ for \emph{reasonable} objective k-tuples of frequencies
$(\TargFreq_1,\ldots,\TargFreq_k)$.


\begin{Definition}[Simple type specification]
	Let $\CombSpecs=\{\CombSpec_i\}$ be a set of standard specifications for a tuple $\CombClasses$ of algebraic (context-free)
	combinatorial classes. \\
	Let $c_{n_1,\ldots,n_{k},r}$ be the number of
	structures of size $n=r+\sum_{i=1}^k n_i$ in a combinatorial class $\CombClass$, having $n_j$ occurrences of
	atom $\At_j$, $ j\in [1,k]$, and $r$ remaining atoms.\\
	Then $\CombSpec$ is said to be \emph{of simple type} if there exists, for each combinatorial class
	$\CombClass\in\CombClasses$, a $k$-dimensional cone $\mathcal{N}_i\subset\mathbb{R}^{k}$ that
	is centered on $0$ and \emph{saturated} such that
	$$\forall(n_1,\ldots,n_{k},r)\in\mathcal{N}_{i}\cap \mathbb{N}^{k+1},\; c^{i}_{n_1,\ldots,n_{k},r}\neq 0.$$
\end{Definition}

\begin{Theorem}[Asymptotics of algebraic specifications~\cite{Drmota97}]\label{th:drmota}
Let $\CombSpec=\{\CombSpec_i\}_{i=1}^m$ be a combinatorial specification for a $m$-tuple $\CombClasses=(\CombClass_1,\ldots,\CombClass_m)$  of combinatorial classes
such that:
\begin{enumerate}
	\item for any $i \in [1,m]$, $\CombClass_i$
is not isomorphic to a rational language.
	\item $\CombSpec$ doesn't use any $\varepsilon$-production.\label{cond:epsilon}
	\item $\CombSpec$ is a \emph{simple type} specification.
	\item $\CombSpec$ is \emph{strongly connected}.
\end{enumerate}
For each $i\in[1,k]$ and $j\in[1,m]$:
\begin{itemize}
\item[-] Let $u_i$ be a random complex variable and $\W_i$ a real valued weight.
\item[-] Let $\SG_j$ be the multivariate generating function for class $\CombClass_j$.
\item[-] Let $\Phi_{j}(t,u_1,\ldots,u_{|\Atoms|},\SG_1,\ldots,\SG_{m})$
be the term obtained from $\CombSpec_j$ by replacing $\At_i$ by
$t\cdot\W_i\cdot u_i$, and $\CombClass_j$ by $\SG_j$.
\end{itemize}
Finally, let $A$ be the Jacobian matrix of $\Phi$, such that $A=\left(\frac{\partial \Phi_i} {\partial \CombClass_j}\right)_{i,j\in[1,|\CombSpec|]}$.\\
Consider the following system:
\begin{equation} \label{eq:drmota}
\left\{  \begin{array}{c}
								\SG_1(t\W_1u_1,\ldots,t\W_{|\Atoms|}u_{|\Atoms|})  =  \Phi_1(t,u_1,\ldots,u_{|\Atoms|},\SG_1,\ldots,\SG_{|\CombSpec|})\\
								\ldots \\
								\SG_{|\CombSpec|}(t\W_1u_1,\ldots,t\W_{|\Atoms|}u_{|\Atoms|})  =  \Phi_{|\CombSpec|}(t,u_1,\ldots,u_{|\Atoms|},\SG_1,\ldots,\SG_{|\CombSpec|})\\
								0  =  \det(\IdM-A)
							\end{array} \right.
\end{equation}
Let $(\rho^*_\Weights,\SG^*_1,\ldots,\SG^*_{|\CombSpec|})$ be a $|\CombSpec|+1$-tuple of functions of ${\bf u} =(u_1,\ldots,u_{|\Atoms|})$, solution of System~\eqref{eq:drmota} such that
$\rho_\Weights^*({\bf 1})\in \mathbb{R}^+$ and is minimal. Then we have:
\begin{equation} \label{eq:finalDrmota}f_\Weights(\At_i,\CombClass,n) =
-\frac{1}{\rho^*_\Weights({\bf 1})}\frac{\partial \rho^*}{\partial u_i}({\bf 1})\; . \; n + \bigO{1}
\end{equation}
\end{Theorem}
The intuition behind the conditions of this theorem is the following:
\begin{itemize}
	\item[-] The \emph{non-rationality} of the corresponding language helps avoiding simple poles, a case where the simplifications presented in section~\ref{genRat} appear.
	\item[-] The \emph{strongly connected} condition ensures that the dominant singularity
	is the same for all functions $\SG_i(t,\ldots,t)$.
	\item[-] Furthermore, adding a \emph{simple type} condition guarantees a
\emph{square-root type} dominant singularities for all generating functions $\SG_i$.
	\item[-] The value $x_\Weights^*(1,\ldots,1)$ is the dominant singularity, necessarily
	positive as we are considering series with positive coefficient (Pringsheim's
	Theorem). 
\end{itemize}

\begin{Remark}
The original formulation of the Theorem~\cite{Drmota97} addresses a wider range of
candidate systems~\eqref{eq:drmota} than the context-free languages, thus it is expected
that some of its most stringent constraints can sometimes be relaxed. For instance, the
coefficients of the equations derived from $\CombSpec$ are positive, which is a real
restriction since the class of context-free languages is not closed under complement.

Also, the \emph{$\varepsilon$-free} condition can be relaxed, since it is a classic
result that any grammar can be transformed into an \emph{$\varepsilon$-free} one
generating the same language.

\begin{figure}[t]
\begin{center}
\begin{tabular}{|m{4.2cm}m{1.7cm}|m{3.4cm}m{1.5cm}|}\hline
$\begin{array}{lrcl}
	\Gram: & S&\to&T\;U                          \\
				 & T&\to&U \TOpenPar T \;U\TClosePar T \\
				 &  &|&\varepsilon \\
				 & U&\to&\TDot U \\
				 & &|&\varepsilon
\end{array}$&
\Fig{\includegraphics[scale=0.5]{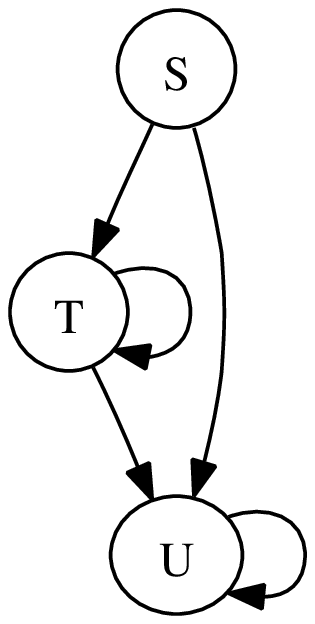}}&
$\begin{array}{lrcl}
 \Gram':&S&\to&\TOpenPar S \TClosePar S\\
				& &|&\TDot S\\
				& &|&\varepsilon
\end{array}$&
 \Fig{\includegraphics[scale=0.5]{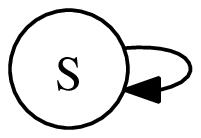}}\\
 \hline
\end{tabular}
\caption{Two equivalent grammars for the Motzkin language along with their dependency graphs.}
\label{fig:depMotzkin}
\end{center}
\end{figure}
Lastly, a property that might be too stringent is the \emph{strong-connectedness},
whose role is to avoid some complicated cases where several concurrent singularities may
interfere, e. g. giving rise to oscillating asymptotic behaviors.
Indeed, many concrete examples show that, as can be verified
through singularity analysis~\cite{flajolet-odlyzko}, correct frequencies can be predicted by mean of the
theorem although their graphs are not strongly connected.

Some of these examples are purely artefactual, a phenomenon illustrated by the two grammars from Figure~\ref{fig:depMotzkin}. In this example, the two grammars have different dependency graphs, and grammar $\Gram$ trivially does not meet the \emph{strong-connectedness} criteria of theorem~\ref{th:drmota}, despite generating the same combinatorial class.
One can even build classes of languages such that the conclusions of theorem~\ref{th:drmota} applies, whereas
the language cannot be generated by any strongly-connected grammar. For instance, one may consider
all sorts of $k$-ary trees whose leaves are sequences of a dedicated axiom.

Therefore it remains to propose a tighter characterization of eligible specifications,
not necessarily based on the structure of the system (not sufficiently informative) or
on properties of associated generating functions (solving some of these systems may be challenging)
but rather on intrinsic properties of the associated combinatorial classes. Such a characterization remains a challenging problem
at the moment.
\end{Remark}

\begin{figure}[ht]
	\centering \Fig{\includegraphics[width=0.75\textwidth]{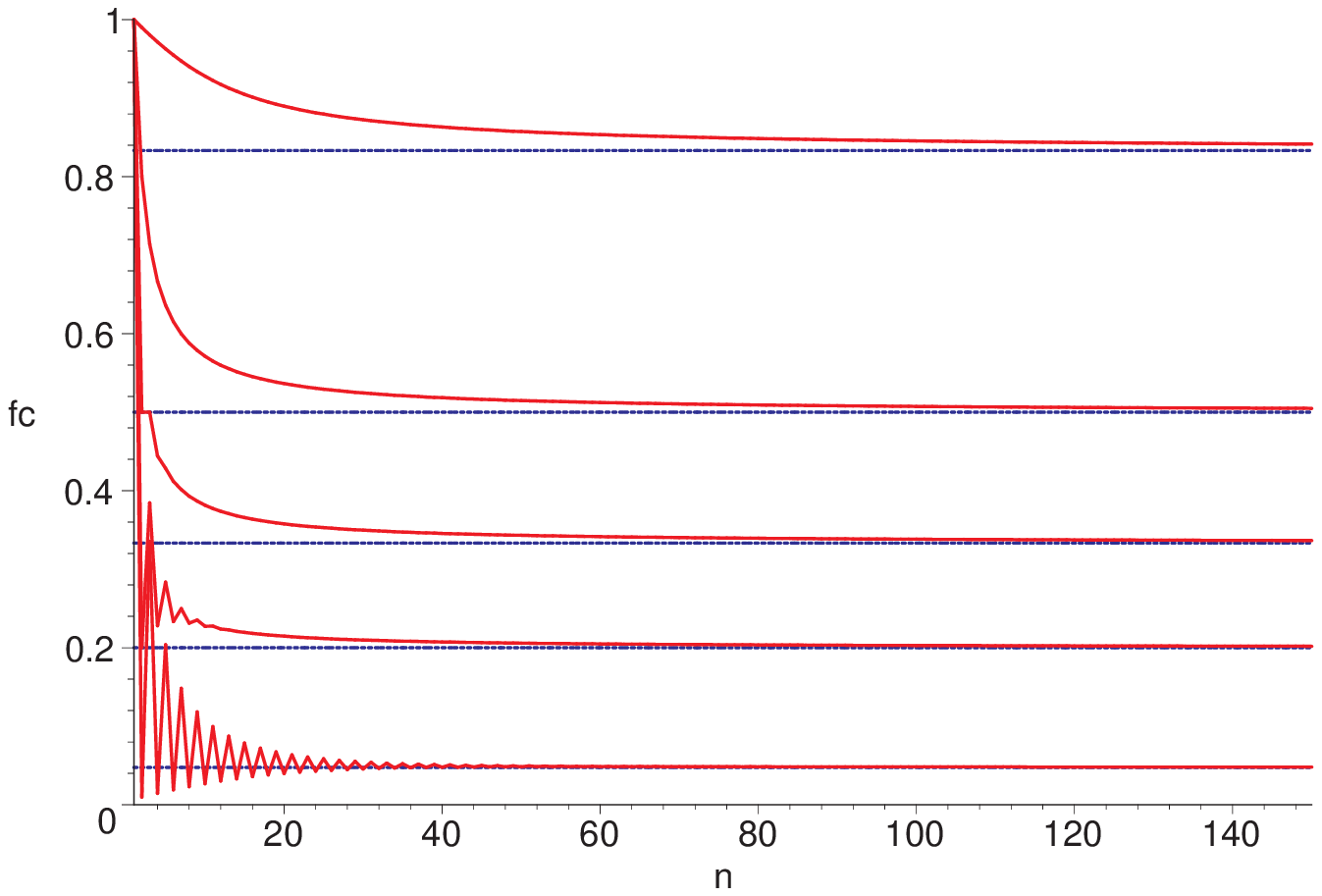}}
\caption{Convergence toward the asymptotic regimes (Dashed lines) of the proportions $f_c$ (Solid lines) of unary nodes among $\W$-weighted unary/binary trees of size $n$. Five values for the couple $(\W,f_c)$ are shown here (From top to bottom): $(10,5/6)$, $(2,1/2)$, $(1,1/3)$, $(1/2,1/5)$, and $(1/10,1/21)$.} \label{fig:ConvAsympt}
\end{figure}

\begin{Example}[Motzkin words/Unary-binary trees]
	Motzkin words are the easiest and most ubiquitous representant
	of the context-free class of languages for which two atoms can occur
	independently. They are also known to be in bijection with the
	rooted trees having nodes of degrees $1$ and $2$. They are generated
	by the following context-free grammar:
	\begin{equation*}
		S \to a\;S\;b\;S \unrule c\;S \unrule \varepsilon
	\end{equation*}
	Through \emph{weighting} the terminal letter $c$ with a real-valued weight $\W$ and \emph{marking} the terminal
	symbol $c$ with a complex variable $u$, we get the following expression for $\Phi_{S_\W}$
	\begin{equation}\label{eq:sysMotzkin1}S_\W(t,tu) = \Phi_{S_\W}(t,u,S_\W) =tS_\W(t,tu)tS_\W(t,tu)+tu\W S_\W(t,tu)+1.\end{equation}
	Since there is only one non-terminal (e.g. combinatorial class) $S$, the Jacobian is reduced to a
	$1\times1$ matrix $A$ such that:
	$$A= 2{t}^{2}S_\W(t,tu)+\W ut  $$
	and
	\begin{equation}\label{eq:sysMotzkin2} \det\left(\IdM-A\right)=1- 2{t}^{2}S_\W(t,tu)-\W ut .\end{equation}
	Putting together equations \ref{eq:sysMotzkin1} and \ref{eq:sysMotzkin2} from above yields the following system
	\begin{equation}\label{eq:sysMotzkin}
		\left\{
			\begin{array}{ccc}
				S_\W(t,tu) & = & tS_\W(t,tu)tS_\W(t,tu)+tu\W S_\W(t,tu)+1\\
				0 & = & 1- 2{t}^{2}S_\W(t,tu)-\W ut
			\end{array}
		\right.
	\end{equation}
	whose solutions for $t$ are
	$$ t^\pm = \frac{1}{\W u \pm 2}.$$
	Taking the positive solution $t^+$ and applying equation~(\ref{eq:finalDrmota}) yields the
	following weight $\W$ that achieves an asymptotic frequency $f_c$ for the terminal symbol $c$
		$$ \W = \frac{2f_c}{1-f_c}.$$

	It is then possible to gain full control over the asymptotic frequency for terminal letters $c$ and $(a,b)$.
	Although in principle this relationship holds only for large values of $n$, a fairly quick convergence toward the asymptotic
	regime is observed, as can be seen in Figure~\ref{fig:ConvAsympt}. Also, the impact of the weight on this convergence, although noticeable,
	does not seem too drastic.
	Alternatively, the three types of atoms can be weighted with a triplet $(\W_a,\W_b,\W_c)$ and the weight/frequency
	relationship remarkably simplifies\OnlyFinal{\footnote{As was pointed out by an anonymous reviewer.}}
into $\W_a=\W_b=f_a=f_b$, and $\W_c=f_c$ with $f_a+f_b+f_c = 1$.

	Since these letters map respectively to unary and binary branches through the classic unary-binary tree
	bijection, we can draw random instances of weighted unary-binary trees.
	We get the \emph{typical} behaviors exhibited in Figure~\ref{fig:Motzkin} for increasing values of $\W$.

\begin{figure}[p]
\begin{center}
\Fig{\includegraphics[scale=\FigRatio]{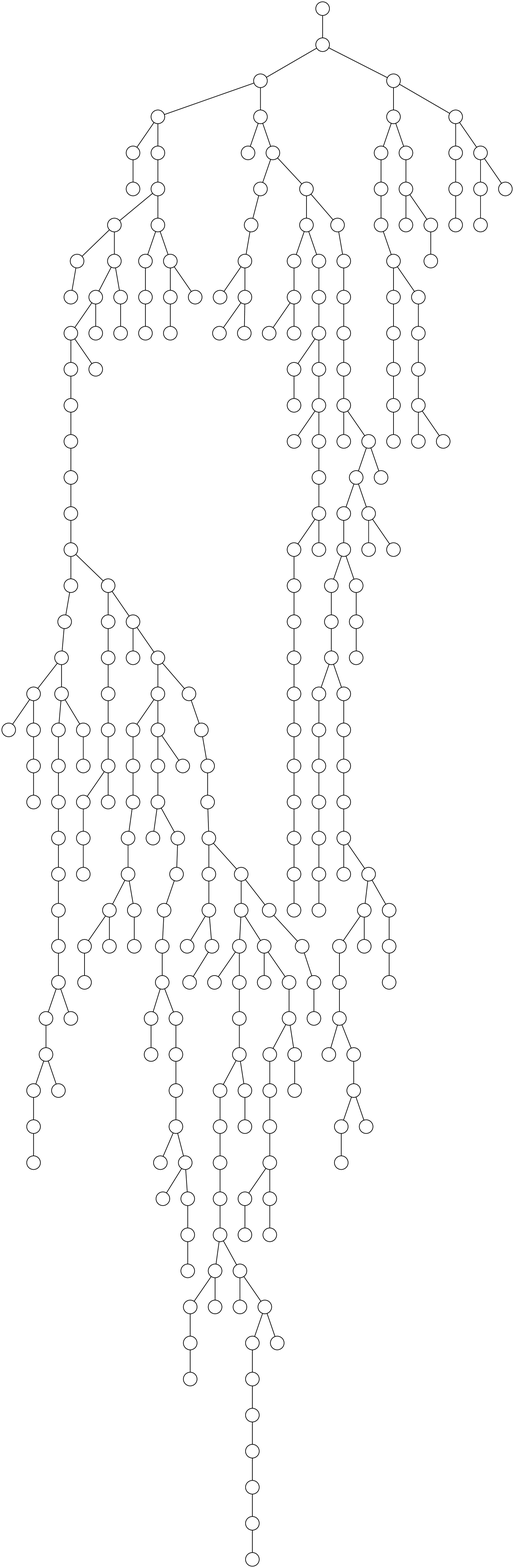}}
\Fig{\includegraphics[scale=\FigRatio]{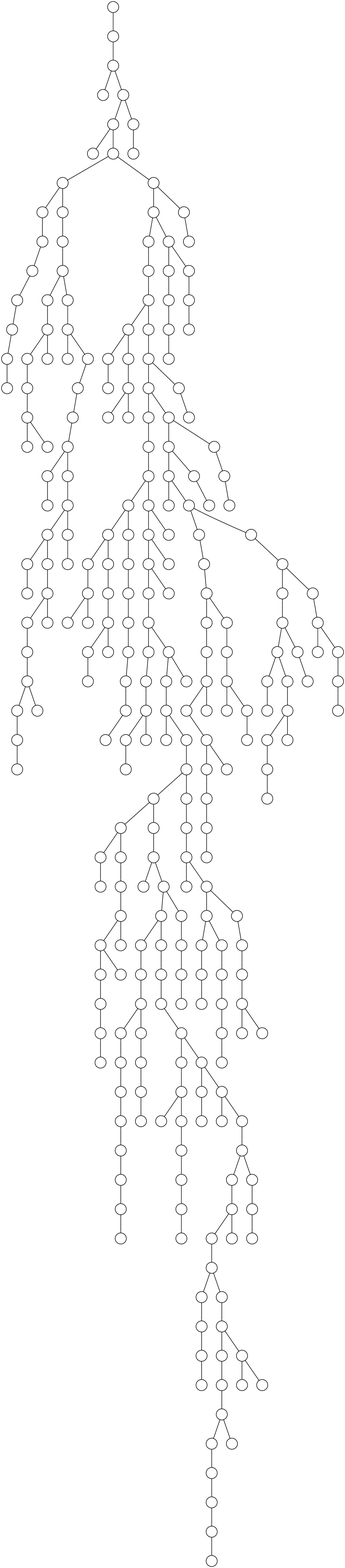}}
\Fig{\includegraphics[scale=\FigRatio]{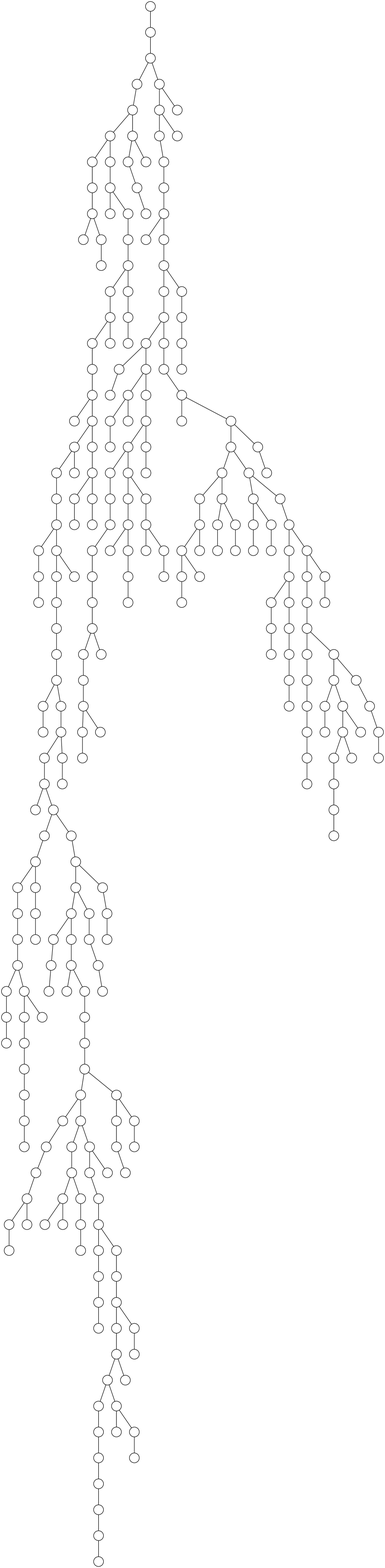}}
\Fig{\includegraphics[scale=\FigRatio]{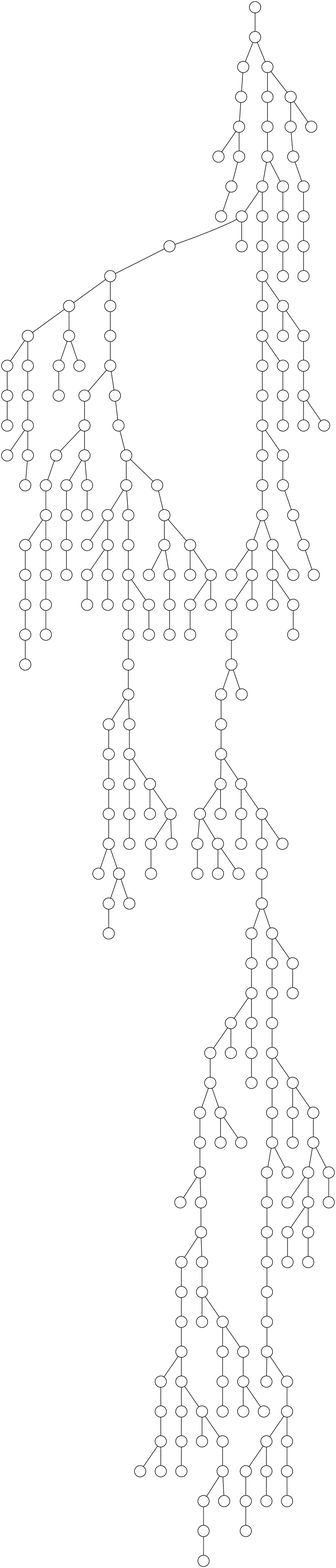}}
\Fig{\includegraphics[scale=\FigRatio]{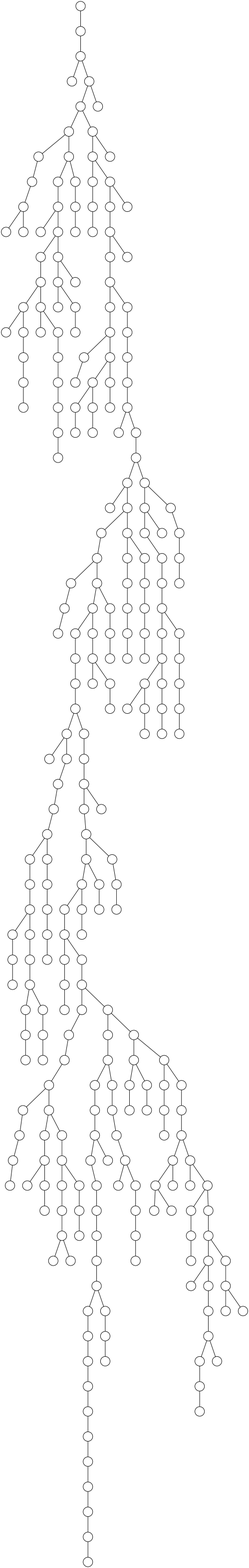}}
\Fig{\includegraphics[scale=\FigRatio]{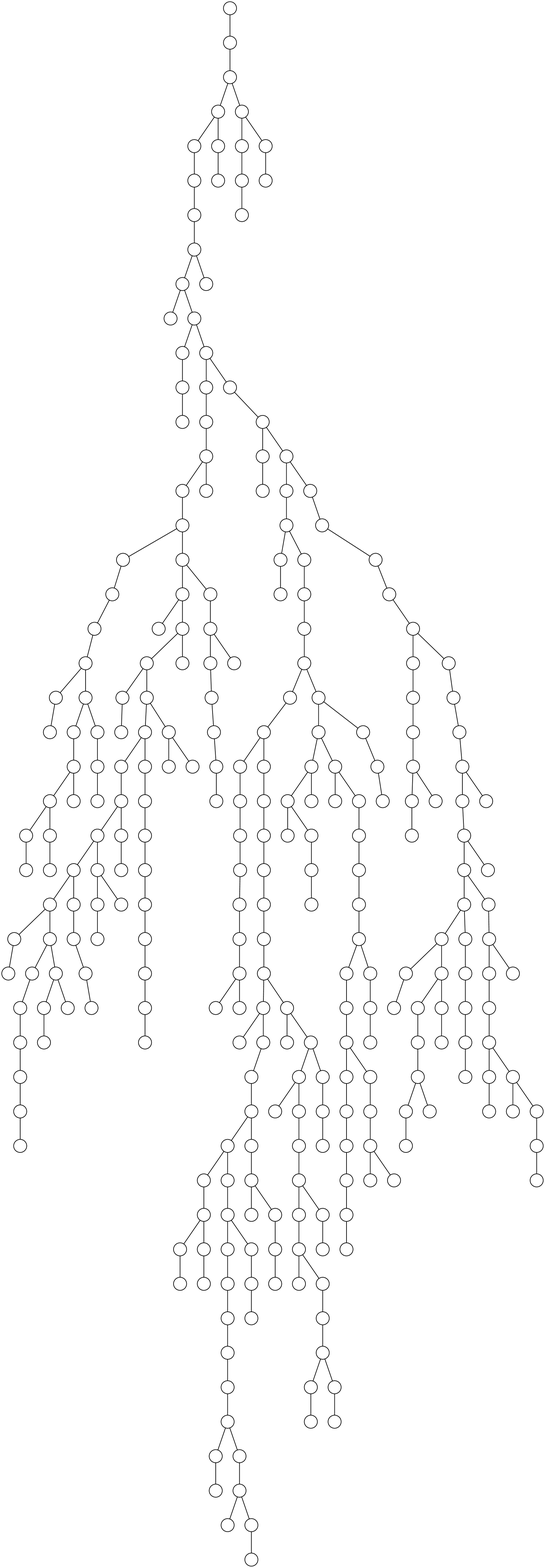}}
\Fig{\includegraphics[scale=\FigRatio]{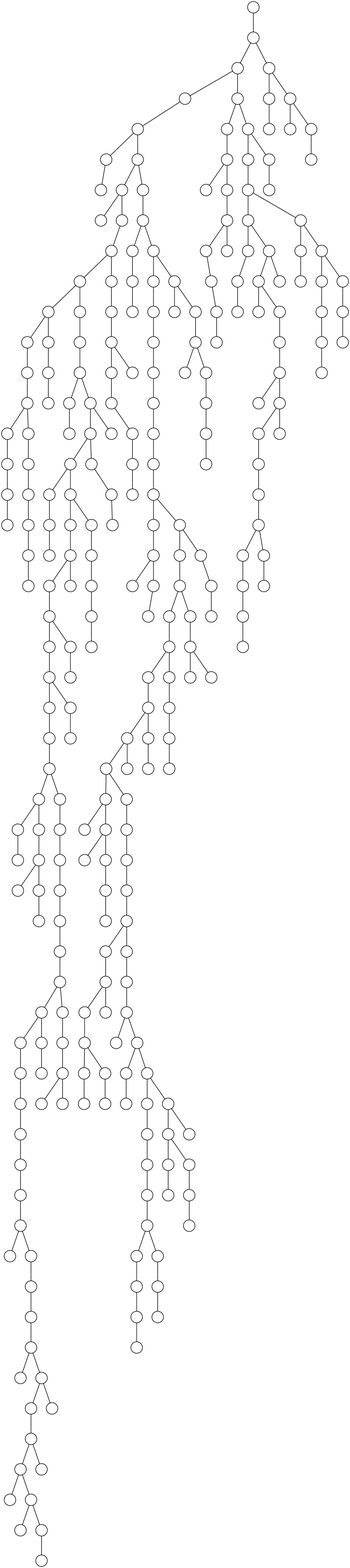}}
\Fig{\includegraphics[scale=\FigRatio]{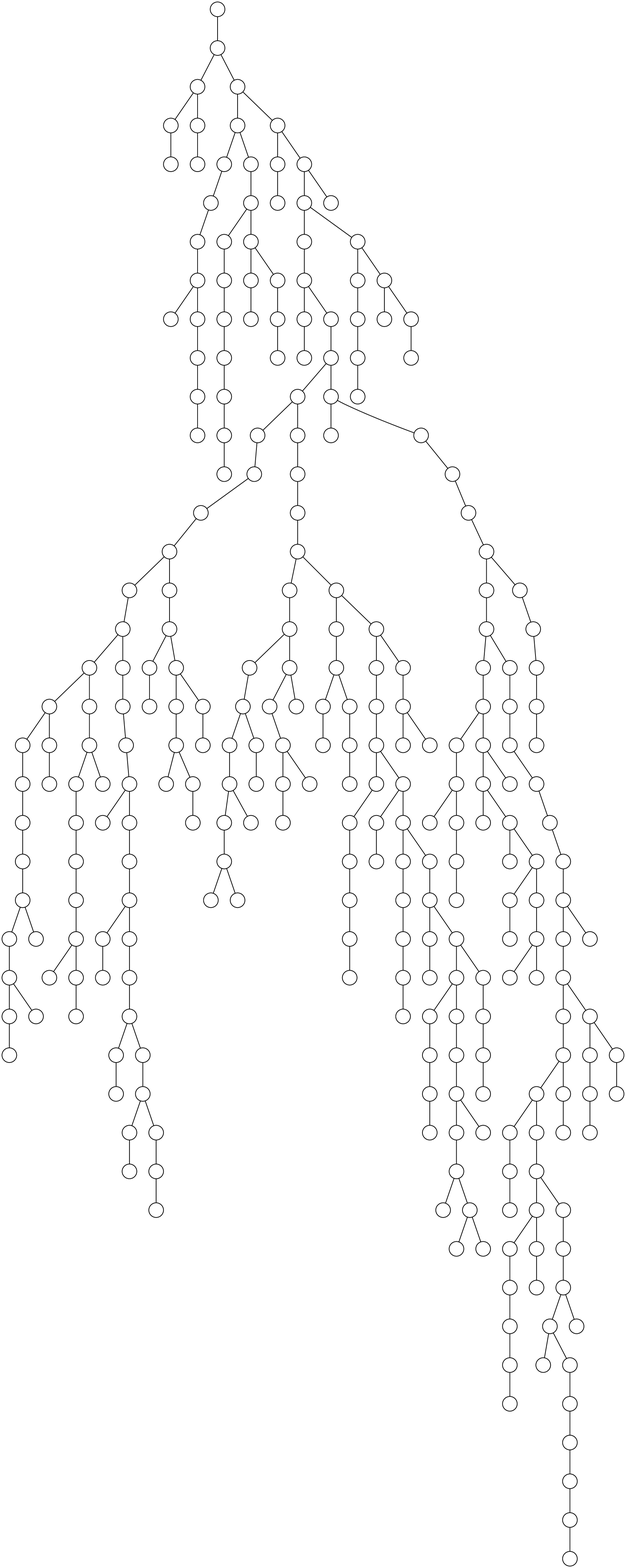}}
\Fig{\includegraphics[scale=\FigRatio]{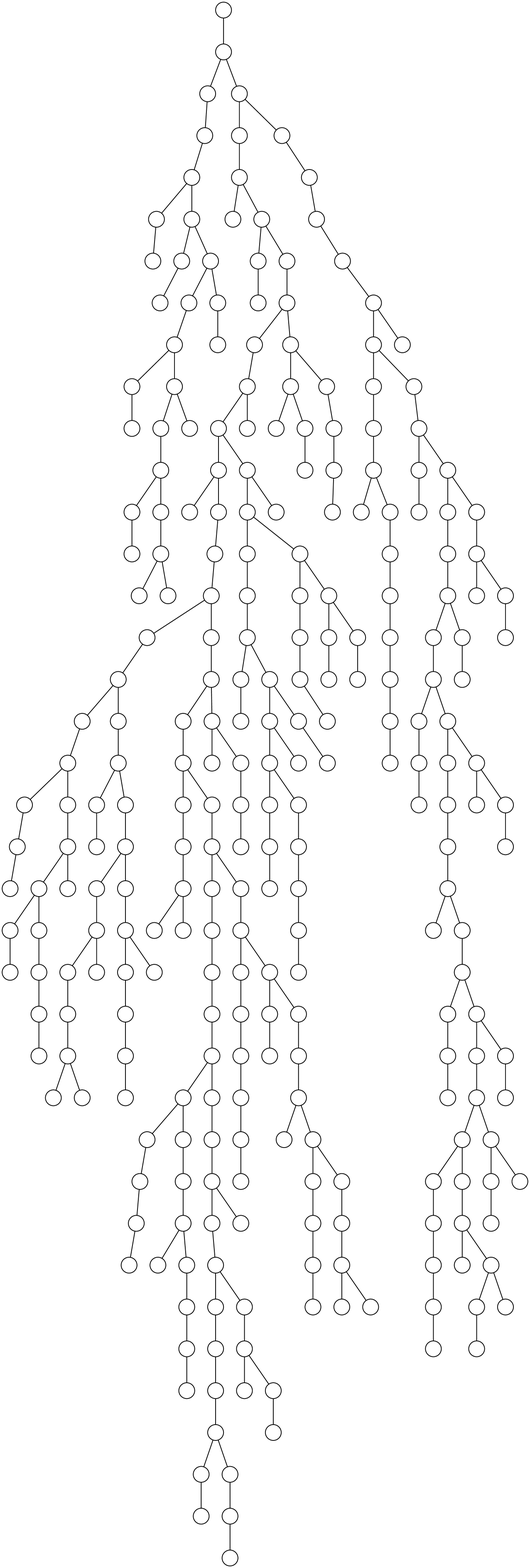}}
\Fig{\includegraphics[scale=\FigRatio]{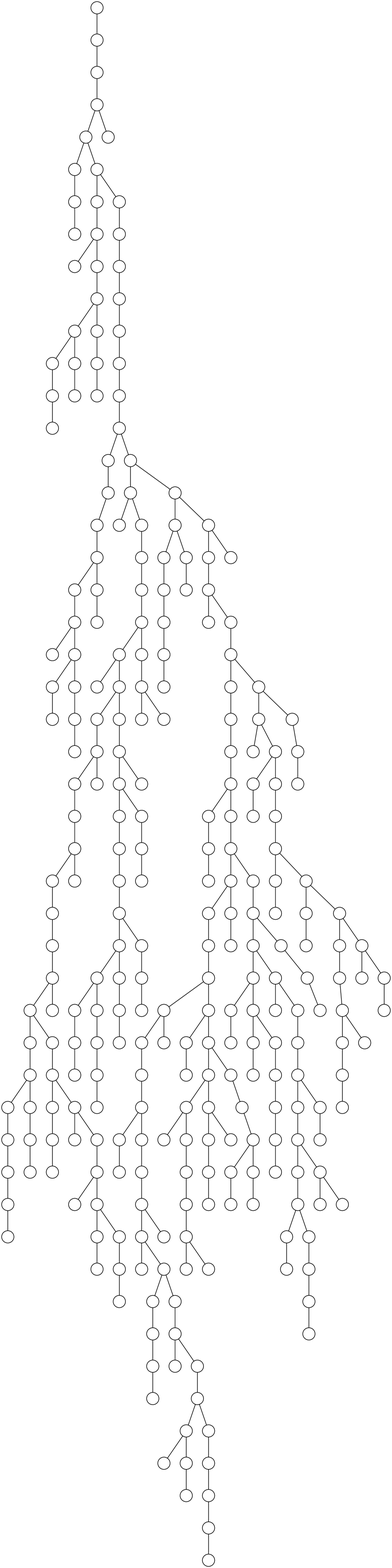}}
\Fig{\includegraphics[scale=\FigRatio]{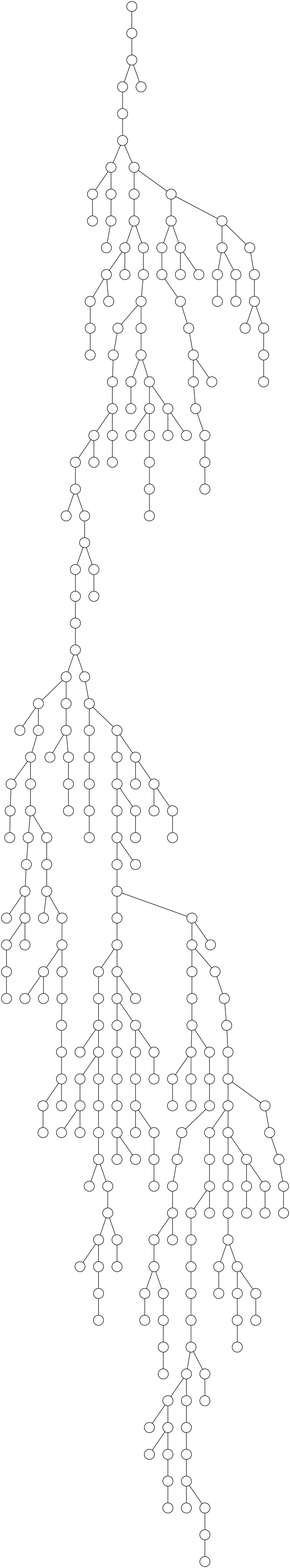}}
\Fig{\includegraphics[scale=\FigRatio]{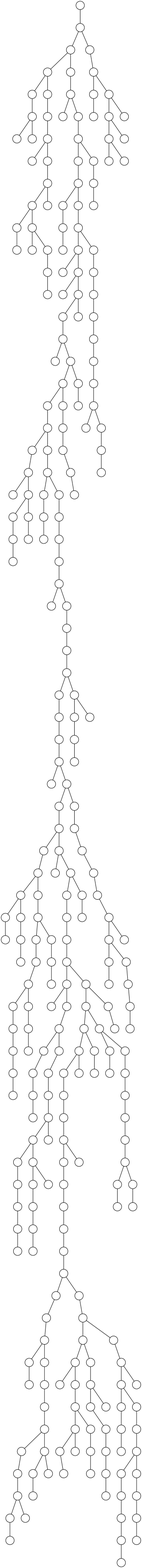}}
\Fig{\includegraphics[scale=\FigRatio]{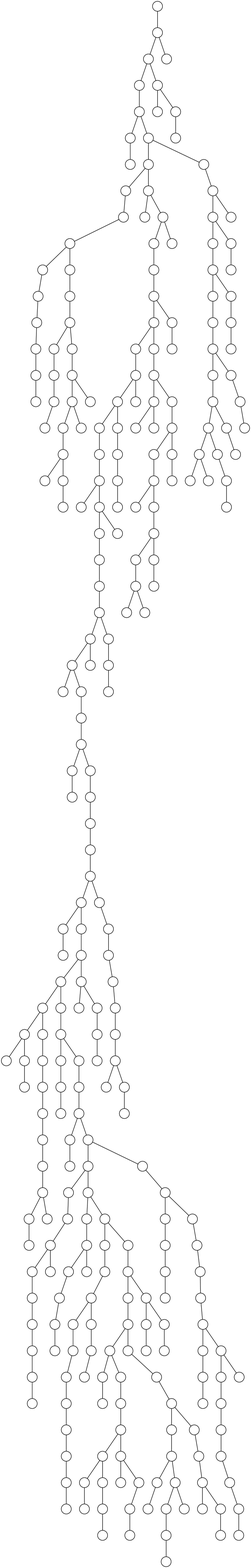}}
\Fig{\includegraphics[scale=\FigRatio]{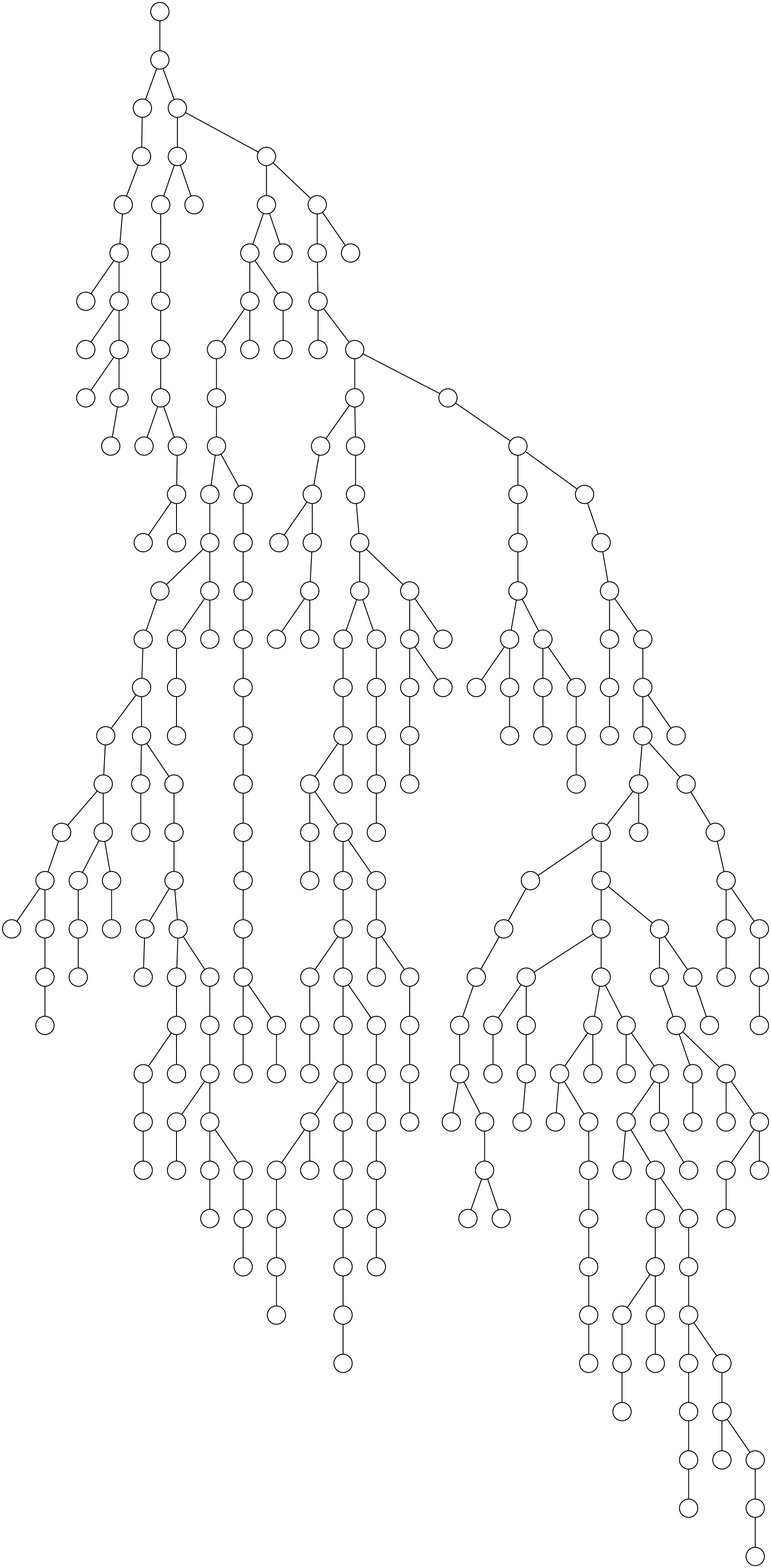}}
\\
$\W=1/4\;\Leftrightarrow\;f_c = 11.11\ldots\%$\\
\Fig{\includegraphics[scale=\FigRatio]{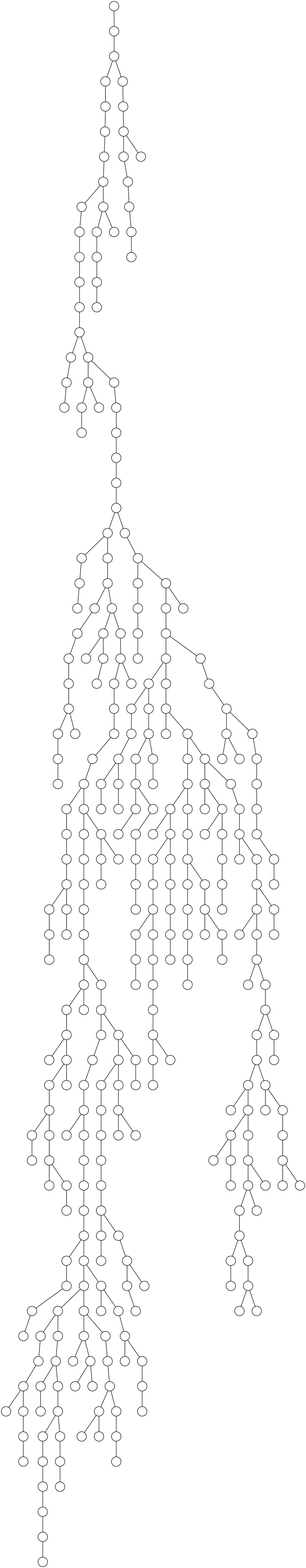}}
\Fig{\includegraphics[scale=\FigRatio]{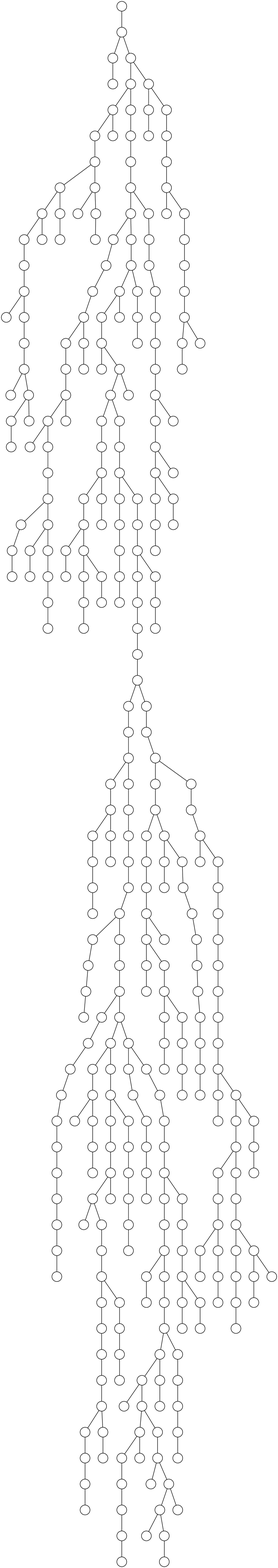}}
\Fig{\includegraphics[scale=\FigRatio]{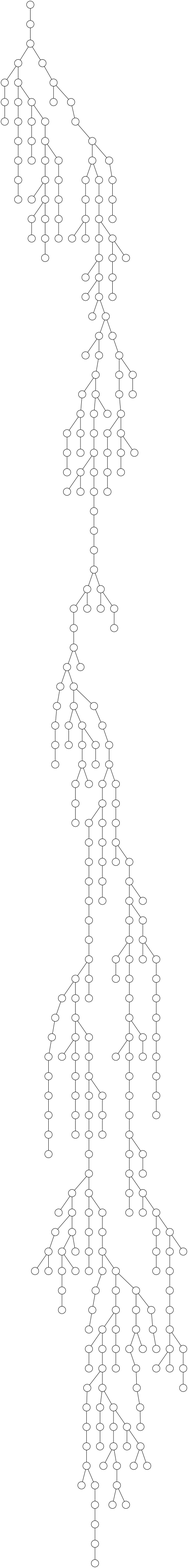}}
\Fig{\includegraphics[scale=\FigRatio]{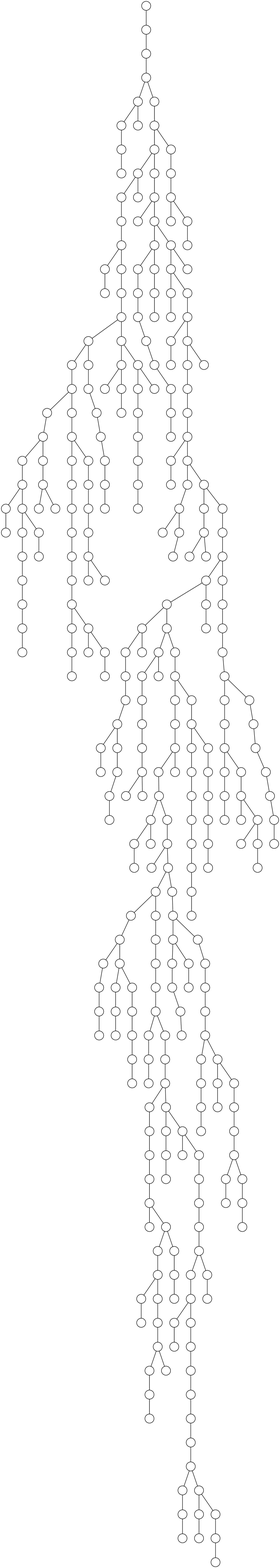}}
\Fig{\includegraphics[scale=\FigRatio]{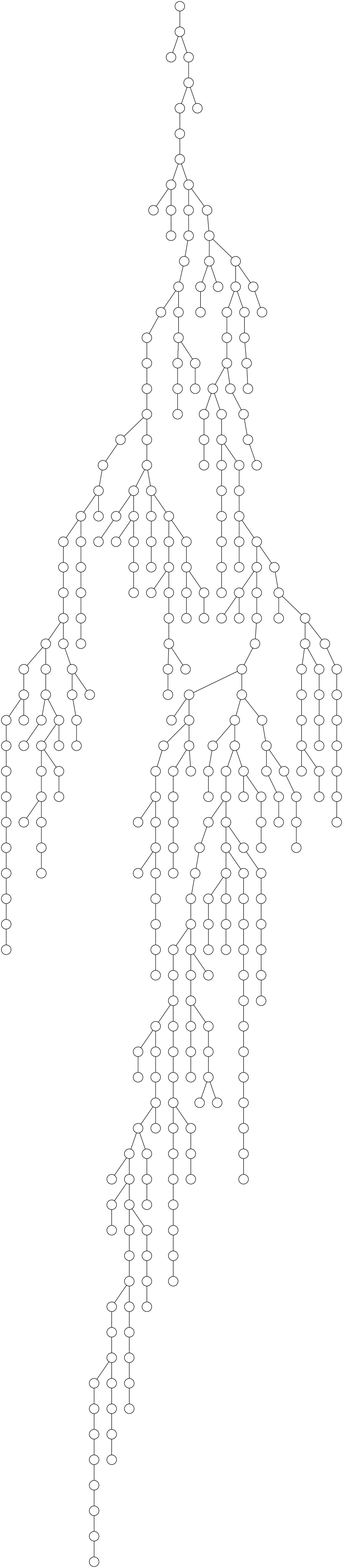}}
\Fig{\includegraphics[scale=\FigRatio]{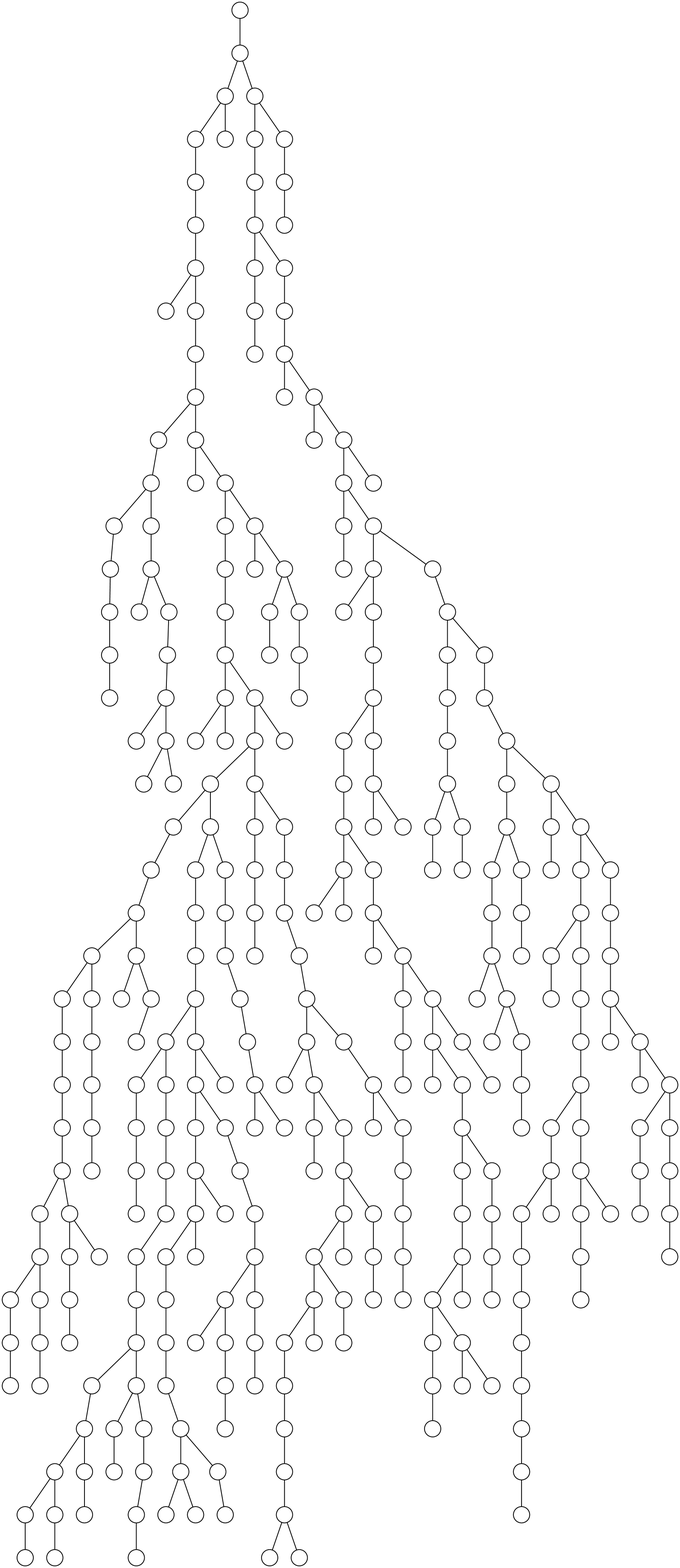}}
\Fig{\includegraphics[scale=\FigRatio]{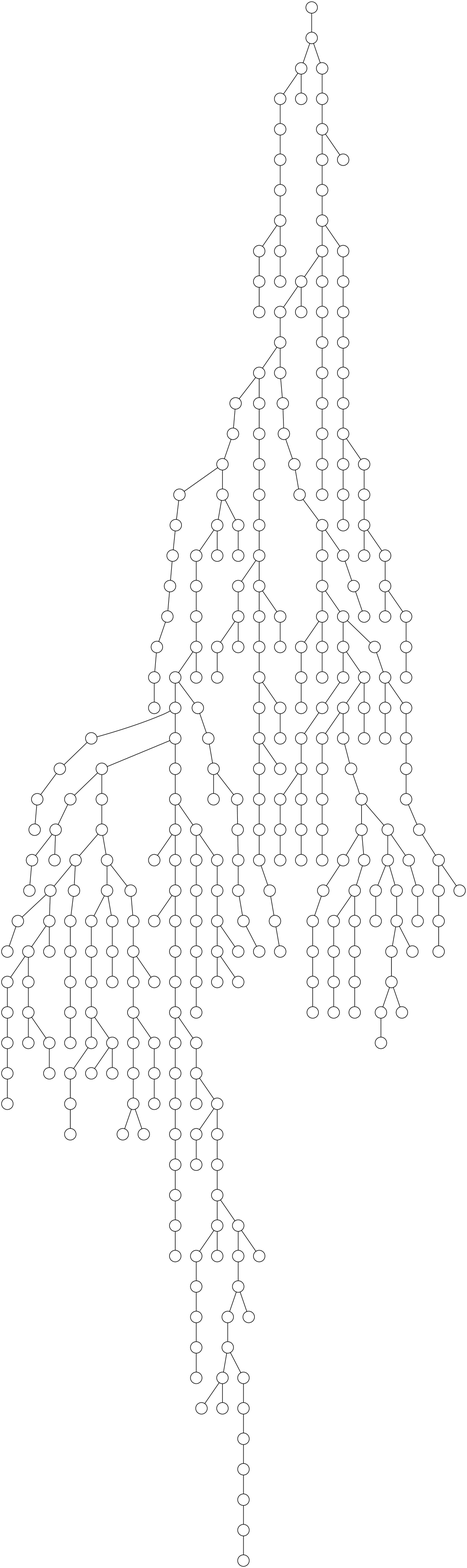}}
\Fig{\includegraphics[scale=\FigRatio]{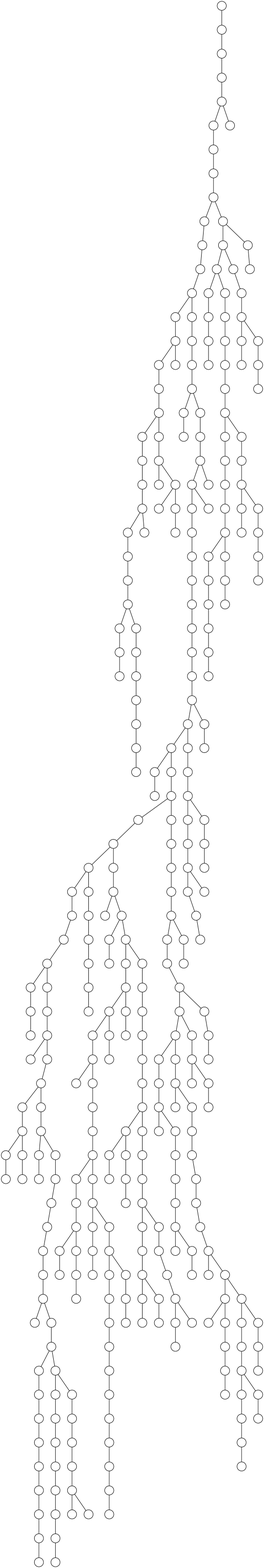}}
\Fig{\includegraphics[scale=\FigRatio]{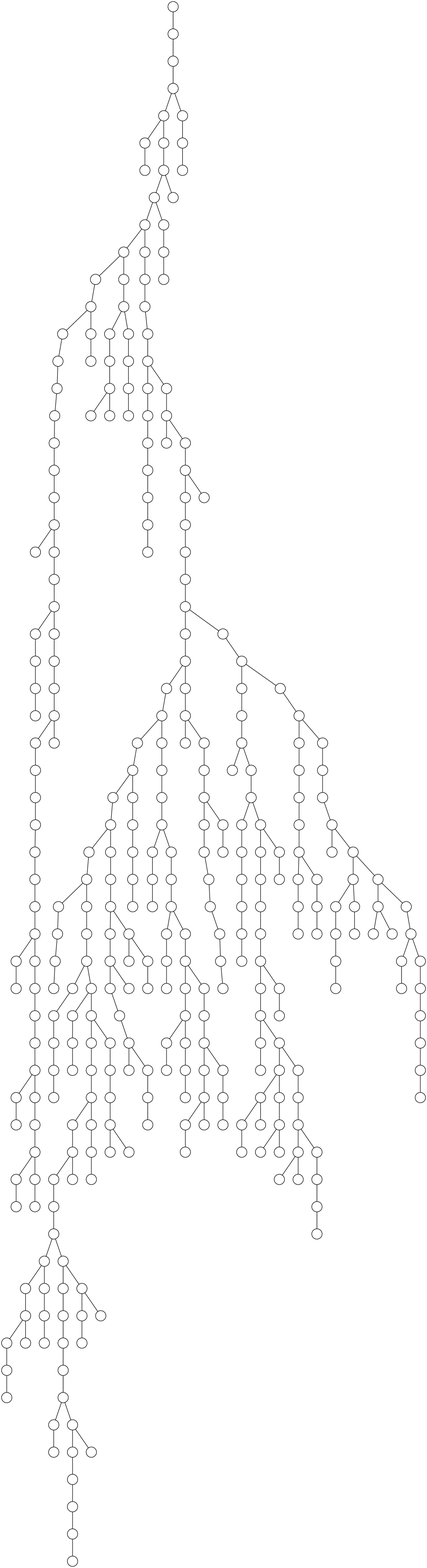}}
\Fig{\includegraphics[scale=\FigRatio]{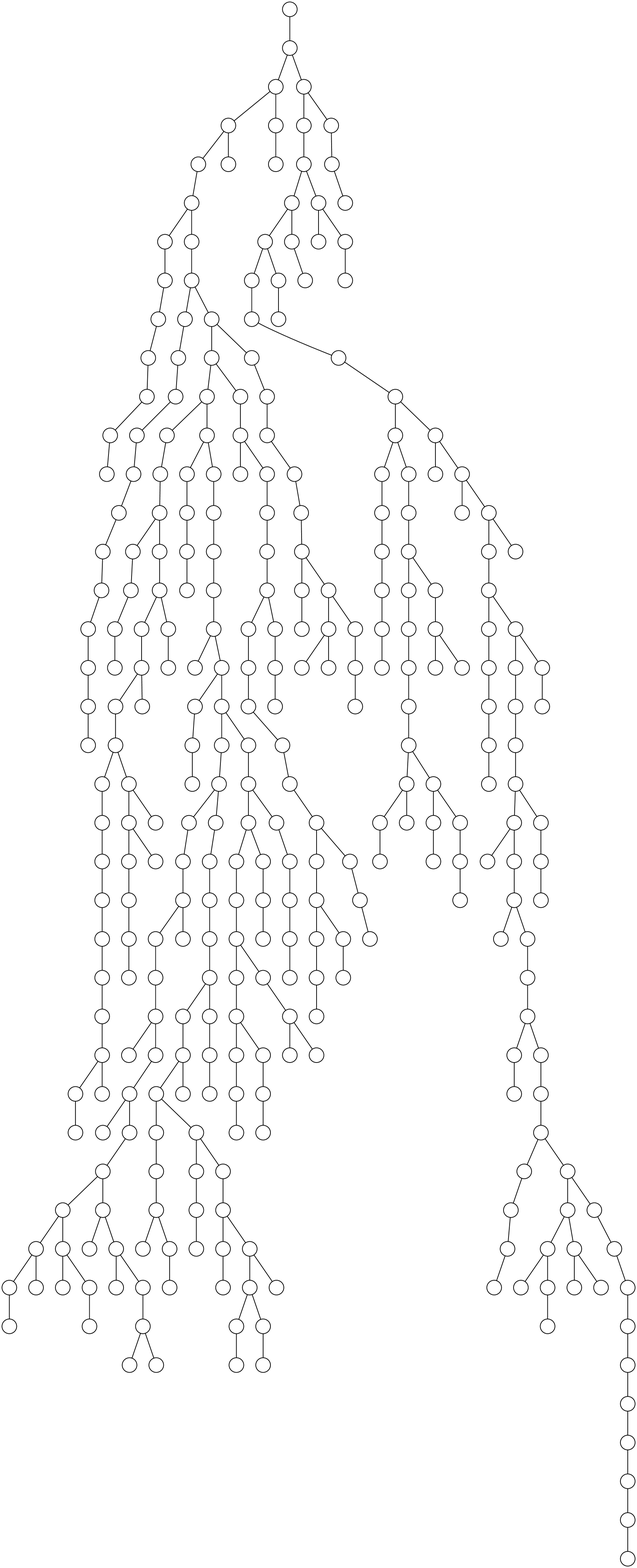}}
\Fig{\includegraphics[scale=\FigRatio]{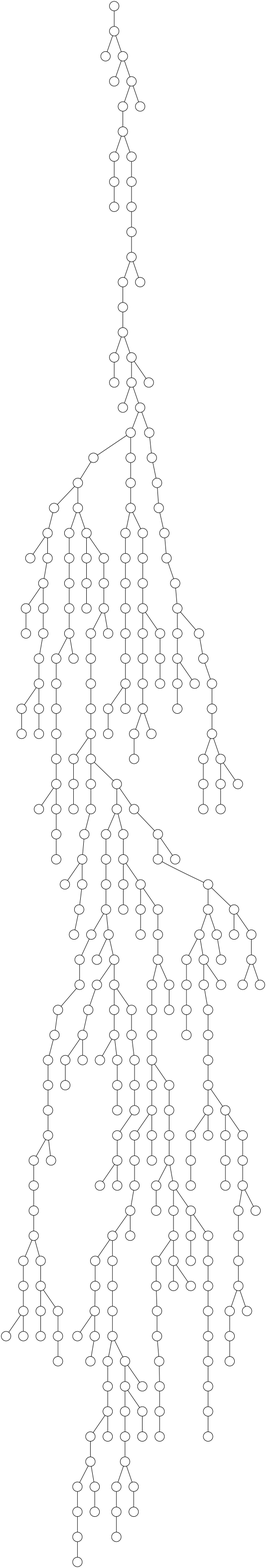}}
\Fig{\includegraphics[scale=\FigRatio]{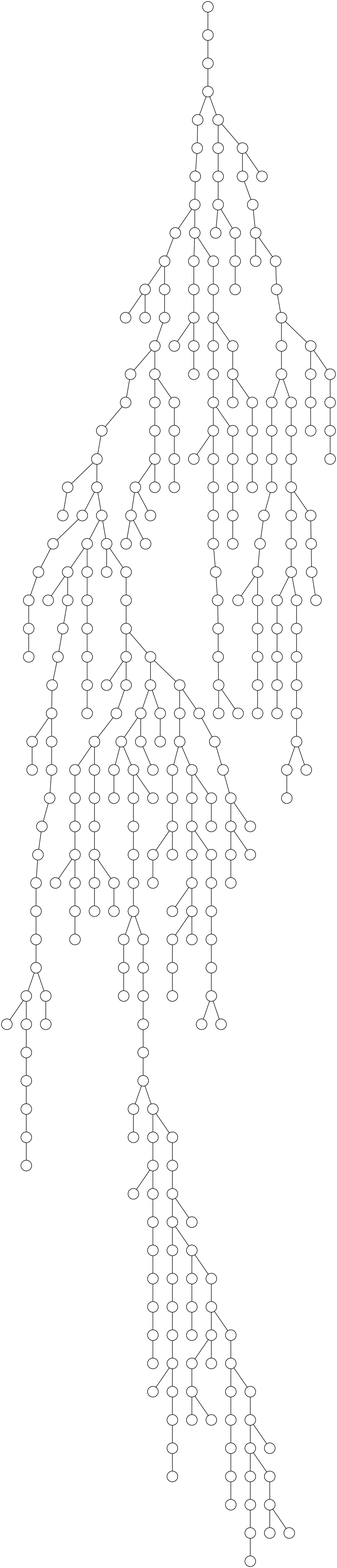}}
\Fig{\includegraphics[scale=\FigRatio]{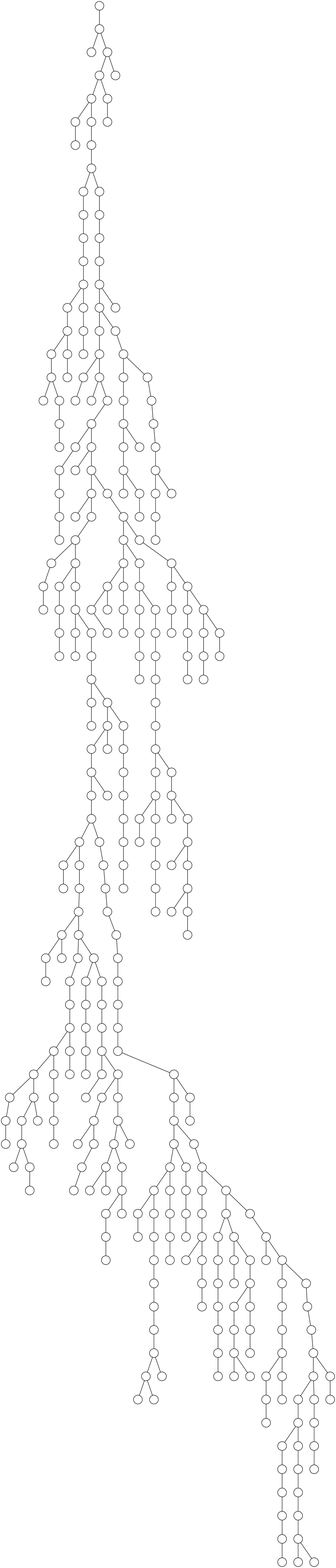}}
\Fig{\includegraphics[scale=\FigRatio]{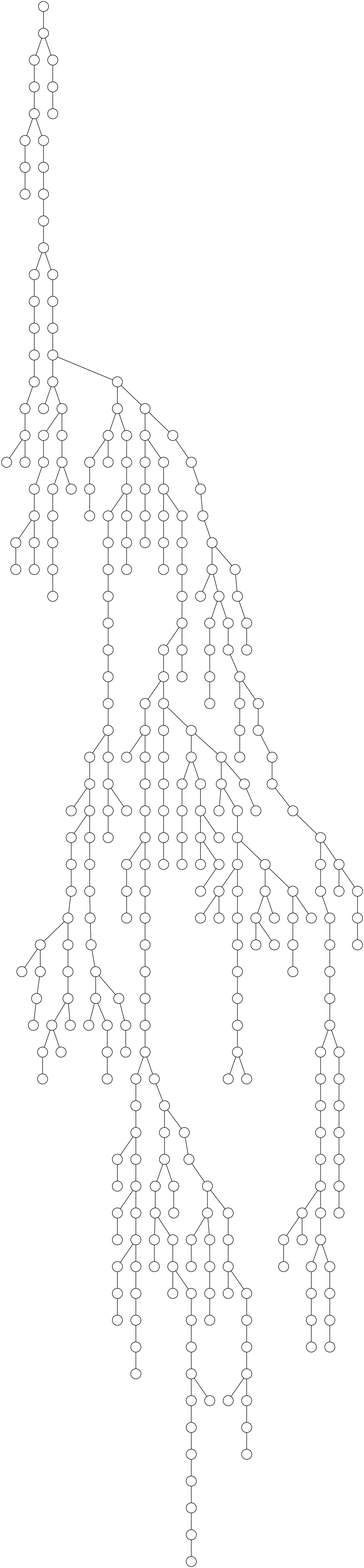}}
\\
$\W=1\;\Leftrightarrow\;f_c = 33.33\ldots\%$\\
\Fig{\includegraphics[scale=\FigRatio]{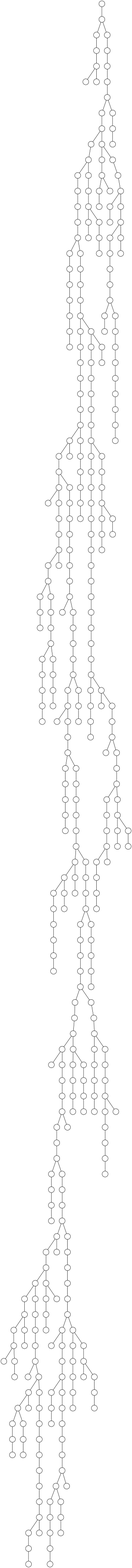}}
\Fig{\includegraphics[scale=\FigRatio]{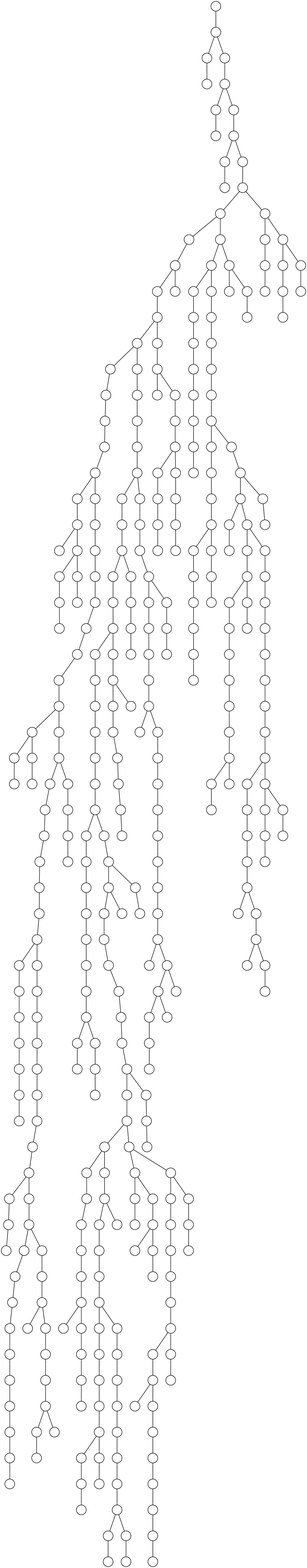}}
\Fig{\includegraphics[scale=\FigRatio]{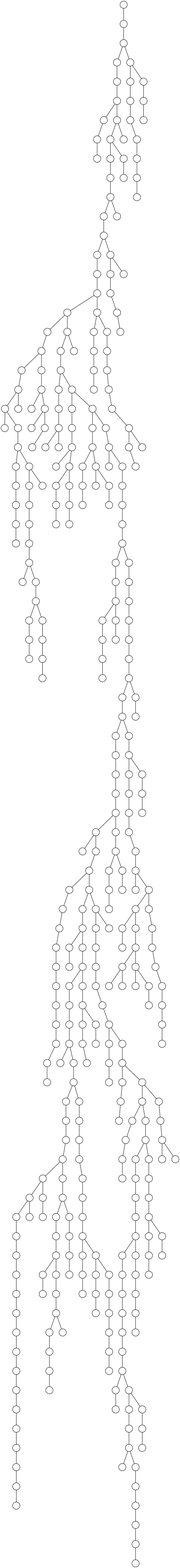}}
\Fig{\includegraphics[scale=\FigRatio]{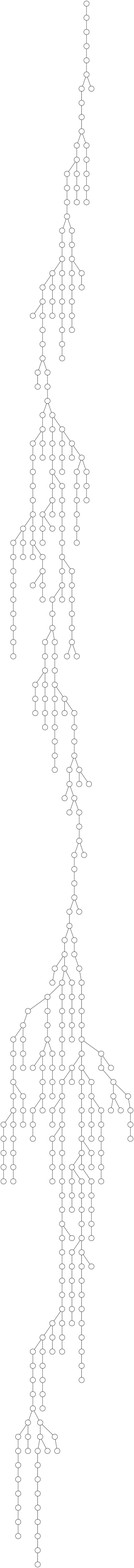}}
\Fig{\includegraphics[scale=\FigRatio]{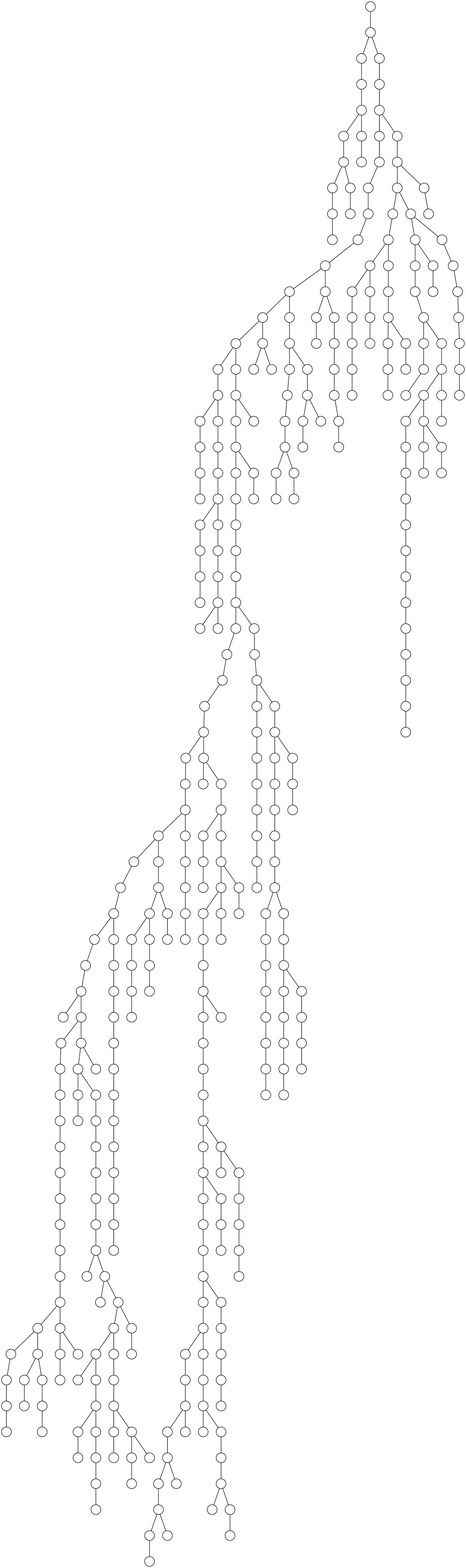}}
\Fig{\includegraphics[scale=\FigRatio]{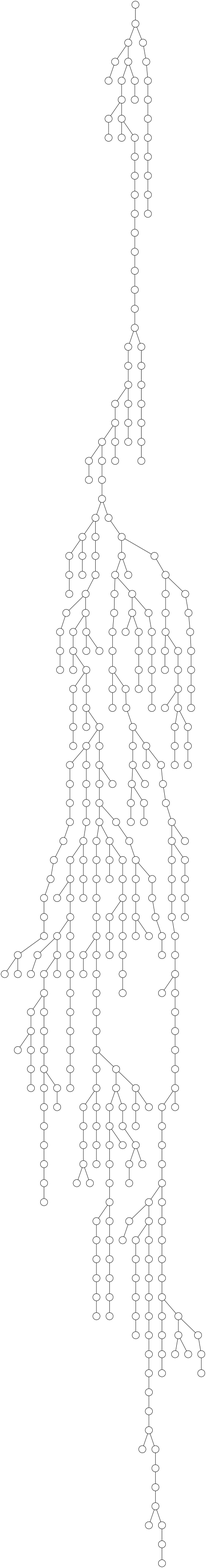}}
\Fig{\includegraphics[scale=\FigRatio]{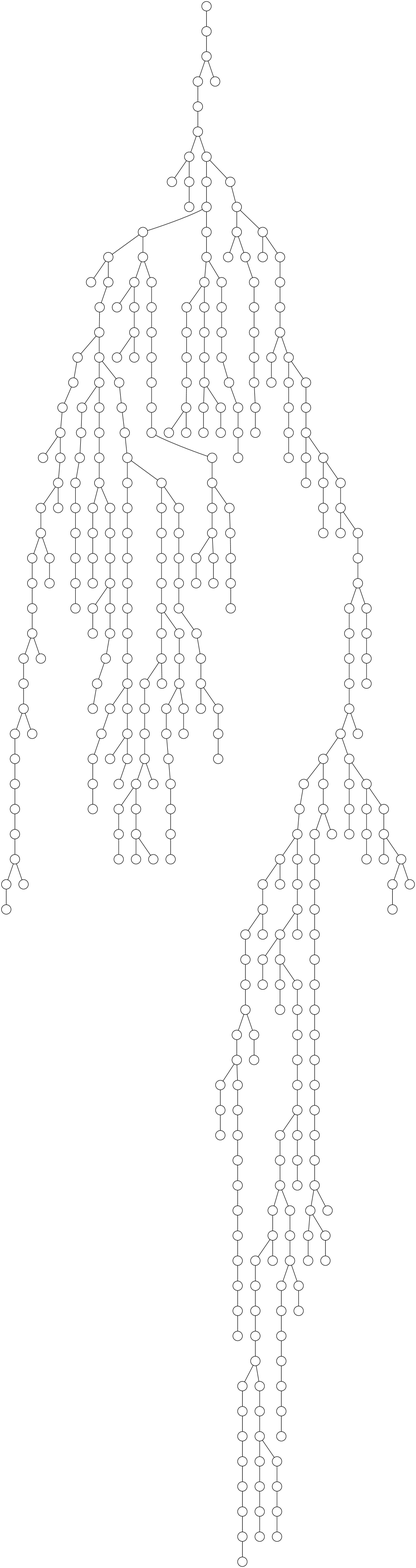}}
\Fig{\includegraphics[scale=\FigRatio]{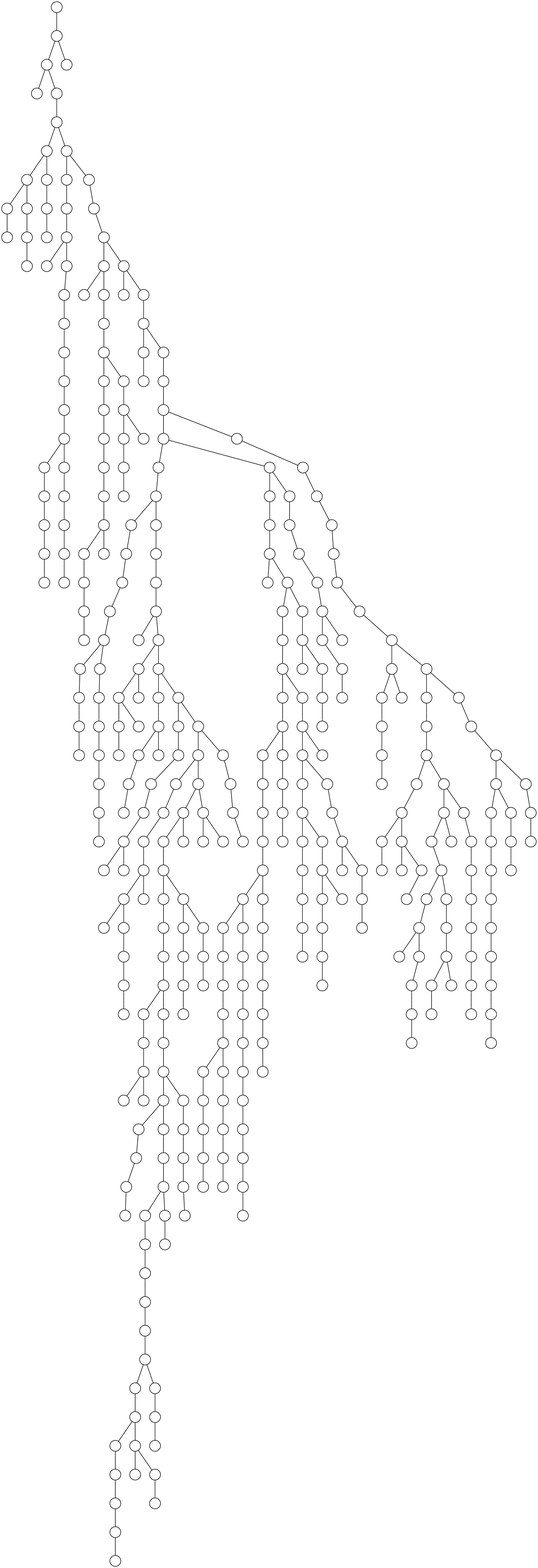}}
\Fig{\includegraphics[scale=\FigRatio]{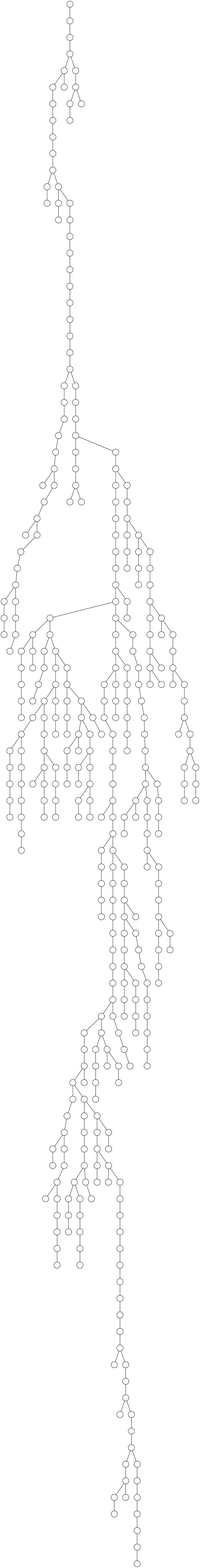}}
\Fig{\includegraphics[scale=\FigRatio]{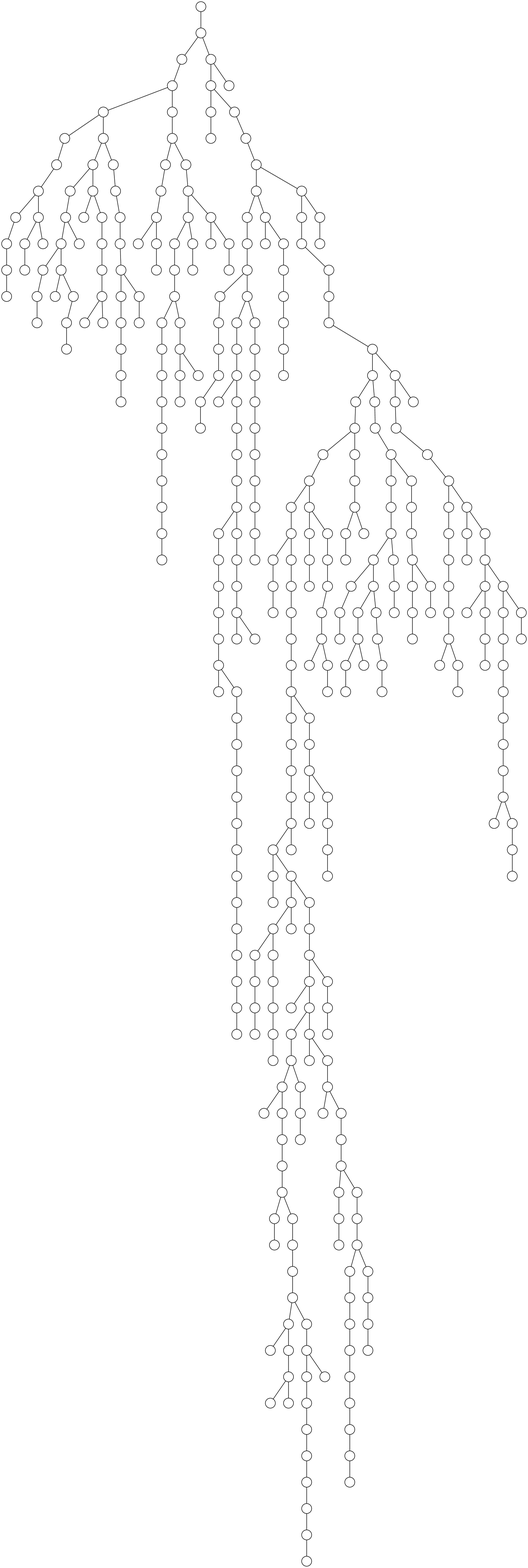}}
\Fig{\includegraphics[scale=\FigRatio]{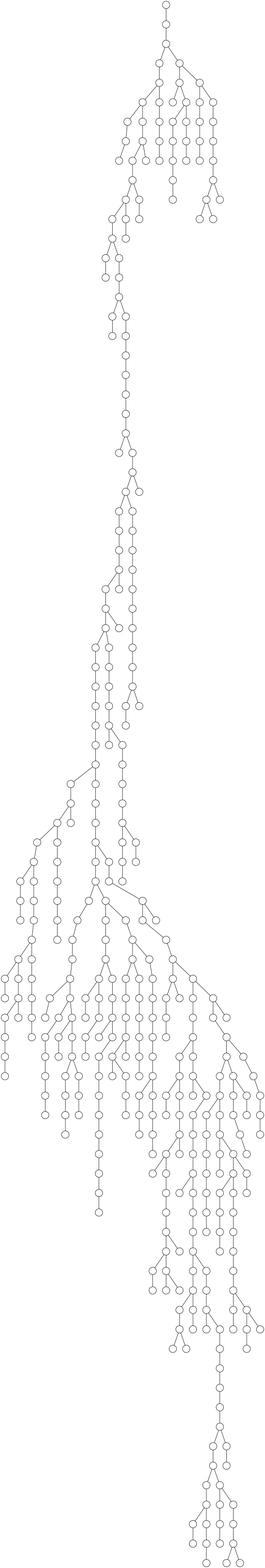}}
\Fig{\includegraphics[scale=\FigRatio]{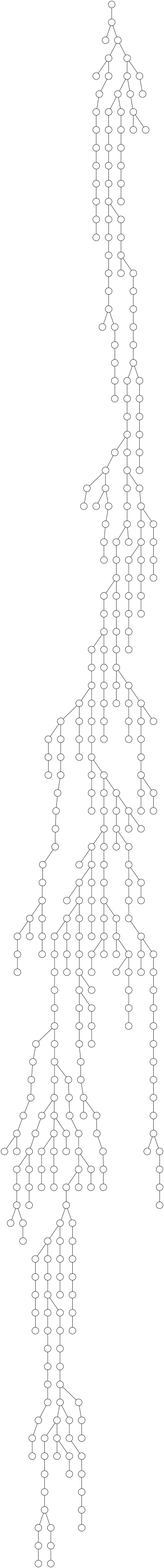}}
\Fig{\includegraphics[scale=\FigRatio]{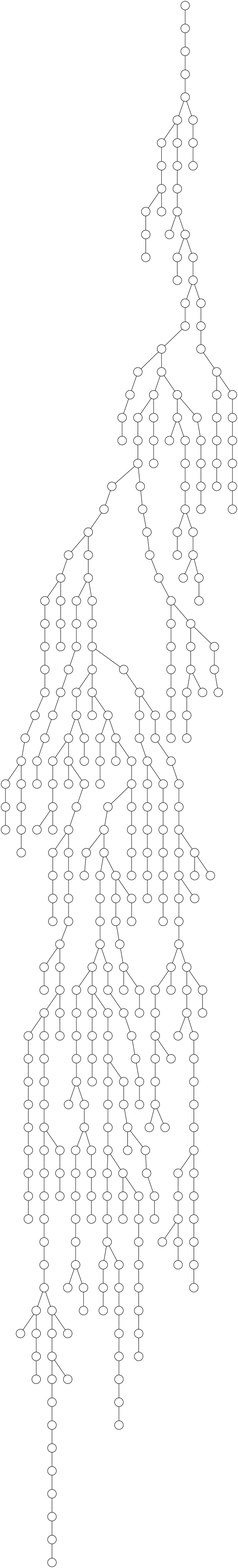}}
\\
$\W=2\;\Leftrightarrow\;f_c = 50\%$\\
\Fig{\includegraphics[scale=\FigRatio]{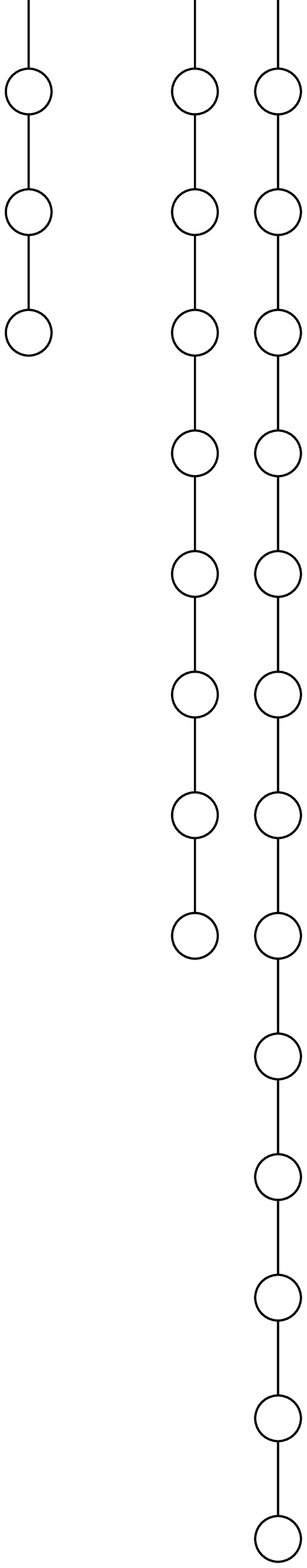}}
\Fig{\includegraphics[scale=\FigRatio]{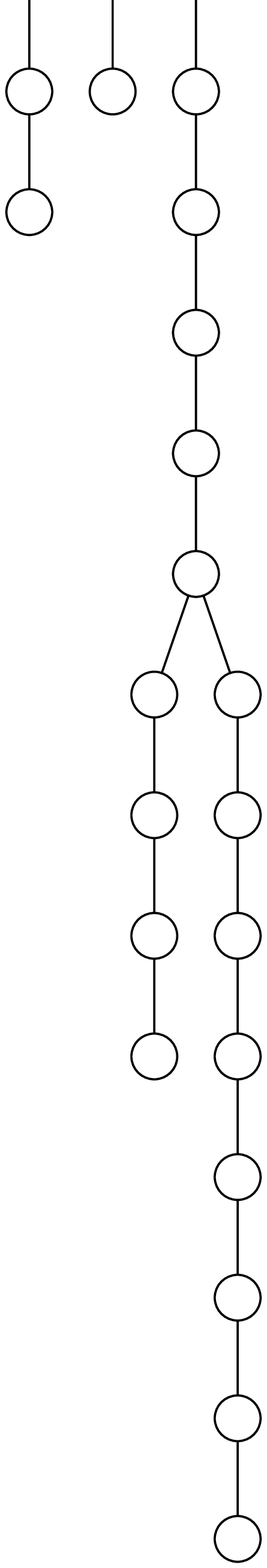}}
\Fig{\includegraphics[scale=\FigRatio]{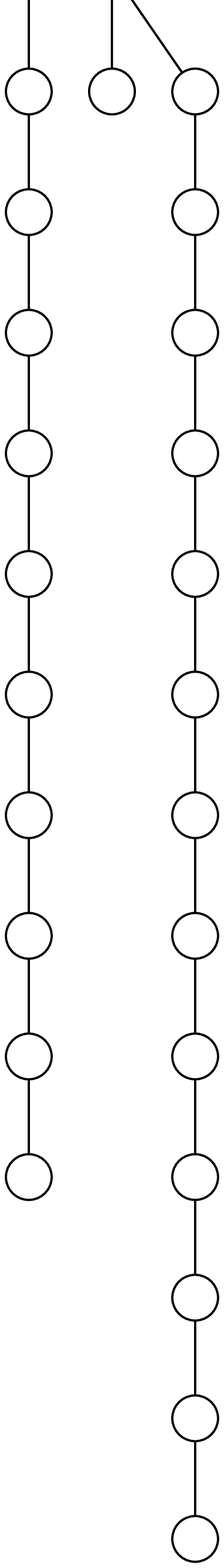}}
\Fig{\includegraphics[scale=\FigRatio]{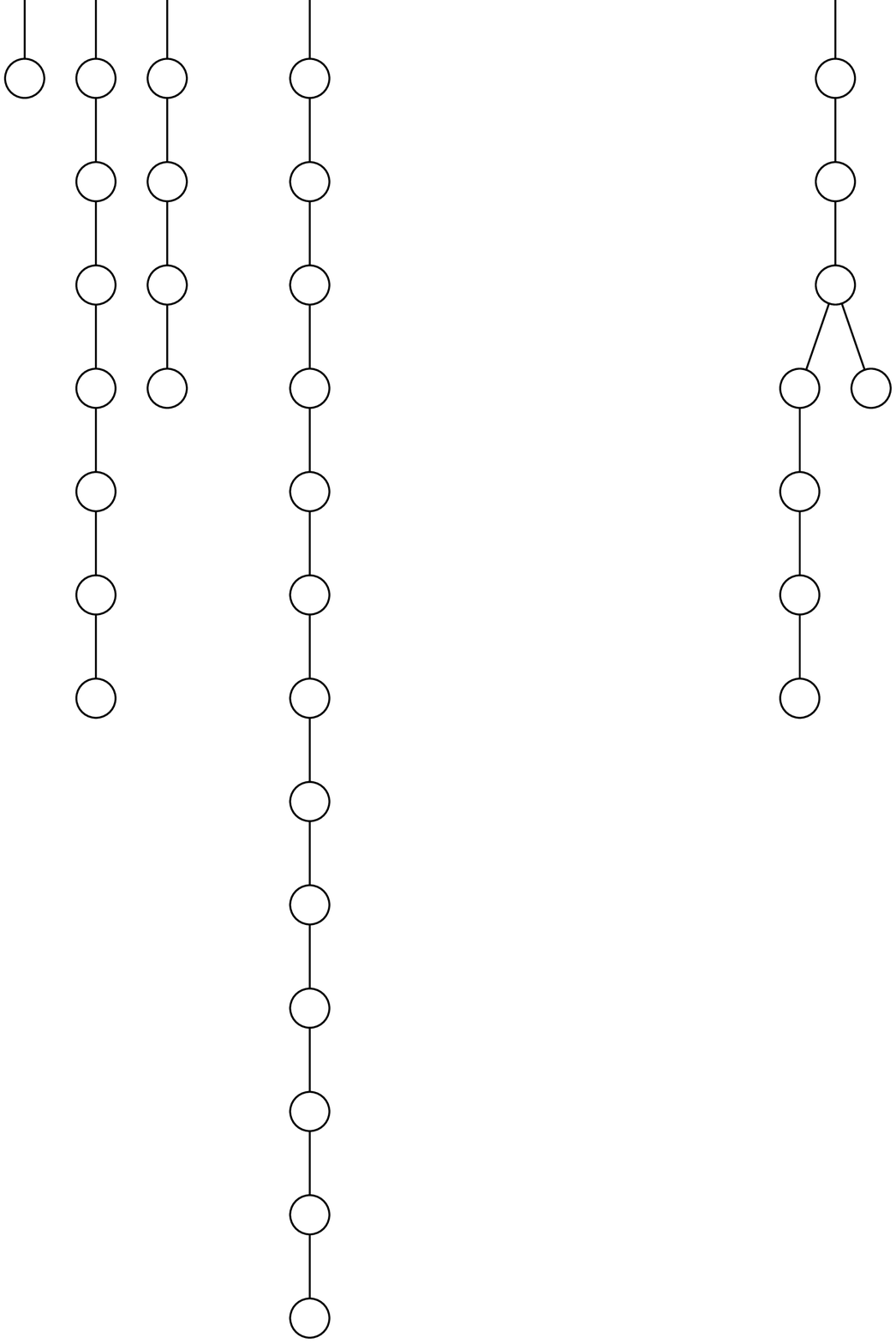}}
\Fig{\includegraphics[scale=\FigRatio]{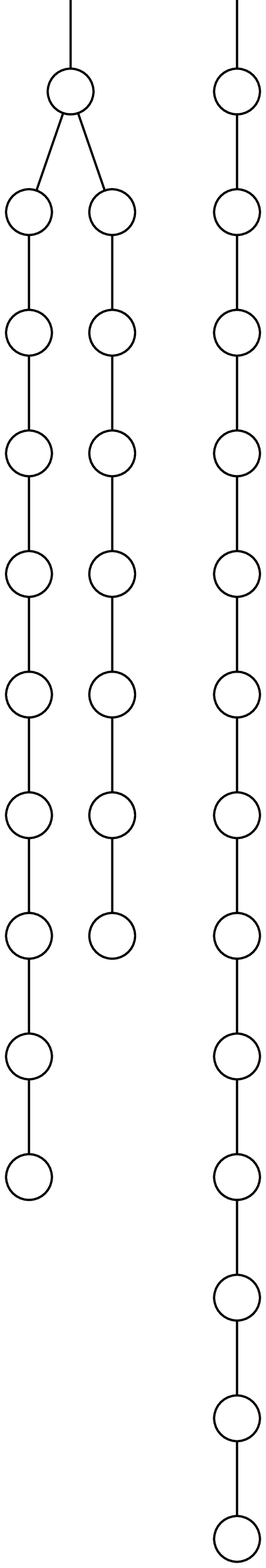}}
\Fig{\includegraphics[scale=\FigRatio]{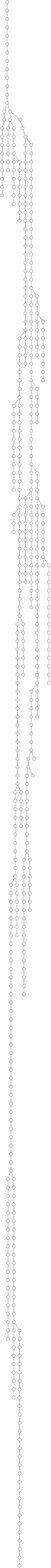}}
\Fig{\includegraphics[scale=\FigRatio]{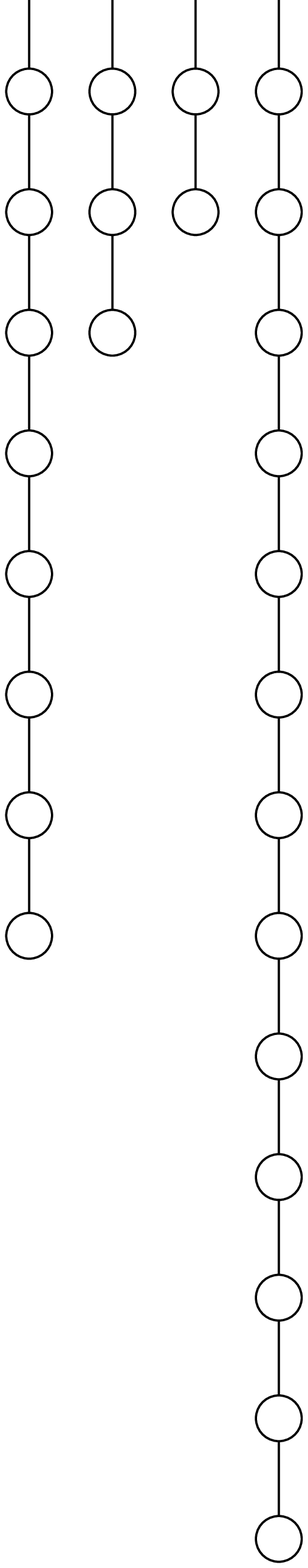}}
\Fig{\includegraphics[scale=\FigRatio]{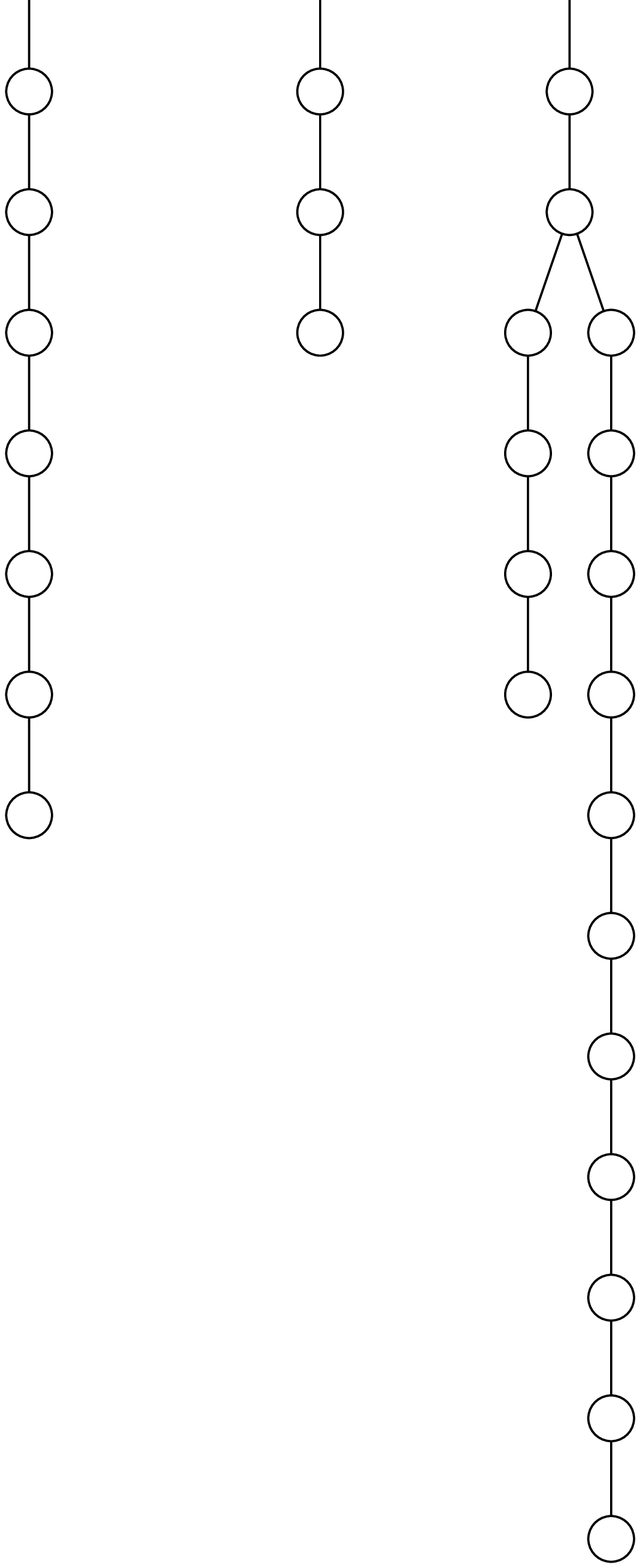}}
\Fig{\includegraphics[scale=\FigRatio]{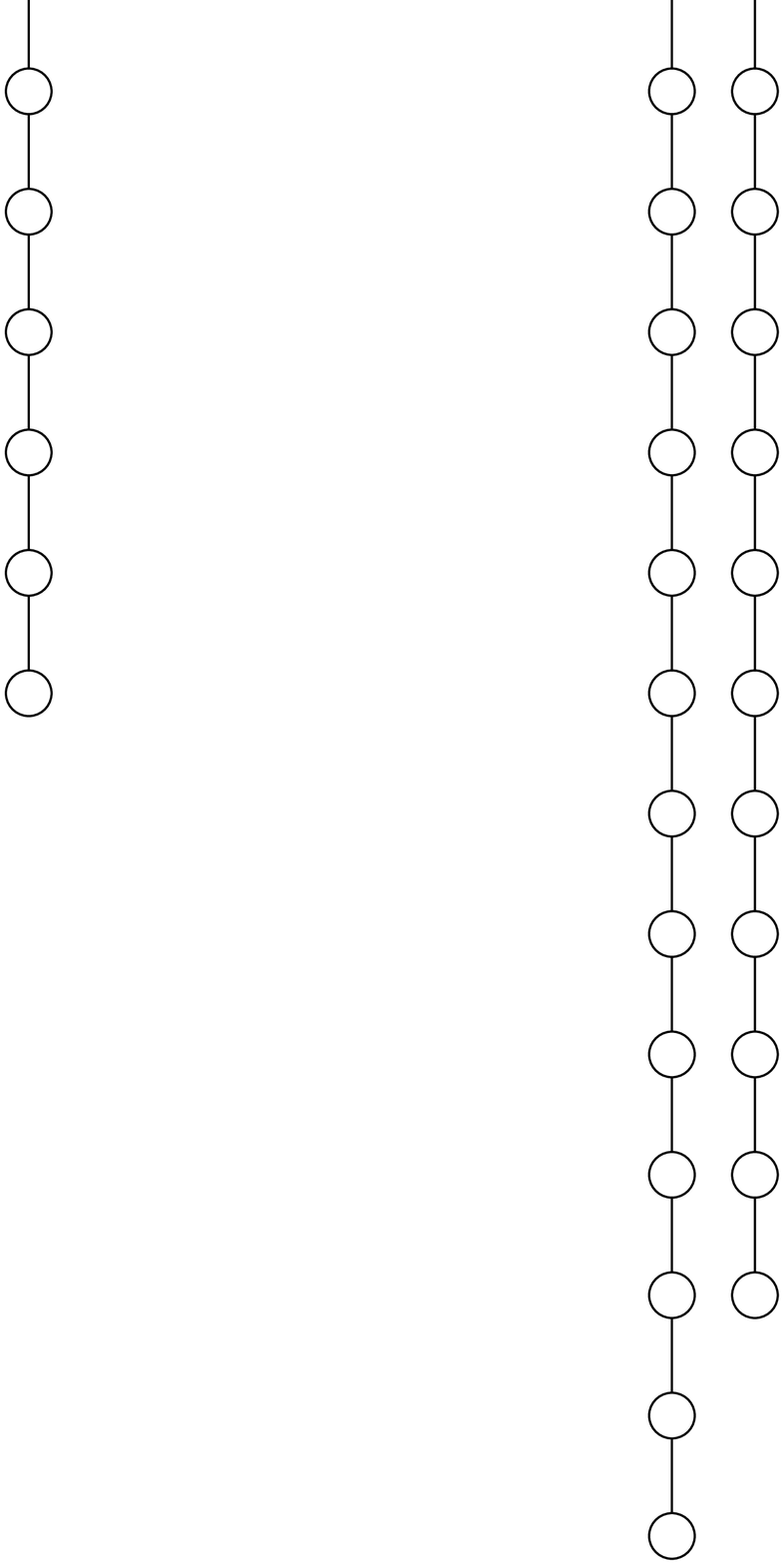}}
\Fig{\includegraphics[scale=\FigRatio]{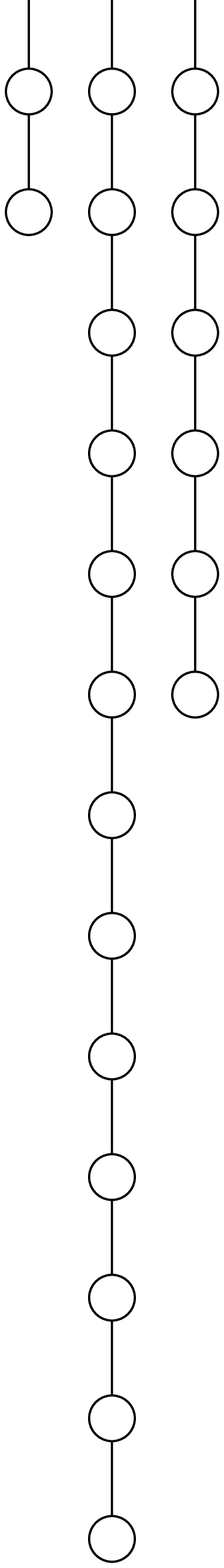}}
\Fig{\includegraphics[scale=\FigRatio]{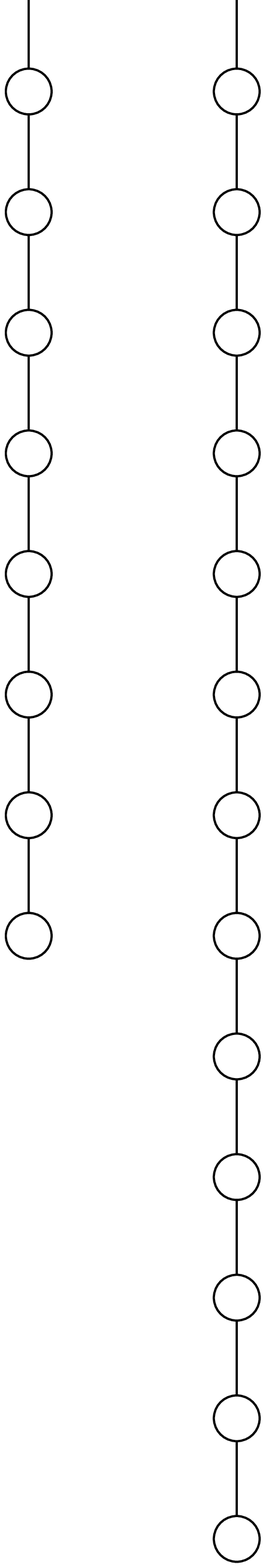}}
\Fig{\includegraphics[scale=\FigRatio]{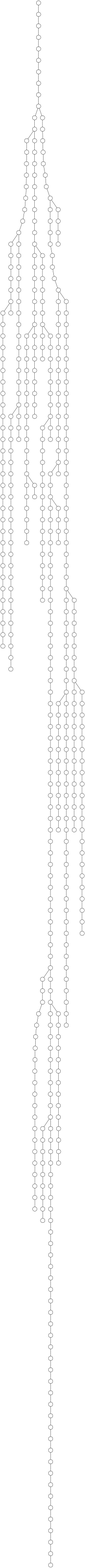}}
\Fig{\includegraphics[scale=\FigRatio]{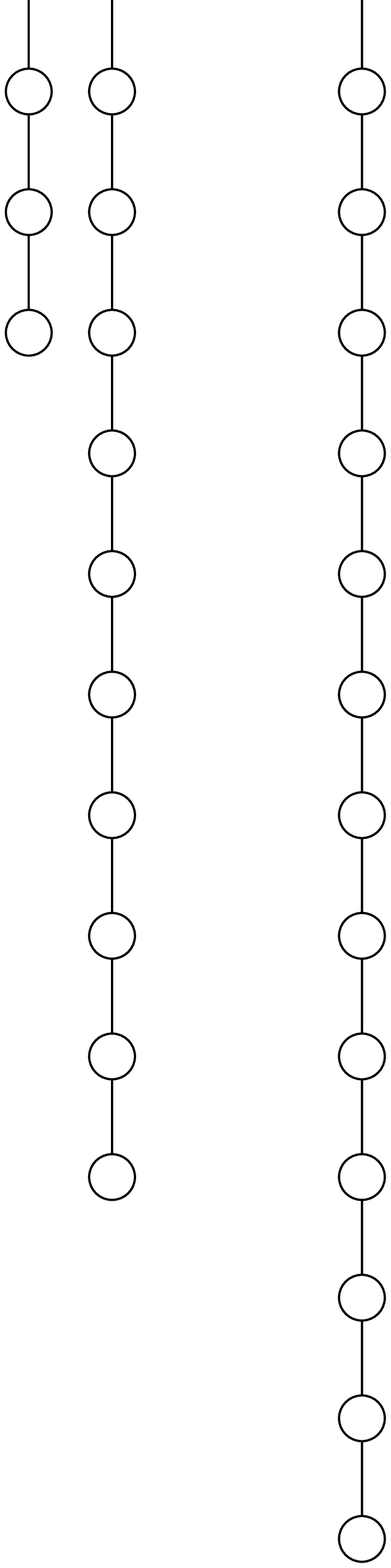}}
\Fig{\includegraphics[scale=\FigRatio]{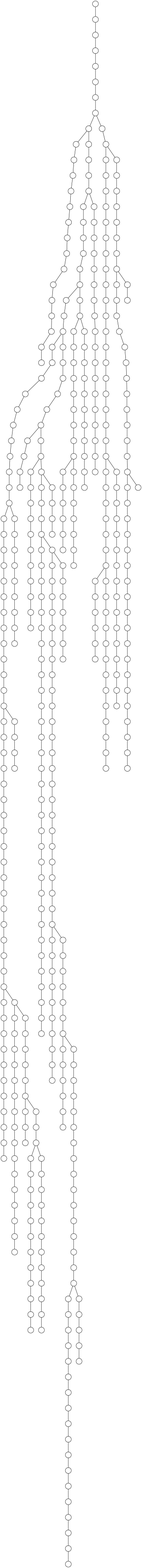}}
\Fig{\includegraphics[scale=\FigRatio]{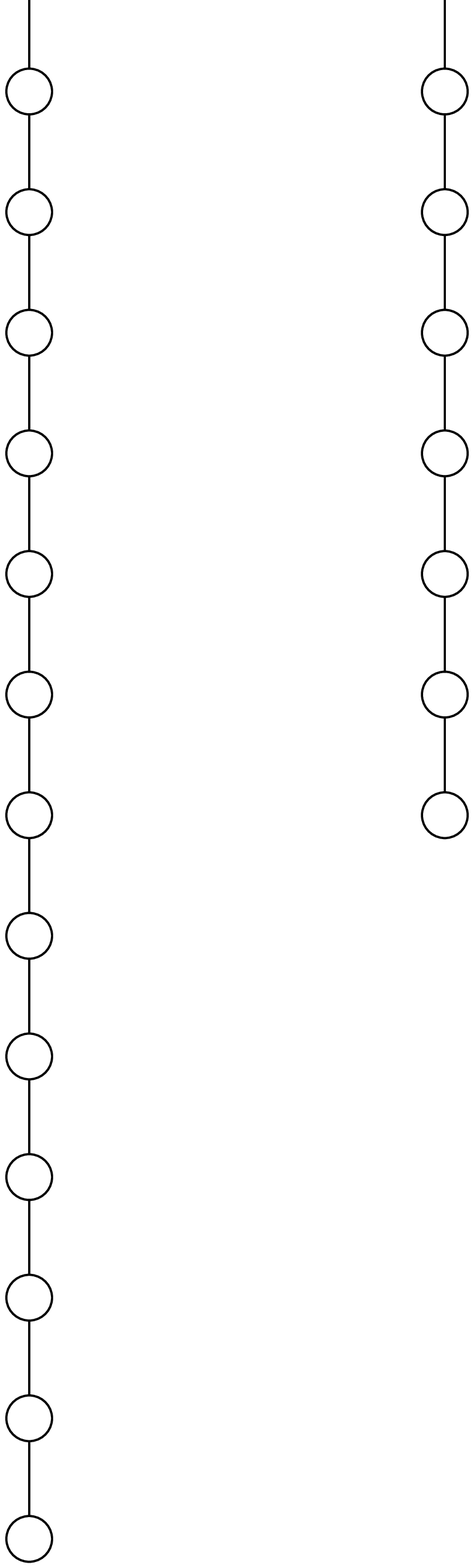}}
\Fig{\includegraphics[scale=\FigRatio]{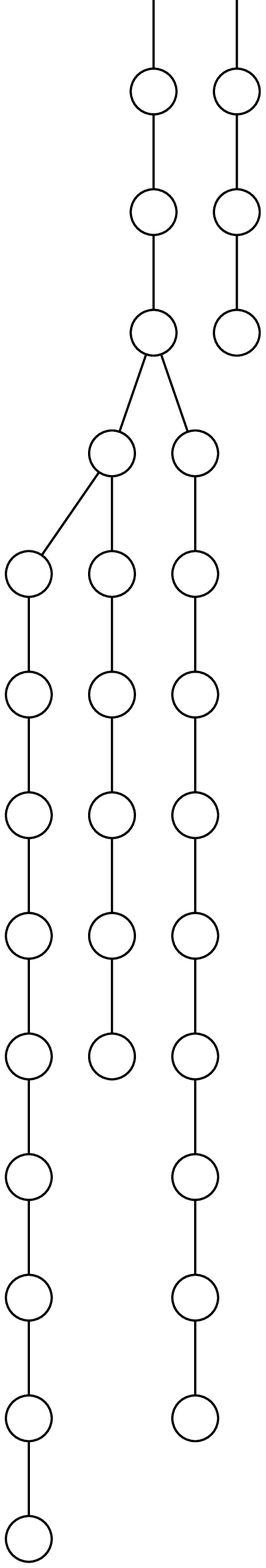}}
\\
$\W=18\;\Leftrightarrow\;f_c = 90\%$\\
\end{center}
\caption{Unary-binary trees associated with weighted Motzkin words of size $500$, for different
values of $\W$ the weight of unary nodes.} \label{fig:Motzkin}
\end{figure} 

\end{Example}
\begin{figure}[ht]
	\centering \Fig{\scalebox{0.8}{\input{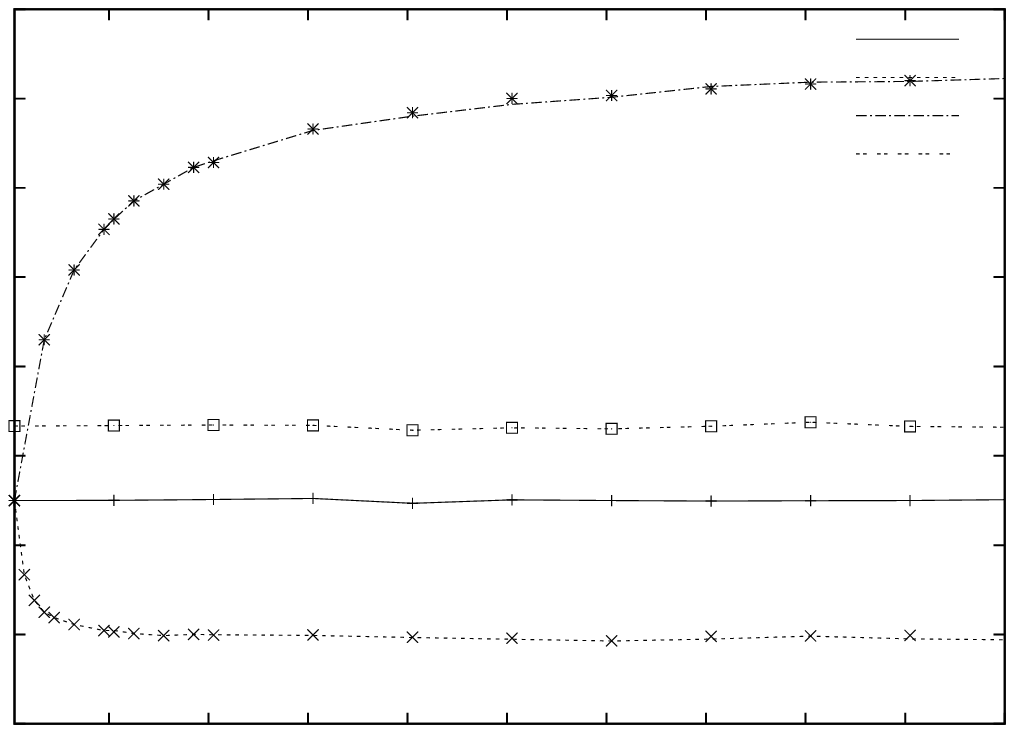}}}
\caption{Average value of an arithmetic expression, computed by generating 100~000 random expression, for various sizes $n$
	and frequencies of symbols $\TPlus$ and $\TOne$. } \label{fig:ResArit}
\end{figure}
\begin{Example}[Binary arithmetic expressions]
	Another class of structures that can be seen as a context-free language is the language
	of arithmetic expressions. We will restrict our operations to the addition and substraction
	and accept only numbers having one binary digit. This yields the following grammar, given
	in polish notation (prefix form) to avoid potential ambiguity:
	\begin{eqnarray*}
		E & \to & \TPlus\;E\;E \unrule \TMinus\;E\;E \unrule N\\
		N & \to & \TZero \unrule \TOne\\
	\end{eqnarray*}
	{\bf Average value of an expression:} Although this problem can probably be solved exactly
	through bivariate generating function
	techniques, we choose a random generation approach to get a rough idea of the influence
	of the number of occurrences of the $\TPlus$ symbol over the average asymptotic
	value of an arithmetic expression. Therefore, we adjoin a weight $\WPlus$ to the atom $\TPlus$
	that will be used to control its frequency $f_{\TPlus}$. Also we define the length $n$ of
	a binary expression to be the length of its encoding, ie its number of terminal symbols.

	As shown previously, the above unambiguous context-free grammar can be translated into a
system of functional equations. Solving the system gives the length generating functions associated
with each non-terminal. In particular for $E$, we have
	$$E_\WPlus(t,u) = \frac {1-\sqrt {1-8\left( 1+u\WPlus \right) {t}^{2}}}{2t\left( 1+{u\WPlus} \right) }$$
	with $u$ and $t$ respectively marking only $\TPlus$ and any atom.

	The above generating function, after some basic singularity analysis, yields
	$$ \WPlus = \frac{2f_{\TPlus}}{1-2f_{\TPlus}}.$$
	Unsurprisingly, it is impossible to find a weight $\WPlus$ such that more than $50\%$ of the symbols
	are $\TPlus$'s, which follows directly from the binary tree-like structure of our expressions.

	One can also adjoin a second weight $\WOne$ to each occurrence of the atom $\TOne$, along with a new complex
	variable $v$. Solving the new system yields the following generating functions:
	$$ E_{\WPlus,\WOne}(t,u,v) = \frac{1-\sqrt{1-4t^2(1+u\WPlus)(1+v\WOne)}}{2t(u\WPlus+1)} $$
	Again it is possible to link the asymptotic frequency $f_\TOne$ (resp. $f_\TPlus$) for $\TOne$ (resp. $\TPlus$)
	with both weights $\WPlus$ and $\WOne$, which yields
	$$ f_\TOne = \frac{\WOne}{2(1+\WOne)} \quad\mbox{ and }\quad f_{\TPlus} = \frac{\WPlus}{2(1+\WPlus)}.$$
	A remarkable property here is the absence of correlation between the frequencies of $\TOne$ and $\TPlus$,
	once again due to the tree-like structure of arithmetic expressions. We can then use these equations
	to estimate the average value of an arithmetic expression having different proportions of $\TOne$ and $\TPlus$'s.
	A random generation of 100~000 expressions for sizes ranging from 1 to 200 allows us to conjecture a
	size-independent average value when $\WPlus=1$ (See Figure~\ref{fig:ResArit}).

\medskip
	\noindent{\bf Exact analysis of the $\WPlus=1$ case :} In the $\WPlus=1$ case, it is an interesting fact that the average
	value $\mathbb{E}(V_n)$ of an expression is in fact independent from $n$. More specifically, it
	can be shown that
	$$\mathbb{E}(V_n)= \frac{\WOne}{1+\WOne}, \forall n\ge 1.$$

	This can be proven by induction on $n$, since
	$$\mathbb{E}(V_1)= \frac{1}{1+\WOne}\cdot0 + \frac{\WOne}{1+\WOne}\cdot1=\frac{\WOne}{1+\WOne}$$
	and that assuming $\mathbb{E}(V_k)=\WOne/(1+\WOne),\;\forall k<n$ yields
	\begin{eqnarray*}
		\mathbb{E}(V_n) & = & \sum_{k\ge 1}^{n-1}p^+_{k,n}\left(\mathbb{E}(V_k)+\mathbb{E}(V_{n-k})\right) + \sum_{k\ge 1}^{n-1}p^-_{k,n}\left(\mathbb{E}(V_k)-\mathbb{E}(V_{n-k})\right)\\
										& = & \sum_{k\ge 1}^{n-1}p^+_{k,n}\frac{2\WOne}{1+\WOne}
	\end{eqnarray*}
	where $p^+_{k,n}$ (resp. $p^-_{k,n}$) is the probability that an expression of size $n$ having root $\TPlus$ (resp. $\TMinus$)
	is composed of two subexpressions having sizes $k$ and $n-k$. Since
	$$\sum_{k\ge 1}^{n-1}p^+_{k,n}+\sum_{k\ge 1}^{n-1}p^-_{k,n}=1, \forall n\ge1$$
	and  $p^-_{k,n}=p^+_{k,n}$ when $\WPlus=1$, then $\sum_{k\ge 1}^{n-1}p^+_{k,n}=1/2$ and the claimed result holds. The results then specializes into
	$\mathbb{E}(V_n) = 1/2$ in the uniform $(\WPlus=1,\WOne=1)$ case, and into $\mathbb{E}(V_n) = 2/3$ in the $(\WPlus=1,\WOne=2)$, both values
	being conjectured from Figure~\ref{fig:ResArit}.
\end{Example}

\subsubsection{The rational case}
\label{genRat}

In this section, we show how to compute a $k$-tuple of weights that is suitable for generating words
according to given frequencies for a non trivial class of rational
languages.
As we will see in some examples below, the result generalizes to
combinatorial classes whose generating functions are rational.

If $C$ is a rational language, then its (weighted) generating
function writes
$$
		S_\Weights(t,{\bf u}) = \frac {P_\Weights(t,{\bf u})} {Q_\Weights(t,{\bf u})}
$$
where {\bf u} stands for $u_1, \ldots, u_k$, and where there
exists $r > 0$ and $\delta_1,  \ldots, \delta_k > 0$
such that $P_\Weights$ and $Q_\Weights$ are analytic in the domain ${\cal D} =
\{(t,{\bf u})\,:\, |t| \leq r, |u_i-1| < \delta_i \forall i\}$.

We establish
a simple formula for the average number of
occurrences of each symbol in the weighted distribution.
Quite noticeably, this formula does not require locating all the actual singularities,
a difficult task as the weights are evolving, but only involves derivatives of
$Q_\Weights$ and $\rho_{\Weights}$ the \emph{unique} dominant singularity.

\begin{Proposition}\label{propcrucialgeneral}
  Let $C$ be a rational language counted by a (weighted) generating function
  $S_{\Weights}(t,{\bf u})=P_{\Weights}(t,{\bf u})/Q_{\Weights}(t,{\bf u})$ such that $S_{\Weights}(t,{\bf u})$ has a \emph{unique} dominant singularity $\rho_{\Weights}\in \mathbb{R}^+$. For any $i\in[1,k]$ and any $k$-tuple $\Weights$ such that $\WF_j \neq 0$, $\forall j\in [1,k]$, we have:
$$
	f_\Weights(\At_i,C,n) = \rho_{\Weights}^{-1}
				{\frac {c_{\Weights,i}(\rho_{\Weights})} {c_\Weights}(\rho_{\Weights})} n + \bigO{1},
$$
where
\begin{eqnarray*}
	c_{\Weights,i}(t) = {\frac{\partial Q_\Weights}{\partial u_i}}(t,{\bf 1})
&\mbox{and}&
	c_{\Weights}(t) =   {\frac{\partial Q_\Weights}{\partial t}}(t,{\bf 1}).
\end{eqnarray*}
and $\rho_{\Weights}$ is the unique real zero of smallest modulus of $Q_\Weights(t,{\bf 1})$.
\end{Proposition}
\Proof For the sake of simplicity, we make the ubiquitous dependency on $\Weights$ implicit by dropping it from our notations.
Let $\alpha\in \mathbb{N}^{+}$ be the multiplicity of $\rho$ as the unique dominant singularity of $S(t,{\bf 1})$.
There exists $\alpha$ roots $\left(\rho_1({\bf u}),\ldots,\rho_\alpha({\bf u})\right)$ of $Q(t,{\bf u})$ such that $\forall j\in[1,\alpha],\; \rho_j({\bf 1})=\rho$. Furthermore there exists a polynom $R(t,{\bf u})$ such that
\begin{equation} Q(t,{\bf u}) = R(t,{\bf u})\cdot\prod_{j=1}^\alpha (1-t/\rho_j({\bf u})) \label{eq:decompositionRoots}\end{equation}
and the function $P(t,{\bf 1})/R(t,{\bf 1})$ is analytic at $t=\rho$, where it takes a positive real value $\kappa$.

As will be shown in Proposition~\ref{propFpi}, we have ${\displaystyle f(\At_i,C,n) = \frac{[t^n]\Diff{S}{u_i}(t,{\bf 1})}{[t^n]S(t,{\bf 1})}}$, and
$$ \Diff{S}{u_i}(t,{\bf 1}) = -\frac{P(t,{\bf 1})}{R(t,{\bf 1})}\frac{t\sum_{j=1}^\alpha \Diff{\rho_j}{u_i}({\bf 1})}{\rho^2(1-t/\rho)^{\alpha+1}} + \frac{\Diff{(P/R)}{u_i}(t,{\bf 1})}{(1-\rho)^\alpha}
$$

Both $S(t,{\bf 1})$ and $\Diff{S}{u_i}(t,{\bf 1})$ are rational generating functions and a generic treatment of such functions (See~\cite{FlSe09})
yields the following asymptotic equivalents:
\begin{eqnarray*}
[t^n]\;S(t,{\bf 1}) & \sim & \kappa\cdot\frac{n^{\alpha-1}}{(\alpha-1)!\rho^{n}} + \bigO{n^{\alpha-2}\rho^{-n}}\\ \,
[t^n]\;\Diff{S}{u_i}(t,{\bf 1}) & \sim &  \kappa\cdot\left(\sum_{j=1}^{\alpha}\frac{-\Diff{\rho_j}{u_i}({\bf 1})}{\rho} \right)\frac{n^{\alpha}}{\alpha!\rho^{n}} + \bigO{n^{\alpha-1}\rho^{-n}}
\end{eqnarray*}
Remark that there exists degenerate cases where the multiplicity of $\rho$ as a pole is decreased (or cancelled) by the derivative on $u_i$.
Therefore the first term of the expansion may cancel but the statement remains valid thanks to the $\bigO{\cdot}$ notation. Taking the ratio, we obtain the following equivalent for $f(\At_i,C,n)$
\begin{equation} f(\At_i,C,n) = -\frac{\sum_{j=1}^{\alpha}\Diff{\rho_j}{u_i}({\bf 1})}{\alpha\rho}n +\bigO{1}. \label{eq:AsymptF}\end{equation}

Now using Equation~\ref{eq:decompositionRoots}, we obtain the following derivatives of $Q$
\begin{eqnarray*}
c_i(t) & = & (1-t/\rho)^{\alpha-1}\left(\frac{\kappa t}{\rho^2}\sum_{j=1}^{\alpha}\Diff{\rho_i}{u_i}({\bf1})+(1-t/\rho)\Diff{R}{u_i}(t,{\bf 1})\right)\\
c(t) & = & (1-t/\rho)^{\alpha-1}\left(-\frac{\kappa\alpha}{\rho}+(1-t/\rho)\Diff{R}{u_i}(t,{\bf 1})\right)
\end{eqnarray*}
and in turn
$$\rho^{-1}\frac{c_i(\rho)}{c(\rho)}n = -\frac{\sum_{j=1}^{\alpha}\Diff{\rho_j}{u_i}({\bf 1})}{\alpha\rho}n$$ where one recognizes the first term of Equation~\ref{eq:AsymptF}.\qed

Now consider that one is given a $k$-tuple $(\TargFreq_1,
 \ldots, \TargFreq_k)$ and aims at finding a $k$-tuple $\Weights$
such that, for any $i$, $f_\Weights(\At_i,C,n) \sim n \TargFreq_i$.
Let
$\W_i$ be the weight of atom $\At_i$ for any $i$.

Under the assumption of a unique dominant singularity in $S_{\Weights}(z,{\bf 1})$,
the following algorithm can solve the problem numerically if such a solution exists:
\begin{itemize}
\item From $Q_\Weights(t,{\bf u})$, compute $c_\Weights(t)$ and the
$c_{\Weights,i}(t)$'s (for $1 \leq i \leq k$) where $t$ and the $\W_i$'s
remain symbolic variables.
\item Build a system of $k$ algebraic equations:
\begin{equation}
\left\{
\begin{array}{lcl}
Q_\Weights(\rho,{\bf 1}) &=& 0\\
\rho^{-1} {\displaystyle \frac {c_{\Weights,1}(\rho)} {c_\Weights}(\rho)} &=& \TargFreq_1\\
&\vdots&\\
\rho^{-1} {\displaystyle \frac {c_{\Weights,k}(\rho)} {c_\Weights}(\rho)} &=& \TargFreq_k\\
\end{array}
\right. \label{eq:SysRat}
\end{equation}
in the unknown variables $\rho,\W_1,\ldots,\W_k$.

Solve the system using numerical techniques (using FGb~\cite{Faugere1999} for example)
\item Among the solutions, take one for which $\rho$ is real and has
the smallest modulus.
\end{itemize}

\begin{Remark}
The prerequisite of Proposition~\ref{propcrucialgeneral} (uniqueness of dominant singularity) is satisfied by specifications associated with strongly connected,
aperiodic automata, where the dominant singularity is known to be unique and has multiplicity $1$ (See~\cite[Theorem~IX-9, p656]{FlSe09}). Such a property also holds for
any specification whose strongly-connected components are aperiodic in the sense that, internally to each component, the greatest common divisor of all cycle length is $1$ (Easily
proved by induction).
\end{Remark}

\begin{Remark}
In the case of multiple dominant singularities, corresponding to periodic automata, Proposition~\ref{propcrucialgeneral} may fail.
However it is worth mentioning that, using partial knowledge of the targeted length $n$, one can transform any rational specification into
an equivalent one meeting the requirement of Proposition~\ref{propcrucialgeneral}.

Let $C$ be a rational specification and $C_{r,D}$ its restriction to objects of any size $n'$ such that $n' \equiv r\;[D]$, respectively counted by
$$S(t,{\bf u}) = \sum_{n\ge 0}\sum_{\bf i\ge {\bf0}} s_{n,{\bf i}}\,t^n \prod_{j=1}^k {u_{j}}^{i_j}\quad\mbox{and}\quad S_{r,D}(t,{\bf u}) = \sum_{N\ge 0}\sum_{\bf i\ge {\bf0}} s_{ND+r,{\bf i}}\,t^N \prod_{j=1}^k {u_{j}}^{i_j}.$$
Notice that, in order to avoid trivial periodicities in $S_{r,D}(t,{\bf u})$, $N$ is no longer the size of counted objects but rather the number of periods.

We rely on the fact that, in any rational generating functions with positive coefficients (See~\cite[Theorem~V-3, p302]{FlSe09}),
there exists a \emph{modulus} $D\in\mathbb{N}^+$ such that, for any \emph{base} $r\in[0,D-1]$, $S_{r,D}(t,{\bf 1})$ has a unique dominant singularity
on the positive real axis. Since any dominant singularity $\rho_j$ is such that $(\rho_j/|\rho_j|)=e^{i\frac{2\pi p_j}{q_j}}$ where
$p_j\in \mathbb{N}$, $q_j\in \mathbb{N}^+$ and $\gcd(p_j,q_j)=1$ (See~\cite[Theorem~IV-3, p267]{FlSe09}), then a suitable value for
$D$ will be the least common multiple of all $q_j$'s.

Then a specification $C_{r,D}$ counted by $S_{r,D}(t,{\bf u})$ can always be built from an automaton for $C$. In short, one starts by intersecting $C$ with the language
denoted by a rational expression $m_{r,D}$ generating all objects of size $n'$ such that $n' \equiv r\;[D]$, given by
$$m_{r,D} = (\At_1 + \ldots+\At_{|\Atoms|})^r((\At_1 + \ldots+\At_{|\Atoms|})^D)^*.$$
The minimal automaton for the intersection language (rational and constructible) only has cycles of lengths that are multiple of $D$. $S_{r,D}(t,{\bf u})$ can then be obtained, either by
only \emph{marking} with the size variable $t$ the atoms occurring at position $p$ such that $p \equiv r+1\;[D]$, or through a variable substitution in the resulting generating function.

Finally, Proposition~\ref{propcrucialgeneral} applies to $S_{r,D}(t,{\bf u})$ such that the weights $\Weights$ and the average proportion $\TargFreq_i$ of an atom $\At_i$
  are interrelated through $\rho^{-1}\frac{c_{\pi,i}}{c_{\pi}}(\rho)= D\TargFreq_i$.
  Reflecting this slight modification into System~\ref{eq:SysRat} and solving the system gives suitable weights for large values of $n$ such that $n\equiv r\; [D]$.
\end{Remark}

\begin{Example}[The Fibonacci language.]
The simple and well known Fibonacci language is defined by the regular
expression $(a+bb)^*$, and admits a strongly connected aperiodic
automaton. Suppose we want to generate words while biasing the average
number of $a$'s. We thus put a weight $\W_a$ on the letter $a$.  The
weighted generating function writes:
$$
	S_{\W_a}(t,u_a,u_b) = \frac{1}{1-\W_a u_a t-u_b^2t^2},
$$
so $Q_{\W_a}(t,u_a,u_b)={1-\W_a u_a t-u_b^2t^2}$.
We have
\begin{eqnarray*}
	c_{\W_a,a}(t,u_a,u_b) = -\W_a t
	&{\rm and}&
	c_{\W}(t,u_a,u_b) = -\W_a u_a - 2 u_b^2 t,
\end{eqnarray*}
which leads to
\begin{eqnarray*}
	f_{\W_a}(a,S,n) &\sim& \rho^{-1} \frac{- \W_a \rho}{-\W_a -2\rho}\, n\\
							 &\sim& \frac{\W_a}{\W_a+2\rho}\, n.
\end{eqnarray*}
Now let $\TargFreq_a$ be the desired asymptotic proportion
of $a$'s in the generated words, we just have to solve
$$
\left\{
\begin{array}{lcl}
	 1 - \W_a \rho - \rho^2 &=& 0\\
	 \displaystyle \frac{\W_a}{\W_a+2\rho} &=& \TargFreq_a
\end{array}
\right.
$$
which gives
$$ \W_a = \frac{2\TargFreq_a}{\sqrt{1-\TargFreq_a^2}}\quad\mbox{ and }\quad\rho = \frac{1-\TargFreq_a}{\sqrt{1-\TargFreq_a^2}}.$$

This gives, for example, $\W_a = 2/\sqrt3 \approx 1.1547$ (and $\rho = 1/\sqrt3 \approx
0.577$) in order to reach $\TargFreq_a = 0.5$, that is an asymptotically
equal proportion of $a$'s and $b$'s in random Fibonacci words. Note
that, in the uniform generation scheme (that is $\W_a=1$), we get
$\TargFreq_a=\frac{1}{\sqrt{5}} \approx 0.447$. Finally, it is worth mentioning
that adding a weight $\W_{bb}$ on each occurrence of $bb$ leads to the
simplification $\W_a = 2\TargFreq_a/(1+\TargFreq_a)$ and $\W_{bb} = 1 - \W_a$.
Figure~\ref{fig:fibWords} shows
random weighted Fibonacci words for different values of $\W_a$.

\begin{figure}[t]
	\begin{center}
		\begin{tabular}{c}
		\begingroup\makeatletter\ifx\SetFigFont\undefined%
			\gdef\SetFigFont#1#2#3#4#5{%
			\reset@font\fontsize{#1}{#2pt}%
			\fontfamily{#3}\fontseries{#4}\fontshape{#5}%
		 \selectfont}%
		\fi\endgroup%
		\newlength{\oldfboxsep}
		\Fig{\scalebox{0.35}{\input{Results0.5.pstex_t}}}\\
		$\W_a=0.5$\\
		\Fig{\scalebox{0.35}{\input{ResultsUnif.pstex_t}}}\\
		$\W_a=1$\\
		\Fig{\scalebox{0.35}{\input{ResultsEq.pstex_t}}}\\
		$\W_a=1.1547$\\
		\Fig{\scalebox{0.35}{\input{Results2.pstex_t}}}\\
		$\W_a=2$\\
		\Fig{\scalebox{0.35}{\input{Results10.pstex_t}}}\\
		$\W_a=10$
		\end{tabular}
	\end{center}
	\caption{Sets of randomly generated Fibonacci words of length $100$
for different values of $\W_a$. White boxes: $a$'s; grey boxes: $b$'s }
	\label{fig:fibWords}
\end{figure}

\end{Example}

\begin{Example}[Motifs in random sequences]
We consider here the number of occurrences of a given motif in a random sequence. This is a
classical issue in bioinformatics. Our approach follows, in some sense, the one
in~\cite{Nicodeme01}, though for a different purpose. Our example is the following: we want to fix
the average number of occurrences of the motif $aug$ in a random RNA sequence, that is a sequence
on the alphabet $\{a,c,g,u\}$. In order to distinguish the $aug$'s, we mark the last $g$, replacing
it with $\bar g$. Hence, in fact we consider words on $\{a,c,g,\bar g,u\}$ where there is no
occurrence of $uag$ and where every occurrence of $\bar g$ is immediately preceded by $ua$.
Obviously, counting the $au\bar g$'s in this language is equivalent to counting the $aug$'s in
$\{a,c,g,u\}^*$. And, in order to generate words in the suitable alphabet, we will just have to
replace each letter $\bar g$ with a letter $g$ during the random generation process.

Our language can be represented by the (strongly connected and
aperiodic) deterministic finite automaton of Figure~\ref{fig:ORF}
\begin{figure}[t]
\begin{center}
	 \Fig{\scalebox{0.45}{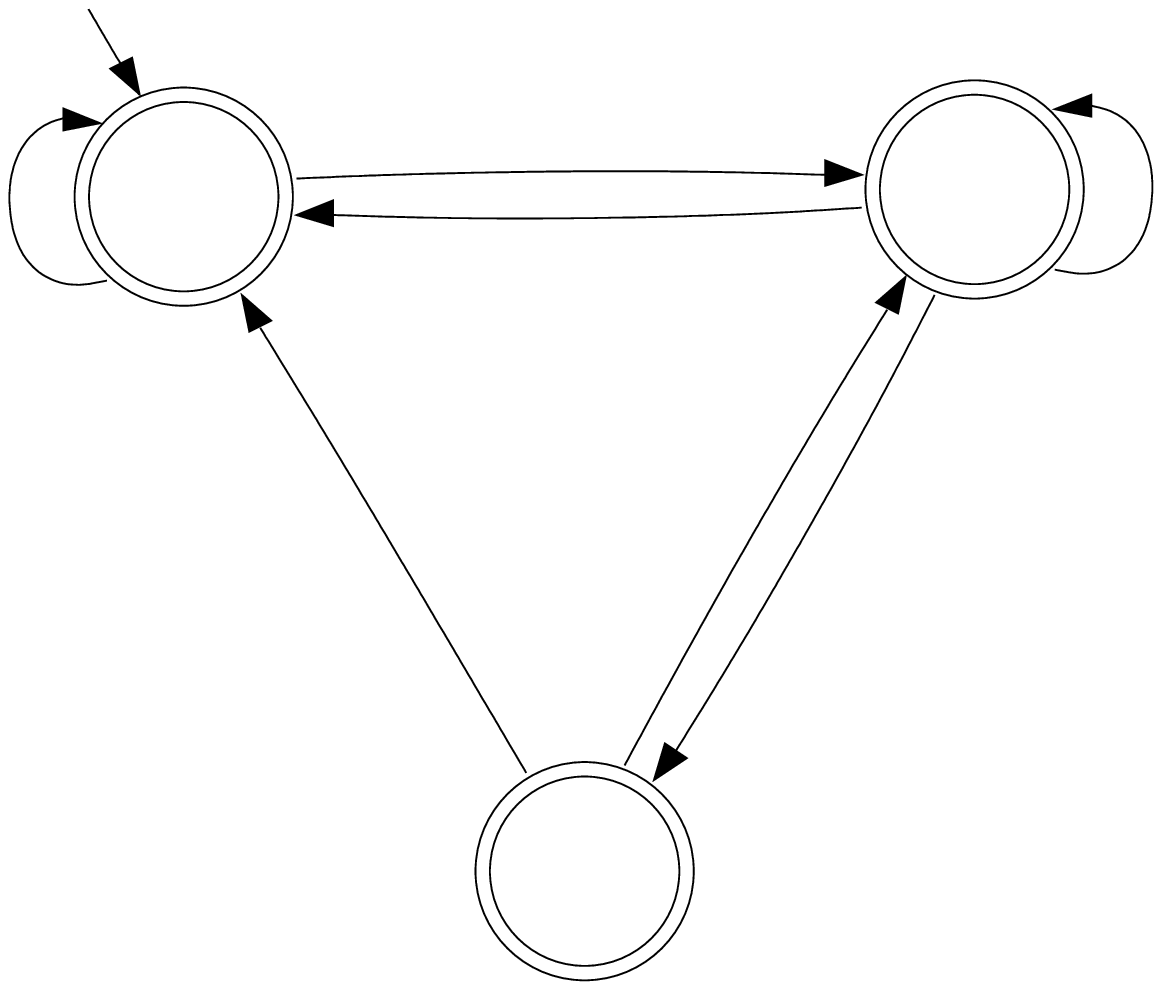}}
\end{center}
\caption{A finite state automaton recognizing the language generated by the grammar.}
\label{fig:ORF}
\end{figure}
or, equivalently, by the following non-ambiguous regular grammar:
\begin{eqnarray*}
S_0 &\rightarrow& \varepsilon \unrule a\;S_1 \unrule c\;S_0 \unrule g\;S_0 \unrule u\;S_0 \\
S_1 &\rightarrow& \varepsilon \unrule a\;S_1 \unrule c\;S_0 \unrule g\;S_0 \unrule u\;S_2 \\
S_2 &\rightarrow& \varepsilon \unrule a\;S_1 \unrule c\;S_0 \unrule {\bar g}\;S_0 \unrule u\;S_0 \\
\end{eqnarray*}
Now by putting a weight $\W_{\bar g}$ on $\bar g$, we are
able to tune the number of occurrences of the motif.
Namely we have:
$$
	S_\W(t,a,c,g,\bar g,u) = \frac{1}{1-t(a+c+g+u)
																			+t^3 a u g
																			-\W_{\bar g} t^3 a u {\bar g}},
$$
thus
$$
	Q_\W(t,a,c,g,\bar g,u) = 1 - t(a+c+g+u) + t^3 a u g
																 - \W_{\bar g} t^3 a u {\bar g}
$$
which gives
\begin{eqnarray*}
	c_{\W_{\bar g},\bar g}(t,a,c,g,\bar g,u) &=& -\W_{\bar g} t^3 u a
\end{eqnarray*}
and
\begin{eqnarray*}
	c_{\W_{\bar g}}(t,a,c,g,\bar g,u) &=& - (a+c+g+u)
												 + 3 t^2 u a g
												 - 3 \W_{\bar g} t^2 u a {\bar g}
\end{eqnarray*}
Hence we find
\begin{eqnarray*}
	f_\W(\bar g,C,n) &\sim&
				\frac{\W_{\bar g} \rho^2}{4 - 3 \rho^2 + 3 \W_{\bar g} \rho^2} n
\end{eqnarray*}
where $\rho$ satisfies the equation
$Q_\W(\rho,1,1,1,1,1)=0$.
Thus we have to solve the system
$$
\left\{
\begin{array}{lcl}
		1-4\rho+(1-\W_{\bar g})\rho^3 &=& 0\\
	 \displaystyle
			 \frac{\W_{\bar g} \rho^2}{4 - 3 \rho^2 + 3 \W_{\bar g} \rho^2}
																			&=& \TargFreq_{\bar g}.
\end{array}
\right.
$$
in order to find the suitable value of $\W_{\bar g}$ that gives
the desired asymptotic ratio $\TargFreq_{\bar g}$ of motifs $atg$ in the
words to be generated. For example, setting $\TargFreq_{\bar g} = 0.1$ gives
$\W_{\bar g} \approx 11.148$ and setting $\TargFreq_{\bar g} = 0.01$ gives
$\W_{\bar g} \approx 0.621$. Note that, in the uniform generation
scheme (that is $\W_{\bar g}=1$), we would have $\TargFreq_{\bar
g}=\frac{1}{64} \approx 0.016$.

Let us take an additional parameter into account. We aim to fix the
(joint) proportion of letters $a$ and $u$ in the sequences, which is
called the ``$a+u$ content'' in bioinformatics.  This is a natural
issue in bioinformatics, where the observed frequencies of nucleotides
have to be taken into account.  To this purpose, let us replace each
letter $a$ or $u$ with a new letter $\alpha$, and let us put the
weight $\W_\alpha$ on this letter. We get
\begin{eqnarray*}
	Q_\W(t,c,g,\bar g,\alpha) &=& 1 - t(2\W_\alpha \alpha+c+g)
														 + \W_\alpha^2 t^3 \alpha^2 g
														 - \W_{\bar g} \W_\alpha^2 t^3 \alpha^2 {\bar g}
\end{eqnarray*}

then
\begin{eqnarray*}
	c_{(\W_\alpha,\W_{\bar g}),\bar g}(t,c,g,\bar g,\alpha) &=& -\W_\alpha^2 \W_{\bar g} t^3 \alpha^2,
\end{eqnarray*}

\begin{eqnarray*}
	c_{(\W_\alpha,\W_{\bar g}),\alpha}(t,c,g,\bar g,\alpha) &=& - 2 \W_\alpha t
											 + 2 \W_\alpha^2 t^3 \alpha g
											 - 2 \W_\alpha^2 \W_{\bar g} t^3 \alpha {\bar g}
\end{eqnarray*}
and
\begin{eqnarray*}
	c_{(\W_\alpha,\W_{\bar g})}(t,c,g,\bar g,\alpha) &=& -(2\W_\alpha \alpha + c + g)
														+ 3 \W_\alpha^2 t^2 \alpha^2 g
														- 3 \W_\alpha^2 \W_{\bar g} t^2 \alpha^2 {\bar g}.
\end{eqnarray*}
Hence
\begin{eqnarray*}
	f_\W(\bar g,C,n) &\sim&
				\frac{\W_\alpha^2 \W_{\bar g} \rho^2}
						 {2 + 2\W_\alpha - 3 \W_\alpha^2 \rho^2
												 + 3 \W_\alpha^2 \W_{\bar g} \rho^2}
				n
\end{eqnarray*}
and
\begin{eqnarray*}
	f_\W(\alpha,C,n) &\sim&
		\frac{2 \W_\alpha (1- \W_\alpha \rho^2 + \W_\alpha \W_{\bar g} \rho^2)}
				 {2 + 2\W_\alpha - 3 \W_\alpha^2 \rho^2
												 + 3 \W_\alpha^2 \W_{\bar g} \rho^2}
				n.
\end{eqnarray*}
Now, adjusting the $a+u$ content and the number of motifs $atg$
reduces to solve a system of three algebraic equations in
$\W_\alpha$, $\W_{\bar g}$, and $\rho$:
$$
\left\{
\begin{array}{lcl}
		1-2\rho(1+\W_{\alpha}) + \rho^3 \W_\alpha^2(1 - \W_{\bar g}) &=& 0\\
		\displaystyle
		 \frac{\W_\alpha^2 \W_{\bar g} \rho^2}
						 {2 + 2\W_\alpha - 3 \W_\alpha^2 \rho^2
											+ 3 \W_\alpha^2 \W_{\bar g} \rho^2} &=& \TargFreq_{\bar g}\\
		\displaystyle
		 \frac{2 \W_\alpha (1- \W_\alpha \rho^2 + \W_\alpha \W_{\bar g} \rho^2)}
				 {2 + 2\W_\alpha - 3 \W_\alpha^2 \rho^2
											 + 3 \W_\alpha^2 \W_{\bar g} \rho^2} &=& \TargFreq_{\alpha}.
\end{array}
\right.
$$
For example, setting $\TargFreq_\alpha=0.7$ and $\TargFreq_{\bar g}=0.1$ gives
$\W_\alpha \approx 2.475$ and $\W_{\bar g} \approx 9.430$ (with
$\rho \approx 0.128$).
\end{Example}

\begin{Example}[RNA multiple stem-loops]
Here we show that Proposition~\ref{propcrucialgeneral} can be sometimes apply in some cases
where the language is not rational.  At first, let us consider the following language :
$L=\{a^{n}c^{m}b^{n}\,: m,n > 0\}$. In molecular biology, this represents what is called a
stem-loop in a RNA secondary structure (see~\cite{waterman78} or~\cite{Vauchaussade85} for details).
Roughly, $a$'s and $b$'s represent paired nucleotides (in the stem), while $c$'s represent
unpaired ones (in the loop). Now let us define the language $L' = d^*(Ld^*)^*$.  that is the
language consisting in series of stem-loops, where each two consecutive stem-loops are possibly
separated by stretches of unpaired nucleotides, represented by $d$'s. Obviously $L$ and $L'$ are not
rational languages, but their generating function are rational. Indeed, there is a straightforward
one-to-one correspondence between the words of $L'$ and the words of the rational language
$d^*((ab)^+c^+d^*)^*$. Additionally, the minimal automaton of this language is aperiodic and
strongly connected, thus Proposition~\ref{propcrucialgeneral} holds.

Suppose we aim to generate words of $L'$ while fixing the average
number of stem-loops and the average number of paired nucleotides. For
the latter, it suffices to put a weight $\W_a$ on each letter
$a$. As regards the number of stem-loops, let us distinguish one letter in
each loop (for example the last one) by changing the $c$ to $\bar
c$. Now our language obeys the following grammar:
\begin{eqnarray*}
S &\rightarrow& D\;T\;S\unrule D \\
T &\rightarrow& a\;T\;b \unrule a\;C\;b\\
C &\rightarrow& c\;C \unrule {\bar c}\\
D &\rightarrow& d\;D \unrule \varepsilon\\
\end{eqnarray*}
The weighted generating function is
$$
	S_\Weights(a,b,c,d) = \frac{1-tc-\W_a t^2ab+\W_a t^3abc}
										 {1-t(c+d) -t^2(\W_a ab-cd)
											 -\W_a t^3(\W_{\bar c}ab{\bar c}-abc-abd)
											 -\W_a t^4abcd}.
$$
Finally we find the following system:
$$
\left\{
\begin{array}{lcl}
		1 - 2\rho + (1-\W_a) \rho^2
				+ (2\W_a - \W_a \W_{\bar c})\rho^3 - \W_a \rho^4  &=& 0
\\
	\displaystyle
	\frac{\W_a \rho (1 + (\W_{\bar c} -2) \rho + \rho^2)}
			 {2 + 2\rho(\W_a -1) +3\rho^2 \W_a (\W_{\bar c}-2) + 4\rho^3 \W_a}
		 &=& \TargFreq_a
\\
	\displaystyle
	\frac{\W_a \W_{\bar c} \rho^2}
			 {2 + 2\rho(\W_a -1) +3\rho^2 \W_a (\W_{\bar c}-2) + 4\rho^3 \W_a}
		 &=& \TargFreq_{\bar c}
\end{array}
\right.
$$
It can be solved symbolically, leading to
$$
\left\{
\begin{array}{lcl}
		\rho &=& \displaystyle
						 \frac{1-2\TargFreq_a-\TargFreq_{\bar c}}{1-2\TargFreq_a+\TargFreq_{\bar c}}
\\
		\W_a &=& \displaystyle
							\frac{(\TargFreq_a-\TargFreq_{\bar c})(1-2\TargFreq_a+\TargFreq_{\bar c})^2}
									 {\TargFreq_a (1-2\TargFreq_a-\TargFreq_{\bar c})^2}
\\
		\W_{\bar c} &=& \displaystyle
										 \frac{4 \TargFreq_{\bar c}^3}
												{(\TargFreq_a-\TargFreq_{\bar c})(1-2\TargFreq_a-\TargFreq_{\bar c})
																	 (1-2\TargFreq_a+\TargFreq_{\bar c})}
\end{array}
\right.
$$
Note that we must have $2\TargFreq_a + \TargFreq_{\bar c} < 1$ since there are as
many $b$'s as $a$'s in the words to be generated, and room must be
left too for $c$'s and $d$'s. For example, setting $\TargFreq_a = 0.4$ (for
80\% of paired nucleotides in average) and $\TargFreq_{\bar c} = 0.1$ (for
$n/10$ stem-loops in average in a structure of size $n$) gives $\W_a
= 27/4$ and $\W_{\bar c} = 4/9$ (with $\rho = 1/3$).
\end{Example}


\subsection{Computing weights for fixed lengths: An heuristic approach.}
\label{sec:opt}

Now we address the problem of finding suitable weights for expected
frequencies in its most general setting. Indeed, it is not always
possible to apply purely analytic methods such a the ones described in
Section~\ref{sec:asympt}, or even only to compute explicitly the
generating function. By contrast, it is always possible to translate
an unambiguous context-free grammar into a recurrence equation, which
allows for an exact evaluation of the numbers of words in the
grammar. Applying this method to the \emph{weighted}
context-free languages gives an algorithm, described in
Subsection~\ref{sec:w2freqs}, for computing the frequencies associated
with given weights. From this, we can use a continuous optimization
algorithm described in Subsection~\ref{sec:heurist}, to obtain
a precise approximation of suitable weights.

\subsubsection{Preliminary: Computing frequencies from weights}
\label{sec:w2freqs}

Let us consider the following generating function:
$$
	S_\Weights(t, {\bf u}) = \sum_{s \in {C}} \WF(s) t^{|s|} u_1^{|s|_{\At_1}}
																										 \ldots
																										 u_k^{|s|_{\At_k}},
$$
where ${\bf u}=(u_1,\ldots,u_k)$. We can write
\begin{equation*}
S_\Weights(t,{\bf u}) =
		\sum_{n,j_1,\ldots,j_k \geq 0}
			 \WF_{n,j_1,\ldots,j_k} t^n u_1^{j_1} \cdots u_k^{j_k},
\end{equation*}
where $\WF_{n,j_1,\ldots,j_k}$ stands for the sum
of weights of the structures of size $n$ having $j_i$ occurrences of atom $\At_i$ for all $i=1, \ldots, k$. The following result holds:
\begin{Proposition}
\label{propFpi}
Let $f_\Weights(\At_i,C,n)$, be the expected number
of occurrences of $\At_i$ in the structures of $\CCS{C}{n}$ generated by the
algorithm. We have:
\begin{equation}
	f_\Weights(\At_i,C,n)  = {\frac
									{[t^n] \frac {\partial  S_\Weights}
															 {\partial u_i} (t,{\bf 1})}
									{[t^n] S_\Weights(t,{\bf 1})}} \label{eq:proportion}
\end{equation}
\end{Proposition}

\Proof
This is a standard result. By definition, we have
\begin{equation*}
	f_\Weights(\At_i,C,n) = {\sum_{s \in \CCS{C}{n}} |s|_{\At_i} \Pr(s)} =
										 \sum_{s \in \CCS{C}{n}} |s|_{\At_i} \frac{\WF(s)}
										 {\WF(\CCS{C}{n})}.
\end{equation*}
from $\Pr(s)= {\frac {\WF(s)} {\WF(\CCS{C}{n})}}$ by Formula~(\ref{condGen4}).
The numerator is obtained from
\begin{equation*}
	{\sum_{s \in \CCS{C}{n}} |s|_{\At_i} \WF(s)}
			 = \sum_{j_1,\ldots,j_k \geq 0} j_i \WF_{n,j_1,\ldots,j_k}\\
			 = [t^n] \frac {\partial  S_\Weights}
											 {\partial u_i} (t,{\bf 1}),
\end{equation*}
while the denominator arises from
\begin{equation*}
	{\WF(\CCS{C}{n})} = \sum_{j_1,\ldots,j_k \geq 0} \WF_{n,j_1,\ldots,j_k}
							 = [t^n] S_\Weights(t,{\bf 1}).
\end{equation*}
\qed

\medskip
This result allows to compute $f_\Weights(\At_i,C,n)$ from the generating
functions $S_\Weights(t, {\bf u})$.
However, computing the partial derivatives requires a closed-form expression of the generating function $S_\Weights$, which
can be hard to obtain for complex grammars. Therefore for practical applications,
we  propose a different approach based on recurrence formulae.

\begin{Proposition}
  The frequencies $f_\Weights(\At_i,C,n)$ associated with all $\At_i$'s
  can be computed in $\bigO{n^4}$ arithmetic operations.
  Moreover, if $C$ uses only the product and union constructs (context-free language), then there
  exists a $\bigO{n^2}$ arithmetic operations algorithm for computing the $f_\Weights(\At_i,C,n)$.
\end{Proposition}

We define $g_\Weights(\At_i,C,n,m)$ to be the sum of
weights for all structures in $\CCS{C}{n}$ featuring $m$ occurrences of
$\At_i$. Then we have:
$$
\begin{array}{lcl}
		C = \At_j & \Rightarrow &g_\Weights(\At_i,C,n,m)=
			\left\{
			 \begin{array}{ll}
				\WF(\At_i)\equiv \W_i & \mbox{if } i=j,\ n=1 \mbox{ and } m=1 \\
				\WF(\At_j)\equiv \W_j & \mbox{if } i\neq j,\ n=1 \mbox{ and } m=0 \\
				0        & \mbox{otherwise}
			 \end{array}
			\right.\\
		C = A + B
					& \Rightarrow
					&g_\Weights(\At_i,C,n,m)=
							 g_\Weights(\At_i,A,n,m) + g_\Weights(\At_i,B,n,m)\\
		C = A \ProdOp B
					& \Rightarrow
					&\displaystyle{g_\Weights(\At_i,C,n,m)=
								\sum_{a=1}^{n-1}\sum_{b=0}^{m}
											g_\Weights(\At_i,A,a,b)\;.\;
											g_\Weights(\At_i,B,n-a,m-b)}\\
    C =  \Theta A
          & \Rightarrow
          &g_\Weights(\At_i,C,n,m) = n\;.\;g_\Weights(\At_i,A,n,m)
\end{array}
$$
and then in turn
$$
	 f_\Weights(\At_i,C,n) = \frac{\sum_{m=0}^{n} m\;.\;g_\Weights(\At_i,C,n,m)}{\sum_{m=0}^n g_\Weights(\At_i,C,n,m)}.
$$

These recurrence relations lead to an algorithm, which needs to compute a table of the values for each
$g_\Weights(\At_i,C,n,m)$. Its size is $\bigO{n^2}$, and each entry needs, at worst, $\bigO{n^2}$
arithmetic operations. Thus the overall worst-case complexity for computing the expected number of
occurrences of any atom $\At_i$ in a structure of size $n$ is
$\bigO{n^4}$.

An alternative way for computing these frequencies in context free grammar specifications is based on
a generalization of the grammar transform associated with the pointing operator (See~\cite{Duchon04} for
examples). Namely, we introduce a partial pointing operator which duplicates objects by marking any occurrences
of a given atom. For context-free languages, we show how to adapt a specification for the partially-pointed
language from the input grammar. Extracting coefficients from the resulting grammars gives us both the
numerators and denominator of equation~\ref{eq:proportion} at the usual cost of coefficient extractions,
effectively improving on the complexity of the previous method.

Let us first define the \emph{partial pointing operator} $\Theta^{\At_i}$, taking a class $C$ and
returning a class $C^{\bullet i}$ whose members are obtained from a member of
$C$ by pointing an occurrence of $\At_i$. Consequently any object $o\in C$ gives rise to a number of
objects in $C^{\bullet i}$ that is equal to its number of occurrences of $\At_i$, and the ordinary
generating function of $C^{\bullet i}$ is therefore clearly $\frac{\partial S_\Weights}{\partial u_i}$.

Based on the obvious combinatorial interpretation of the partial pointing operator, it is possible to build
a grammar $\mathcal{G}^{\bullet i}$ for partially pointed language from the rules of an initial context free
grammar $\mathcal{G}$. Generalizing from the rules used for the general pointing operator~\cite{Duchon04}, we
obtain
$$\begin{array}{lcl}
		C \to A\; |\; B & \Rightarrow & C^{\bullet i} \to A^{\bullet i}\; |\; B^{\bullet i}\\
		C \to A \cdot\ProdOp B & \Rightarrow & C^{\bullet i} \to A^{\bullet i} \cdot B\; |\; A \cdot B^{\bullet i} \\
		C \to \At_j & \Rightarrow & C^{\bullet i} \to \left\{\begin{array}{cl} \At_j^{\bullet i} & \mbox{If }i=j\\ \emptyset & \mbox{Otherwise.}  \end{array}\right.\\
\end{array}$$
The $\emptyset$ symbol tags as non-productive a non-terminal $C$, which can be
eliminated through an iterated post-treatment. However non-necessary, this may decrease
the constants involved in the complexity of this approach, since the complexity
of our enumeration algorithm depends, in a somewhat hidden fashion, on the number of non-terminals.

Using counting rules from Table~\ref{tabSS}, we can then evaluate the number $g^{\bullet i}_n$ of
words of size $n$ in $\mathcal{G}^{\bullet i}$. Since the generating function $S^{\bullet
i}_\Weights(t,\mathbf{u})$ of $\mathcal{G}^{\bullet i}$ is such that $S^{\bullet i}_\Weights(t,\mathbf{u}) =
u_i\cdot\frac{\partial S_\Weights(t,\mathbf{u})}{\partial u_i}$, then we have
$$ [t^n] \frac{\partial S_\Weights}{\partial u_i}(t,\mathbf{1}) = [t^n] S^{\bullet i}_\Weights(t,\mathbf{1}) = g^{\bullet i}_n $$
The expression of Proposition~\ref{propFpi} for $f_\Weights$  can then be rephrased as follows :
$$f_\Weights(\At_i,\mathcal{G},n) = \frac{g^{\bullet i}_n}{g_n}$$

Since both $g^{\bullet i}_n$ and $g_n$ are numbers (resp. total weights in weighted specifications) of words in a context-free grammar, they can be
computed in $\bigO{n^2}$ arithmetic operations and in $\Theta(n^3)$ space complexity and so can
$f_\Weights(\At_i,\mathcal{G},n)$. These can be lowered to $\bigO{n}$ arithmetic operations and
$\Theta(n^2)$ space complexity by using the linear recurrences obtained for any grammar by symbolic
methods (\texttt{GFun}~\cite{SalZim94}). Although this approach could in principle be adapted to
general standard specifications, it is unclear at the moment how some of the partial/general
pointing/unpointing combinations may interact, and we favored the former approach in our implementation
despite its higher theoretical complexity.

\subsubsection{Assessing suitable weights through an optimization heuristic}
\label{sec:heurist}

Remember we want to find a $k$-tuple of weights $\Weights=(\W_i)_{i\in[1,k]}$
that achieves \Def{targeted} frequencies $(\TargFreq_1, \ldots, \TargFreq_k)$
associated with our $k$ distinguished atoms $(\At_1,\ldots,\At_k)$. To
that purpose, we reformulate our problem as an optimization one.

Let $\WeightsToFreqs:
\mathbb{R}^{k}\times\mathbb{N}\to
\mathbb{R}^{k}$ be the function that takes a $k$-tuple of weights $\Weights=(\W_1,\ldots,\W_k)$
and a length $n\in\mathbb{N}$, and returns the $k$-tuple of frequencies
$(\ObsFreq_i)_{i\in[1,k]}$ observed among words of length $n$. We described in
Section~\ref{sec:w2freqs} two methods to compute the function $\WeightsToFreqs$ which, in addition
to an expected smoothness of the function $\WeightsToFreqs$, allows us to foresee an efficient
optimization approach for the \emph{inversion} of $\WeightsToFreqs$. More specifically, we want to
find weights that achieves \Def{targeted} frequencies
$\TargFreqs=(\TargFreq_i)_{i\in[1,k]}$. To that purpose we reformulate our problem as an
optimization problem by defining an \emph{objective function}
$\ObjFun:\mathbb{R}^{k}\times\mathbb{N}\to \mathbb{R}$ such that
$$\ObjFun(\W_1,\ldots,\W_{k},n)=\sqrt{\sum_{i=1}^{k}\left(\frac{\ObsFreq_i-\TargFreq_i}{\ObsFreq_i}\right)^2}.$$
We point out the fact that
$$ \left(\ObjFun(\W^*_1,\ldots,\W^*_{k},n) = 0 \right)\quad \Rightarrow \quad \left(\WeightsToFreqs(\W^*_1,\ldots,\W^*_{k},n) = (\TargFreq_1,\ldots,\TargFreq_{k})\right) $$
so that solving the former yields a solution for the latter. Also, it is worth noticing that, thanks to the partial pointing described above, $\ObjFun$ can be computed
in $\bigO{k\cdot n}$ arithmetic operations.

\begin{figure}[t]
\begin{center}
\Fig{\scalebox{0.8}{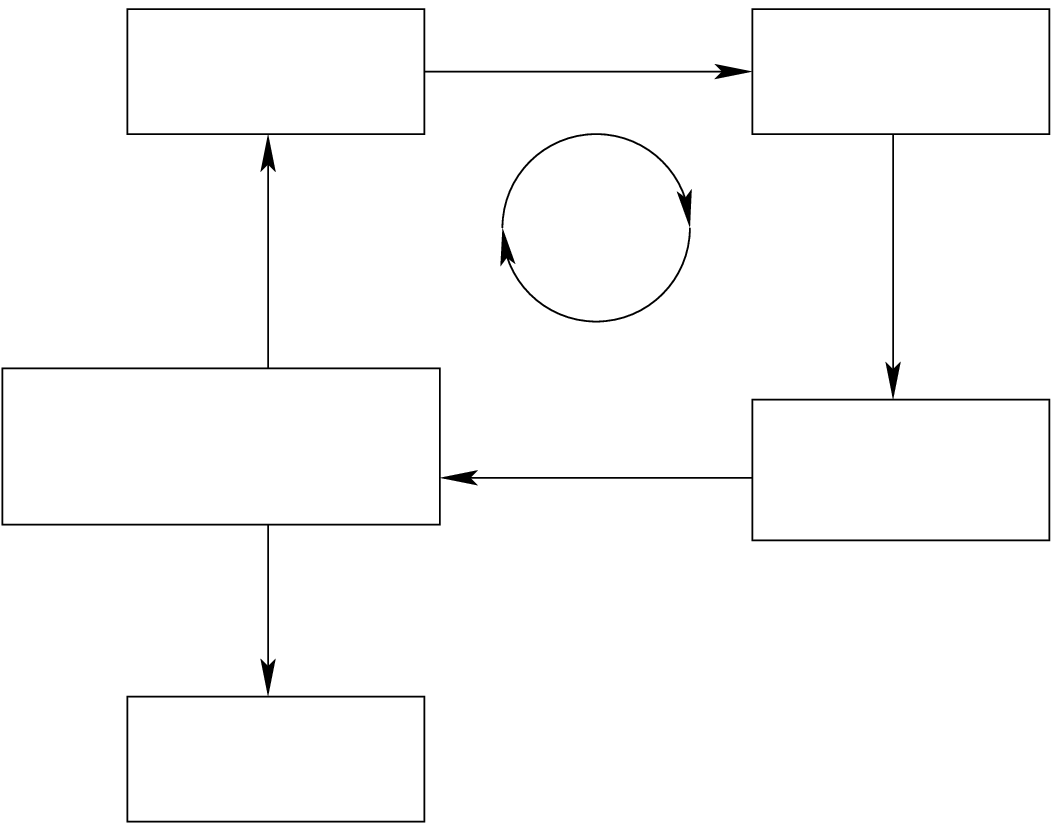}}
\end{center}
\caption{General principle of our heuristic approach to the problem of computing weights $\Weights$ that
achieve targeted frequencies $\TargFreq$.} \label{fig:opt}
\end{figure}

{\tt CONDOR} is a continuous optimization algorithm, developed and implemented by Vanden Berghen
\emph{et al}~\cite{condor05}. It attempts at finding the values for a set of parameters that
minimizes an objective function. It proceeds by building a local approximation of $\ObjFun$ around
a given point, as a polynomial of degree two and uses it to perform an analog of a steepest descent
while maintaining a trust regions. We used a C++ implementation of the {\tt CONDOR}  algorithm, downloaded
from F. Vanden Berghen's website. We implemented the \emph{partial pointing algorithm} described in
Section~\ref{sec:w2freqs} for the computation of $\WeightsToFreqs$, using the C++ arbitrary
precision library {\tt apfloat} created by M. Tommila. We combined these three components into
a software {\GRGFreqs}, which takes as input a grammar formatted as a {\tt
GenRGenS}~\cite{PoTeDe06} description file with additional \emph{target frequencies} for the
terminal symbols, and iteratively finds a set of weights that achieves such frequencies.

By contrast to the analytic approach, which relies on the assumption that the asymptotic regime has
been reached, this approach  works for fixed, potentially small, values of $n$. Moreover it is fully
automated and does not require any interaction with a computer algebra system. This allows for a computation
of suitable weights, even for complex grammars for which solving the associated systems of functional equations
by computer algebra is challenging.
Finally it is also possible to use sophisticated methods inspired by~\cite{Denise00} to achieve \emph{exact} values for
$\ObjFun$, or just to take advantage of the numerical stability of our algorithm and set the
precision of the mantissa to a large fixed value. Since the CONDOR algorithm uses real numbers
internally, this allows for a reasonably accurate computation of suitable weights, as illustrated
by the following application.

\medskip
\begin{Remark}\label{rem:precision}
As pointed out by one of the referees, one can bound the error made
on targeted frequencies when using fixed-precision reals for
computing the weights. Let $\W^*_1, ..., \W^*_k$ be the exact solution, i.e. a set of weights that generates the atoms with the targeted probabilities
$\TargFreq_1, ..., \TargFreq_k$. Now suppose that floating point
approximations $\W_1, ..., \W_k$ are used instead of exact weights, then
one can define the relative errors $\varepsilon_i$ as $\W_i = (1+\varepsilon_i) \W^*_i$.
Consider the maximal and minimal relative errors $M_{\varepsilon} = \max_i (\varepsilon_i)$ and $m_\varepsilon = \min_i(\varepsilon_i)$,
then one has
$$ (1+m_{\varepsilon})^n \W^*(s) \leq \;\;\W(s) \equiv  \W^*(s)\cdot \prod_{1\leq i \leq k} (1+\varepsilon_i)^{|s|_{\At_i}} \;\;\leq (1+M_{\varepsilon})^n \W^*(s)$$
and similar bounds hold for $\pi(C_n)$ the cumulated weights of structures of size $n$.
By construction, each structure is generated with probability $\Pr(s) = \frac {\W(s)}
{\W(C_n)}$  therefore we have
\[ (1/q)\cdot \Pr^*(s) \leq \Pr(s)  \leq  q\cdot\Pr^*(s),\quad\text{with } q:=\left(\frac{1+M_{\varepsilon}}{1+m_{\varepsilon}}\right)^n. \]

Let us now use floating point arithmetics with a binary
mantissa of a given fixed size $b$. Assuming that the method converges
toward the closest expressible approximation of $\W^*$, one has $m_{\varepsilon}=-2^{1-b}$ and
$M_{\varepsilon}=2^{1-b}$.  One can then compute a precision $b$ such that the sampling probability $\Pr(s)$ for any structure deviates from the targeted one $\Pr^*(s)$ by less
than some $\varepsilon\in [0,1[$:
\[ (1-\varepsilon)\cdot\Pr^*(s) \leq \Pr(s)  \leq  {(1+\varepsilon)}\cdot\Pr^*(s). \]

It can be easily shown that $q\le 1+\varepsilon$ implies $1/q\ge 1-\varepsilon$,%
so we are left to find a precision $b$ such that
\[ \left(\frac{1+2^{1-b}}{1-2^{1-b}}\right)^n \le 1+\varepsilon.\]
Applying the natural logarithm on both sides, one obtains
$$ n\left(\log(1+2^{1-b})-\log(1-2^{1-b})\right) \le \log(1+\varepsilon)$$
Taylor expansions can be used for both logarithms, simplifying into
$$\log(1+X)-\log(1-X) = 2 X + X\cdot\sum_{k\ge 1}\frac{2X^{2k}}{2k+1} \le 3 X,\quad \forall 0\le X\le1/2.$$
Here $X=2^{1-b}$ and the $X\le1/2$ condition holds for any $b\ge 2$, so any $b$ such that
$$b \ge 1 + \frac{\log 3+\log(n)-\log\log(1+\varepsilon)}{\log2}$$
will achieve a relative error less than $\varepsilon$.
\end{Remark}

\medskip
Future directions for this research will aim at replacing the current optimization
scheme with a numerical iteration, following the pioneering work of Pivoteau~\emph{et al}~\cite{Pivoteau2008}
for computing the so-called Boltzmann oracle.

\subsubsection{Application 1: Altering the node degree distribution for quadtrees}
\begin{figure}[t]
  \centering \Fig{\includegraphics{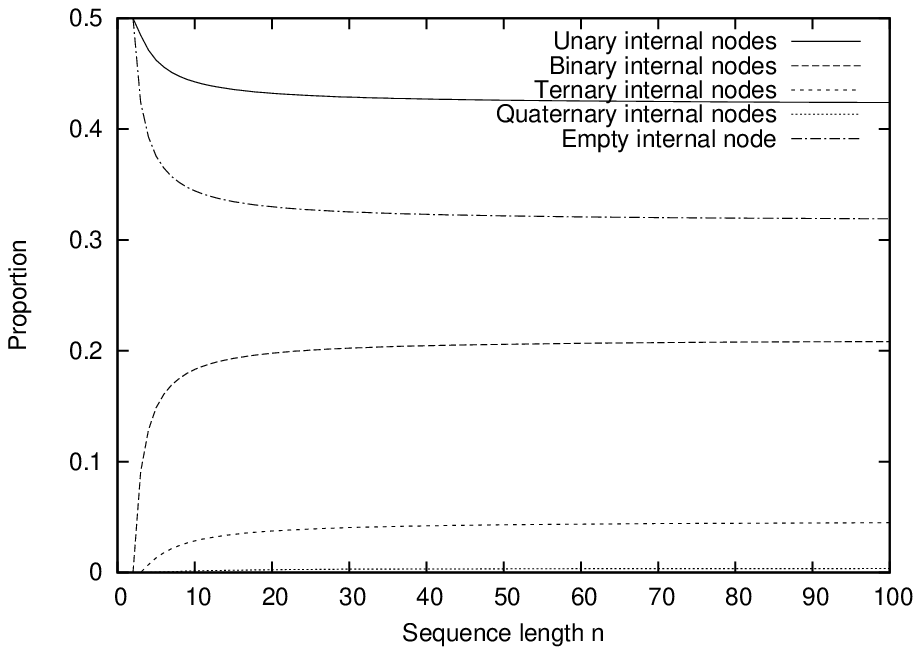}}
  \caption{Evolution of the node degree distribution for trees of increasing size in the uniform model. The asymptotic proportions of nodes of degree $(0,1,2,3,4)$ are respectively $(81/256,27/64,27/128,3/64,1/256)$.} \label{fig:FreqsUnif}
\end{figure}

Quadtrees are data structures, mostly used in computer graphics to partition
the view plane, thus helping in determining which parts are obfuscated,
or which  geometrical objects are in collision. Considered as a combinatorial object, a quadtree
can be recursively defined as either an empty tree, or a tree having four
children, denoted by their orientations (Northern-eastern, southern-eastern, southern-western and northern-western). This definition gives rise to the
following context-free grammar
$$ S\to a\;S\;b\;S\;c\;S\;d\;S\unrule \varepsilon $$
which generates all quadtrees through an encoding similar to that of Dyck words for binary trees.
More specifically, it can be shown that the number of words of length $4n$ generated by
this grammar is exactly the number of quadtrees having $n$ internal nodes.

Now, we defines the degree of a node to be the number of its non-empty children.

The grammar above can then be altered in such a way that each production will create a
node of known degree $i$, marked by an occurrence of a distinctive letter $a_i$:
\begin{eqnarray*}
S &\to & T\;|\;\varepsilon \\
T &\to & a_4\;T\;b\;T\;c\;T\;d\;T \\
  & |  & a_3\;b\;T\;c\;T\;d\;T \unrule a_3\;T\;b\;c\;T\;d\;T \unrule a_3\;T\;b\;T\;c\;d\;T \unrule a_3\;T\;b\;T\;c\;T\;d \\
  & |  & a_2\;b\;c\;T\;d\;T \unrule a_2\;b\;T\;c\;d\;T \unrule a_2\;b\;T\;c\;T\;d \unrule a_2\;T\;b\;c\;d\;T \\
  & |  & a_2\;T\;b\;c\;T\;d \unrule a_2\;T\;b\;T\;c\;d \\
  & |  & a_1\;T\;b\;c\;d \unrule  a_1\;b\;T\;c\;d \unrule  a_1\;b\;c\;T\;d \unrule  a_1\;b\;c\;d\;T \\
  & |  & a_0\;b\;c\;d
\end{eqnarray*}
Computing the proportions of symbols $\{a_0,\ldots,a_4\}$, which can be done for instance by
one of the algorithms from Subsection~\ref{sec:w2freqs}), yields the distribution of node degrees for increasing lengths
plotted in Figure~\ref{fig:FreqsUnif}. This distribution shows uneven proportions of each types of
nodes.

\begin{figure}
  \centering\Fig{\includegraphics[width=0.49\textwidth]{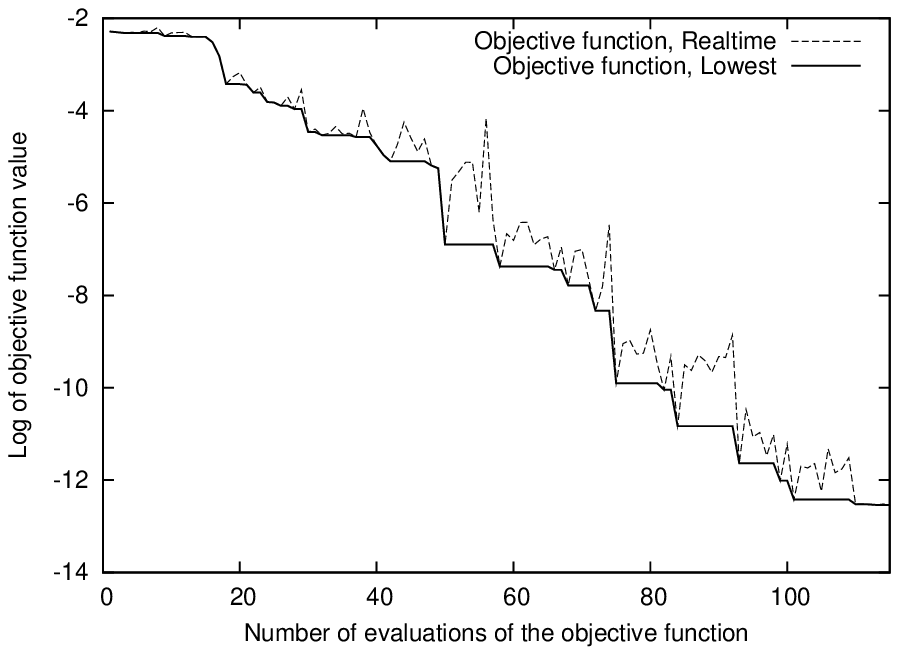}}\Fig{\includegraphics[width=0.49\textwidth]{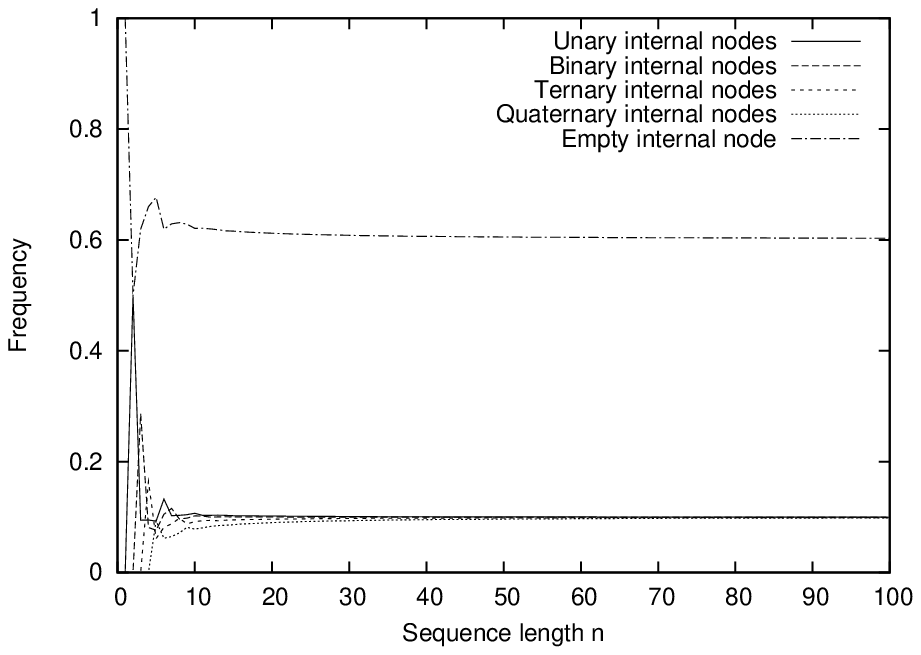}}
  \caption{{\bf Left:} Weight optimization for weighted quadtrees of size $201$.
  The targeted proportions are $121/201$ (resp. $20/201$) for nodes of degree $0$ (resp.
  $1$, $2$, $3$ and $4$). \newline{\bf Right:} Node degree distributions for weighted quad trees of increasing size in our weighted model.
  Although formally the computed weights only work for size $201$ structures, a good approximation of the
  targeted distribution is already observed for smaller sizes.}
  \label{fig:OptEqFreqs}
\end{figure}
Assume we want to draw quadtrees at random in a weighted model, chosen such that the proportions of nodes of
degree $1$, $2$, $3$ and $4$ are equal, while leaving out nodes of degree $0$ as a necessary degree of freedom.
Furthermore, we want to make sure that there exists a quadtree that achieves the target frequencies.
Let $\{n_0,\ldots,n_4\}$ be the numbers of nodes of respective degrees $\{0,\ldots,4\}$ in a quadtree,
then our quadtrees must obey the following constraints:
\begin{itemize}
\item The number of nodes $n$ in any tree is related to the sum of degrees.
\item The numbers $n_i$ of nodes of different degrees have to sum to $n$.
\item Nodes having degrees $1$ to $4$ have to
be equally represented.
\end{itemize}
These constraints translate into the following system
$$ \left\{\begin{array}{rcl}
  0n_0+1n_1+2n_2+3n_3+4n_4&=&n-1 \\
  n_0+n_1+n_2+n_3+n_4&=&n \\
  n_1=n_2=n_3=n_4 &=& k
  \end{array}\right. $$
Solving the system yields the following values in $n_0$ and $k$:
$$ \left\{\begin{array}{rcl}
  n_0& = &\frac{3n+2}{5} \\
  k & = & \frac{n-1}{10}
  \end{array}\right. $$
A corollary is that our set of constraints can only be fulfilled by trees of size
equal to $1$ modulo $10$.

\begin{figure}[p]
\begin{center}
{
\begin{tabular}{ccccccc}
  \ShowTree{QuadsUnif}{1}&
  \ShowTree{QuadsUnif}{2}&
  \ShowTree{QuadsUnif}{3}&
  \ShowTree{QuadsUnif}{4}&
  \ShowTree{QuadsUnif}{5}&
  \ShowTree{QuadsUnif}{6}&
  \ShowTree{QuadsUnif}{7}\\
  \ShowQuad{QuadsUnif}{1}&
  \ShowQuad{QuadsUnif}{2}&
  \ShowQuad{QuadsUnif}{3}&
  \ShowQuad{QuadsUnif}{4}&
  \ShowQuad{QuadsUnif}{5}&
  \ShowQuad{QuadsUnif}{6}&
  \ShowQuad{QuadsUnif}{7}\\
  \ShowTree{QuadsUnif}{11}&
  \ShowTree{QuadsUnif}{12}&
  \ShowTree{QuadsUnif}{13}&
  \ShowTree{QuadsUnif}{14}&
  \ShowTree{QuadsUnif}{15}&
  \ShowTree{QuadsUnif}{16}&
  \ShowTree{QuadsUnif}{17}\\
  \ShowQuad{QuadsUnif}{11}&
  \ShowQuad{QuadsUnif}{12}&
  \ShowQuad{QuadsUnif}{13}&
  \ShowQuad{QuadsUnif}{14}&
  \ShowQuad{QuadsUnif}{15}&
  \ShowQuad{QuadsUnif}{16}&
  \ShowQuad{QuadsUnif}{17}\\
\end{tabular}}\\
Uniformly generated quad trees
\\[0.5cm]
\begin{tabular}{ccccccc}
  \ShowTree{QuadsFinal}{1}&
  \ShowTree{QuadsFinal}{2}&
  \ShowTree{QuadsFinal}{3}&
  \ShowTree{QuadsFinal}{6}&
  \ShowTree{QuadsFinal}{7}\\
  \ShowQuad{QuadsFinal}{1}&
  \ShowQuad{QuadsFinal}{2}&
  \ShowQuad{QuadsFinal}{3}&
  \ShowQuad{QuadsFinal}{6}&
  \ShowQuad{QuadsFinal}{7}\\
  \ShowTree{QuadsFinal}{11}&
  \ShowTree{QuadsFinal}{12}&
  \ShowTree{QuadsFinal}{13}&
  \ShowTree{QuadsFinal}{16}&
  \ShowTree{QuadsFinal}{17}\\
  \ShowQuad{QuadsFinal}{11}&
  \ShowQuad{QuadsFinal}{12}&
  \ShowQuad{QuadsFinal}{13}&
  \ShowQuad{QuadsFinal}{16}&
  \ShowQuad{QuadsFinal}{17}\\
\end{tabular}\\
Generation using calculated weights
\end{center}
\caption{Typical sets of randomly generated quad trees of size $201$ in the uniform model (Top) and
using weights output by our optimizer, whose objective was to balance the numbers of nodes
for each degree (Bottom). We show here the tree representation of quad trees in addition to the classic
square one, since the latter tends to overemphasize nodes of low depth.\label{fig:quads}}
\end{figure}

For instance, any quadtree of size $201$ that meets the three conditions above
will necessarily contain $121$ nodes of degree $0$ and $20$ nodes
of each other degree. Figure~\ref{fig:OptEqFreqs}--Left illustrates a run of our software
\texttt{GrgFreqs} using such proportions as target ($121/201$ for nodes of degree $0$ and
$20/201$ otherwise). After about $100$ evaluation of the objective function, a
$k$-tuple $\Weights$ of candidate weights for symbols $a_i$, giving rise to a value $3.6\,10^{-6}$ for the objective
function, was found. 
From Remark~\ref{rem:precision}, the weights can be safely truncated to 6 decimal
digits to ensure a $10^{-3}$ precision in each frequency, thus we obtain
  \begin{equation*}
  \begin{array}{cccccc}
  \text{Letter } a_i & a_0 & a_1 & a_2 & a_3 & a_4\\
  \text{Weight } \WF(a_i)          &1.0 & 0.0711964 & 0.0819891 & 0.212971 & 1.47891\\
  \text{Frequency }\ObsFreq_i \text{ (\%)}  &60.19949 & 9.94975 & 9.95000 & 9.95024 & 9.95049
 \end{array}
  \end{equation*}

Using these weights, it is then possible to replot the average
frequencies for these symbols for sizes between $1$ and $100$ (Figure~\ref{fig:OptEqFreqs}--Right).
The modification of the average profile resulting from adding such weights is illustrated by random
instances drawn in Figure~\ref{fig:quads}.

Finally, as pointed out by one of the referees, there
  also exists a simple and efficient {\it ad hoc} way to generate
  quadtrees that obeys to an {\it exact} degree distribution.  This
  can be done through a well-known bijection between the set of trees
  having nodes of degree less than a given $k$ and the Lukasiewicz
  language on the alphabet $\{a_0, a_1, \ldots,
  a_k\}$~\cite{viennot94}.  The letter $a_i$ in the Lukaciewicz word
  corresponds to a node of degree $i$ in the left to right depth-first
  traversal of the tree.  For adapting this bijection to quadtrees, we
  set $k=4$, and each letter $a_i$ must be colored to differentiate
  the children's positions of a node. For example, there will be $6$
  different colors for $a_2$ since there are $6$ ways to choose two
  leaves within the four possible nodes.  Thus, to generate a tree
  with the node degree distribution $(n_0, n_1, n_2, n_3, n_4)$, it
  suffices to generate a random word with $n_0$ occurrences of the
  $a_0$ symbol, $n_1$ symbols $a_1$ (with 4 possible colors), $n_2$
  symbols $a_2$ (6 colors), $n_3$ symbol $a_3$ (4 colors), $n_4$
  symbol $a_4$; Then use the Cyclic Lemma~\cite{DZ90} to change this
  word into a Lukaciewicz word, which corresponds to a quadtree, and
  finally build the quadtree for a total $O(n)$ complexity.


\newcommand{\ARN}{\ensuremath{S}}

\newcommand{\Term}{\ensuremath{\mbox{\bf t}}}
\newcommand{\term}{\ensuremath{\mbox{t}}}
\newcommand{\Bulge}{\ensuremath{\mbox{\bf b}}}
\newcommand{\bulge}{\ensuremath{\mbox{b}}}
\newcommand{\Interior}{\ensuremath{\mbox{\bf i}}}
\newcommand{\interior}{\ensuremath{\mbox{i}}}
\newcommand{\Multi}{\ensuremath{\mbox{\bf m}}}
\newcommand{\multi}{\ensuremath{\mbox{m}}}
\newcommand{\Hairpin}{\ensuremath{\mbox{\bf h}}}
\newcommand{\hairpin}{\ensuremath{\mbox{h}}}
\newcommand{\Unpaired}{\ensuremath{\mbox{u}}}

\newcommand{\TermLoop}{\ensuremath{T}}
\newcommand{\BulgeLoop}{\ensuremath{B}}
\newcommand{\InteriorLoop}{\ensuremath{I}}
\newcommand{\MultiLoop}{\ensuremath{M}}
\newcommand{\MultiLoopAux}{\ensuremath{M''}}
\newcommand{\HairpinLoop}{\ensuremath{H}}
\newcommand{\UniformModel}{\ensuremath{\mathcal{M}_{0}}}
\newcommand{\HelicesModel}{\ensuremath{\mathcal{M}_{\mathcal{H}}}}
\newcommand{\LoopsModel}{\ensuremath{\mathcal{M}_{\mathcal{L}}}}
\newcommand{\HelicesWeight}{\ensuremath{\pi_{\mathcal{H}}}}
\newcommand{\LoopsWeight}{\ensuremath{\pi_{\mathcal{L}}}}
\newcommand{\smalltitle}[1]{\quad\\[0.2cm]\noindent{\bf #1.}\;}

\subsubsection{Application 2: Realistic RNA secondary structures}

\begin{figure}[t]
\begin{center}
	\Fig{\includegraphics[width=6cm]{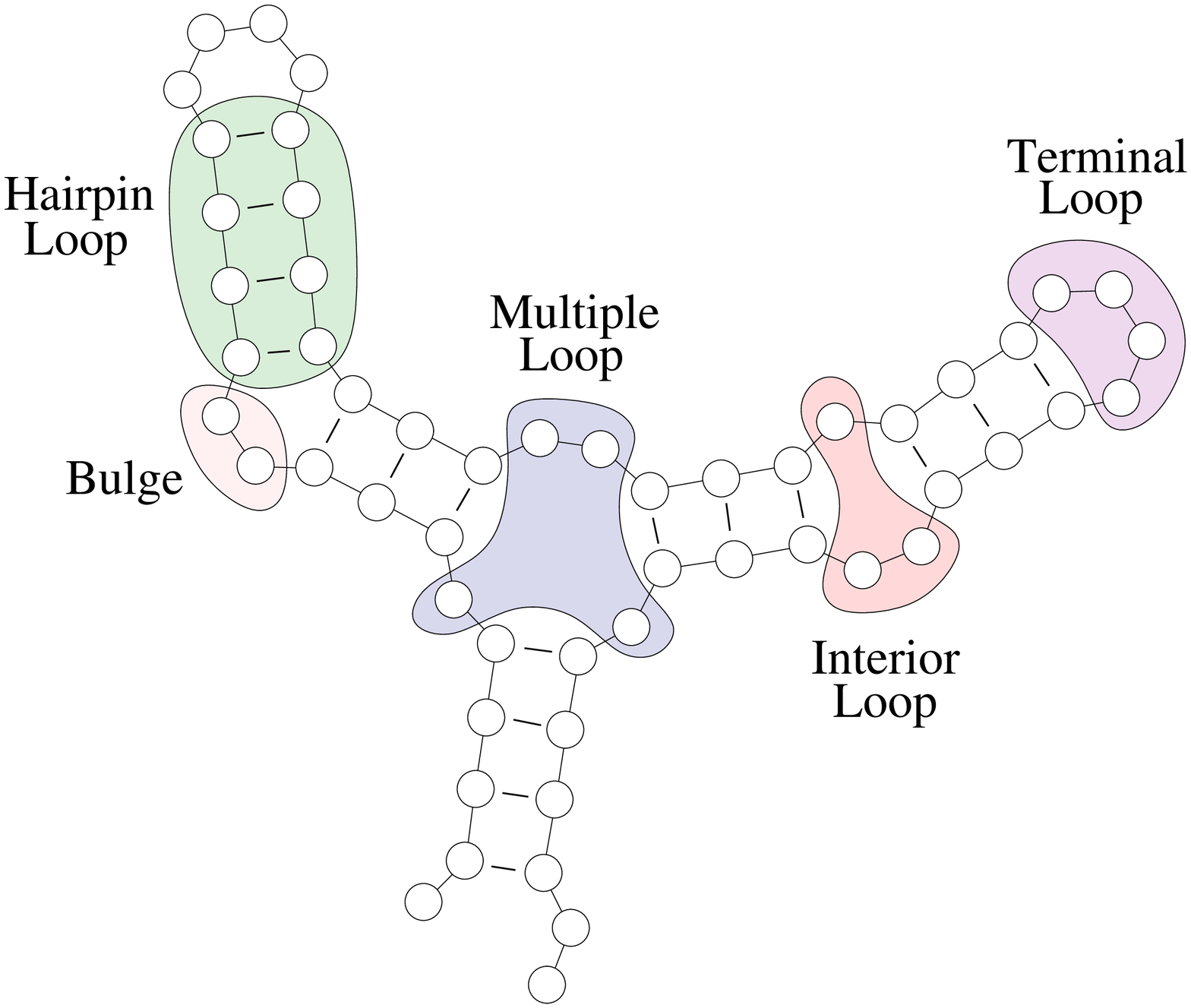}}
	\hspace{0.3cm}\Fig{\includegraphics[width=5.4cm]{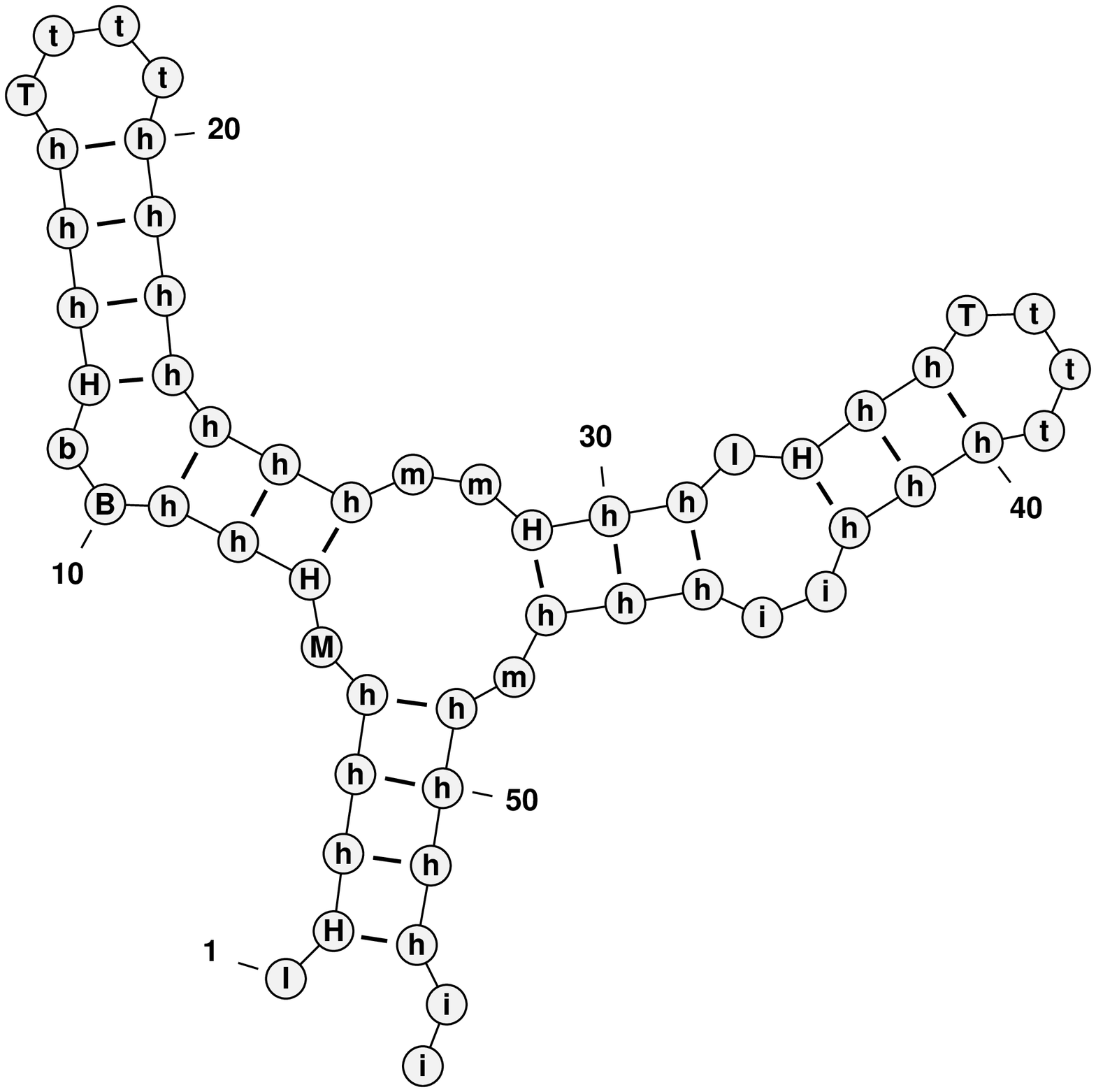}}\\[0.5cm]
\begin{tabular}{rc}
	Structure:&\texttt{.((((.(((..((((....)))))))..(((.(((....)))..))).))))..}\\
	Annotation:&\texttt{IHhhhMHhhBbHhhhTttthhhhhhhmmHhhIHhhTttthhhiihhhmhhhhii}
\end{tabular}
\caption{Different types of loops in an RNA secondary structure (Left), principles of our structure annotation (Right)
and result of the annotation (Bottom). }
\label{fig:RNAAnnotate}
\end{center}
\end{figure}

\smalltitle{Features of a realistic model}
The combinatorial properties of RNA structures have been thoroughly
studied~\cite{waterman78,Vauchaussade85,Ne01,Fontana1993,hofacker99a,Jin2008c}. The asymptotical
analysis of the uniform model~\cite{Ne01,Po03} shows striking
dissimilarities between the structural features of the uniform model and those experimentally
observed. By structural features, one understands:
\begin{itemize}
	\item Proportions of paired and unpaired bases
	\item Numbers and average size of hairpin, bulge, interior, and terminal loops
\end{itemize}
Figure~\ref{fig:RNAAnnotate} (upper-left) illustrates the principle of a loop decomposition, underlying
the so-called Turner model of energy~\cite{turner}.
We show how weighted
grammars provide in such a case with an elegant way to build a model that captures observed
properties.

\smalltitle{Annotation of existing structures} First, we evaluate our features on a database of known RNA secondary structures~\cite{Mathews2004},
previously used to benchmark thermodynamics based approaches for the ab-initio folding problem. To that purpose, we {\bf annotate} these secondary structures as
follows:
\begin{itemize}
	\item[-]Replace each base with a character depending on the type of loop it belongs to: Hairpin (\hairpin), Bulges (\bulge), Terminal loops (\term),
	Interior loops (\interior) or Multiple loops (\multi).
	\item[-]Bold characters (\Hairpin, \Bulge, \Term, \Interior, and \Multi) are used for the first element of each loop.
\end{itemize}
The result of this  process is illustrated by Figure~\ref{fig:RNAAnnotate}.
Through a carefully designed recursive scheme, this operation can be performed in linear time. We
get the following frequencies for each characters among the whole database of secondary structures:
\begin{center}
\begin{tabular}{|c|cccccccccc|}\hline
Feature                 &\Bulge& \bulge & \Interior & \interior & \Multi & \multi & \Term & \term & \Hairpin& \hairpin\\
\hline \hline
{\bf Target} freq. (\%) &1.5   & 2.3    & 1.9       & 11.2      & 1.1    & 9.0    & 2.6    & 16.6 & 4.8     & 48.9\\
\hline
\end{tabular}
\end{center}

\smalltitle{Structural features of the uniform model}
Then, we use a general grammar, independently proposed by one of the authors~\cite{Po03}
and M. Nebel~\cite{Ne04}, from which these features can be distinguished:
\begin{eqnarray*}
	\ARN & \to & \TermLoop\unrule \HairpinLoop\;\unrule \BulgeLoop\;\HairpinLoop\unrule \HairpinLoop\;\BulgeLoop \unrule
	\Interior\;\InteriorLoop\;\HairpinLoop\;\InteriorLoop\;\interior \unrule \MultiLoop \unrule \varepsilon\\
	\TermLoop & \to & \Term\;\term^{\tau-1} \unrule \TermLoop\;\term \\
	\BulgeLoop & \to & \Bulge \unrule \BulgeLoop\;\bulge \\
	\InteriorLoop & \to &  \varepsilon \unrule  \InteriorLoop\;\interior\\
	\HairpinLoop & \to & \Hairpin\;\HairpinLoop' \;\hairpin\\
	\HairpinLoop' & \to & \hairpin\;\HairpinLoop' \;\hairpin \unrule \TermLoop \unrule \BulgeLoop\;\HairpinLoop \unrule \HairpinLoop\;\BulgeLoop \unrule
	\Interior\;\InteriorLoop\;\HairpinLoop\;\InteriorLoop\;\interior \unrule \MultiLoop\\
	\MultiLoop & \to & \HairpinLoop\;\MultiLoop \unrule \Multi\;\MultiLoopAux\;\HairpinLoop\;\MultiLoop'\\
						 & | & \Multi\;\MultiLoopAux\;\HairpinLoop\;\MultiLoopAux\;\HairpinLoop\;\MultiLoopAux \\
						 & | & \HairpinLoop\;\Multi\;\MultiLoopAux\;\HairpinLoop\;\MultiLoopAux \\
						 & | & \HairpinLoop\;\HairpinLoop\;\Multi\;\MultiLoopAux \\
	\MultiLoop'  & \to & \MultiLoopAux\;\HairpinLoop\;\MultiLoop'\\
						 & \to & \MultiLoopAux\;\HairpinLoop\;\MultiLoopAux\;\HairpinLoop\;\MultiLoopAux \\
	\MultiLoopAux &\to & \MultiLoopAux\;\multi \unrule \varepsilon
\end{eqnarray*}
This grammar ensures that at least $\tau$ unpaired bases are found in each terminal loop.
Additionally, this grammar requires at least one unpaired base to be found in each multiple loop, since
we need to \emph{mark} each occurrence of a multiple loop with a character $\Multi$.

A combinatorial \emph{validation} for this complex grammar can be found in the following way: Set
$\tau=1$; Replace $\MultiLoop$ by $\MultiLoop'$ in the right hand sides of the grammar; Translate
the grammar into a system of functional equations on the univariate generating functions associated
with each non-terminal; Solve the algebraic system. We obtain the generating function of RNA secondary structures as first counted by
Waterman~\cite{waterman78}. It is worth noticing that doing the same with $\tau=0$ gives the Motzkin numbers.
Therefore we claim that the restrictions imprinted in our grammar only induce a \emph{controlled} and
\emph{biologically relevant} loss of generality.

In the rest of this study, we will focus on RNA structures having 300 nucleotides. We use {\GRGFreqs} to evaluate the exact
expected frequencies for each of the terminal symbols in the uniform model \UniformModel, and obtain the following frequencies:
\begin{center}
\begin{tabular}{|c|cccccccccc|}\hline
Feature                  &\Bulge& \bulge & \Interior & \interior & \Multi& \multi & \Term & \term & \Hairpin& \hairpin\\ \hline \hline
{\UniformModel}  (\%)&7.2  & 5.6   & 2.8      & 7.3      & 3.7  & 7.6    & 5.2   & 14.5  &{\bf 18.6}&{\bf 27.5}\\ \hline \hline
{\bf Target} &1.5   & 2.3    & 1.9       & 11.2      & 1.1    & 9.0    & 2.6    & 16.6 &{\bf 4.8}&{\bf 48.9}\\
\hline
\end{tabular}
\end{center}

\begin{figure}[t]
\begin{center}
	\Fig{\includegraphics[angle=270,width=6.2cm]{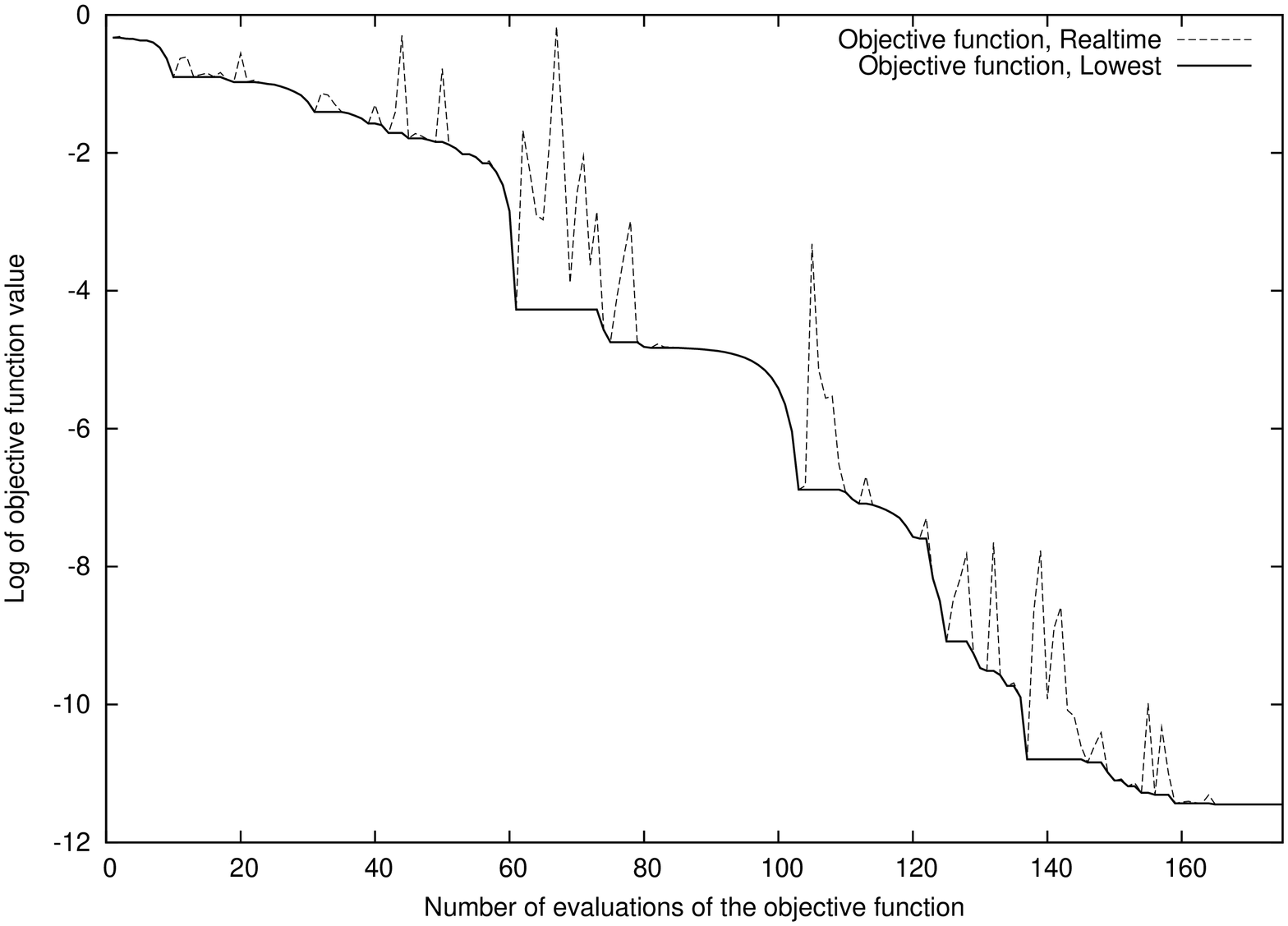}}
	\Fig{\includegraphics[angle=270,width=6.2cm]{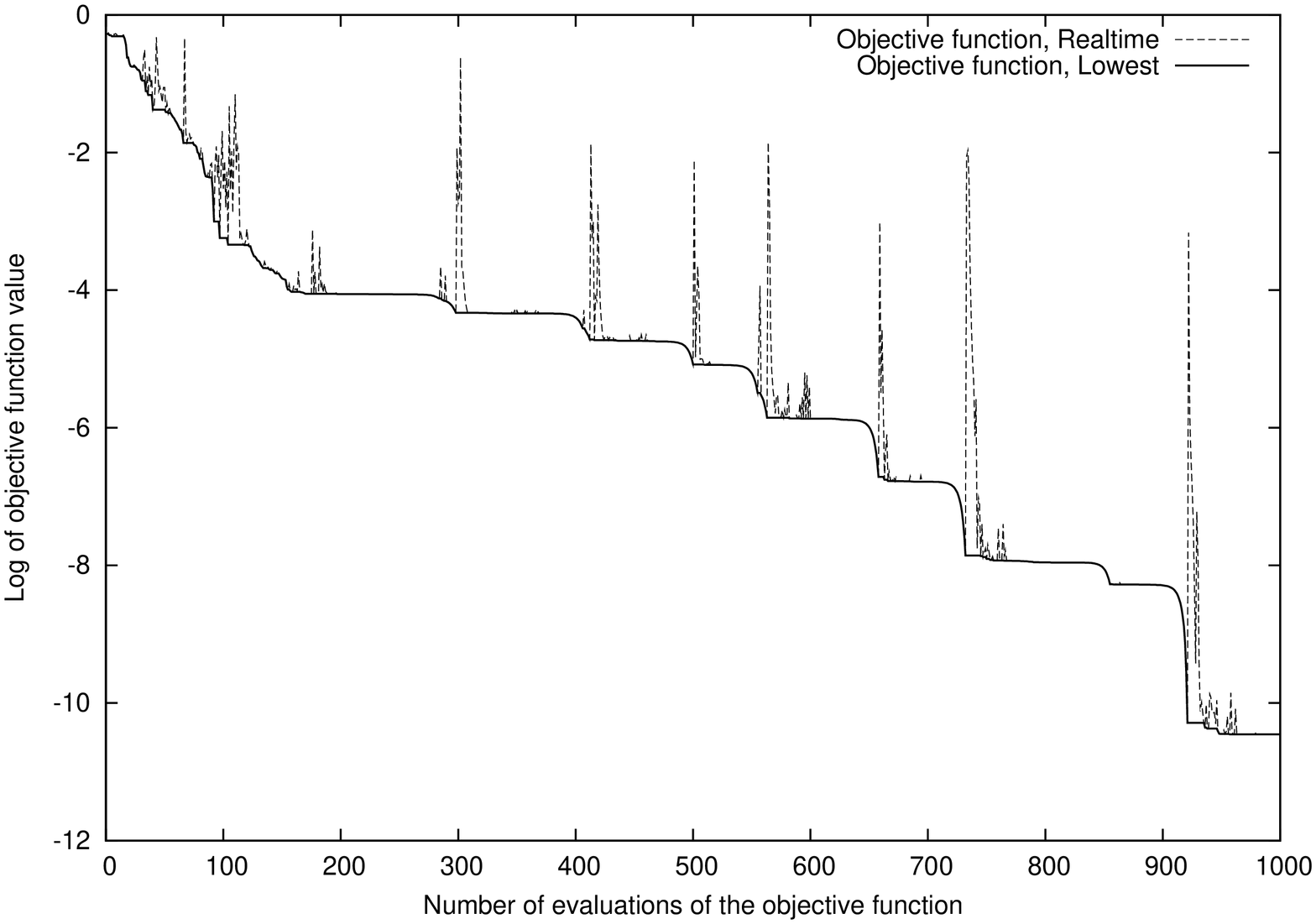}}
\caption{Minimization of the objective functions in the {\bf Helices} (Left) and {\bf Loops} (Right) models. A logarithmic
	scale is used for the value of the objective function (Y-axis).}
\label{fig:OptScenario}
\end{center}
\end{figure}
\smalltitle{Adequate weights for hairpins}
Since the optimizer complexity empirically grows quickly with the number of variables, we will first focus on hairpin features,
for which the highest discrepancy is observed between the uniform model and real structures.
Namely, we will build an {\bf Helix model} \HelicesModel, that achieves average expected lengths and frequencies for
hairpins similar to that of real structures.
We slightly alter the general grammar in order to \emph{anonymize} all symbols for
which we do not need a specific weight to be computed (\Bulge, \bulge, \Interior, \interior, \Multi, \multi, {\Term} and \term),
replacing them with a generic letter \Unpaired. The respective targeted frequencies $(\TargFreq_{\Unpaired},\TargFreq_{\Hairpin},\TargFreq_{\hairpin})$
for $\Unpaired$, $\Hairpin$ and $\hairpin$ are then such that
$$ \TargFreq_{\Unpaired} = 46.3\quad \quad \TargFreq_{\Hairpin} = 4.8 \quad \quad \TargFreq_{\hairpin} = 48.9$$
We run {\GRGFreqs} with these settings, and observe the optimization scenario from Figure~\ref{fig:OptScenario} (Left part).
After only $150$ evaluations of $\ObjFun$, a candidate set of weights for $\Unpaired$, $\Hairpin$ and $\hairpin$ is found such
that associated frequencies only deviate by less than $e^{-11}\approx 1.6\;10^{-5}$ from the target frequencies. Namely, we get
$$ \WF_{\Unpaired}^{\mathcal{H}} = 1.0\quad \quad \WF_{\Hairpin}^{\mathcal{H}} \approx 3.6036391\,10^{-3} \quad \quad \WF_{\hairpin}^{\mathcal{H}} \approx 1.1359318$$
Using these weights, we can exactly compute the frequencies for the full set of atoms in the {\bf Helix} model \HelicesModel:
\begin{center}
\begin{tabular}{|c|cccccccccc|}\hline
Features           &\Bulge& \bulge & \Interior & \interior & \Multi& \multi & \Term & \term & \Hairpin& \hairpin\\ \hline \hline
{\HelicesModel} (\%)      &0.6  & 2.3     & 1.2      & 10.4       & {\it 1.8}   & {\it 15.5}   & 2.2   & 13.0  & {\bf 4.8}& {\bf 48.9}\\ \hline \hline
{\bf Target}         &1.5   & 2.3     & 1.9      & 11.2       & {\it 1.1}   & {\it 9.0}    & 2.6   & 16.6  & {\bf 4.8}      & {\bf 48.9}\\
\hline
\end{tabular}
\end{center}

\smalltitle{Adding constraints to multiple loops}
From the values just above, we can see that the biggest divergence between the model {\HelicesModel} and real data
resides in multiple loops. Since these act indirectly on the connectivity of the \emph{tree backbone} of sampled
structures, it may be useful to further constraint associated features (Characters {\Multi} and \multi).
Therefore we propose a {\bf loop model} $\LoopsModel$ which adds {\Multi} and {\multi} to the constraints of the previous model helix model:
$$ \TargFreq_{\Unpaired} = 37.3\quad \quad \TargFreq_{\Multi} = 1.1 \quad \quad \TargFreq_{\multi} = 9.0 \quad \quad \TargFreq_{\Hairpin} = 4.8 \quad \quad \TargFreq_{\hairpin} = 48.9$$
Running {\GRGFreqs} with these new settings yields a set of weights $\WF_\mathcal{L}$, that scores less than $e^{-10.5} \approx 2.76\cdot 10^{-5}$,
after about 1000 evaluations of the objective function.
$$ \WF_{\multi}^{\mathcal{L}} = 1.0 \quad \WF_{\Unpaired}^{\mathcal{L}} \approx 1.138626\quad \WF_{\Multi}^{\mathcal{L}} \approx 2.168521 \quad
\WF_{\Hairpin}^{\mathcal{L}} \approx 3.422990\,10^{-3} \quad \WF_{\hairpin}^{\mathcal{L}} \approx 1.246468$$

\begin{center}
\begin{tabular}{|c|cccccccccc|}\hline
Feature                  &\Bulge& \bulge & \Interior & \interior & \Multi         & \multi       & \Term & \term    & \Hairpin     &  \hairpin\\ \hline\hline
{\LoopsModel} (\%)       &0.6   & 3      & 1.5       & 15.9      & {\bf 1.1}      & {\bf 9.0}    & 1.9   & 13.2     & {\bf 4.8}    & {\bf 48.9}\\ \hline\hline
{\bf Target}             &1.5   & 2.3    & 1.9       & 11.2      & {\bf 1.1}      & {\bf 9.0}    & 2.6   & 16.6     & {\bf 4.8}    & {\bf 48.9}\\
\hline
\end{tabular}
\end{center}
\begin{figure}[p]
\begin{center}
\fbox{
\begin{tabular}{c}
\Fig{\includegraphics[width=2cm]{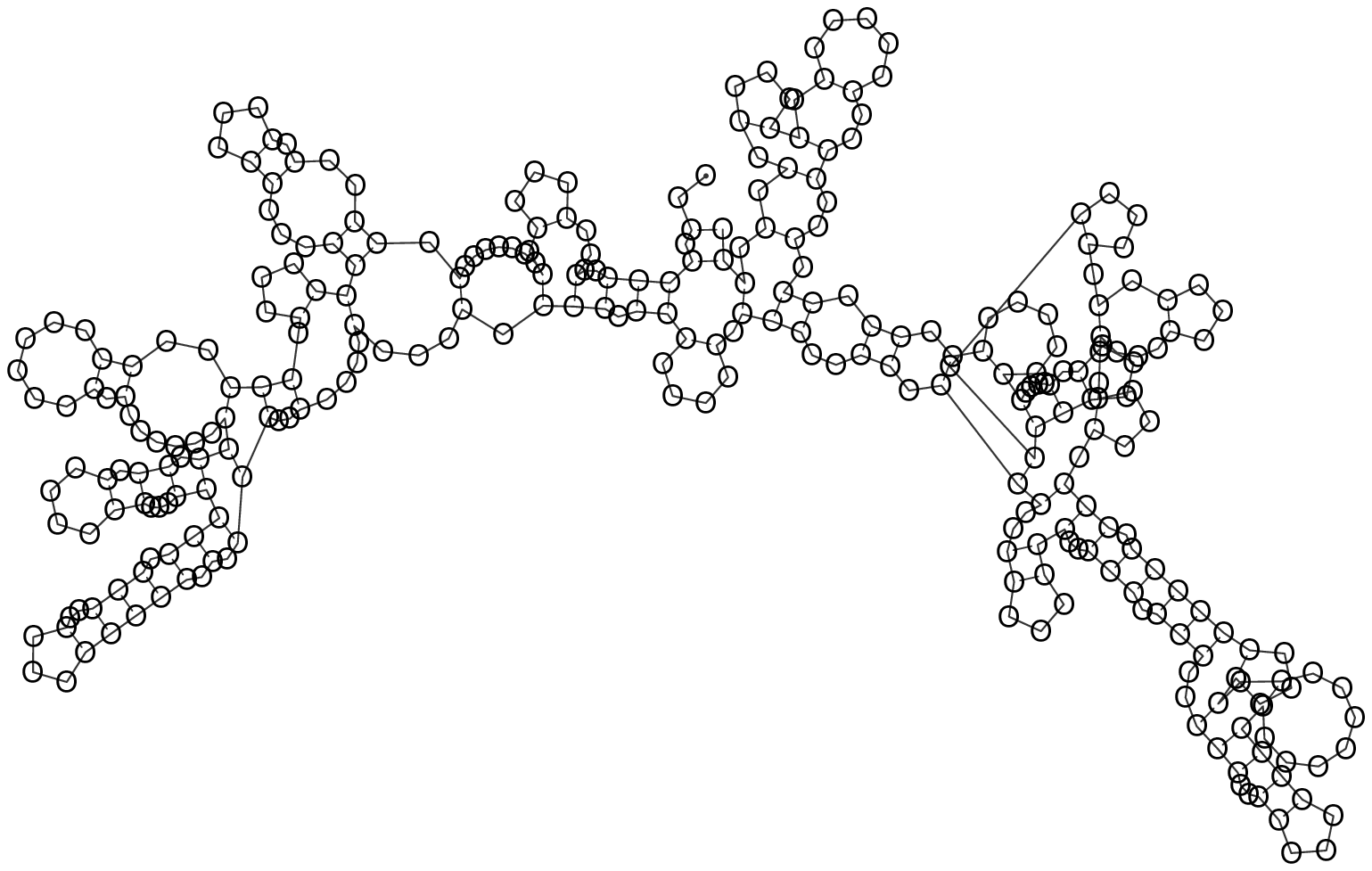}}
\Fig{\includegraphics[width=2cm]{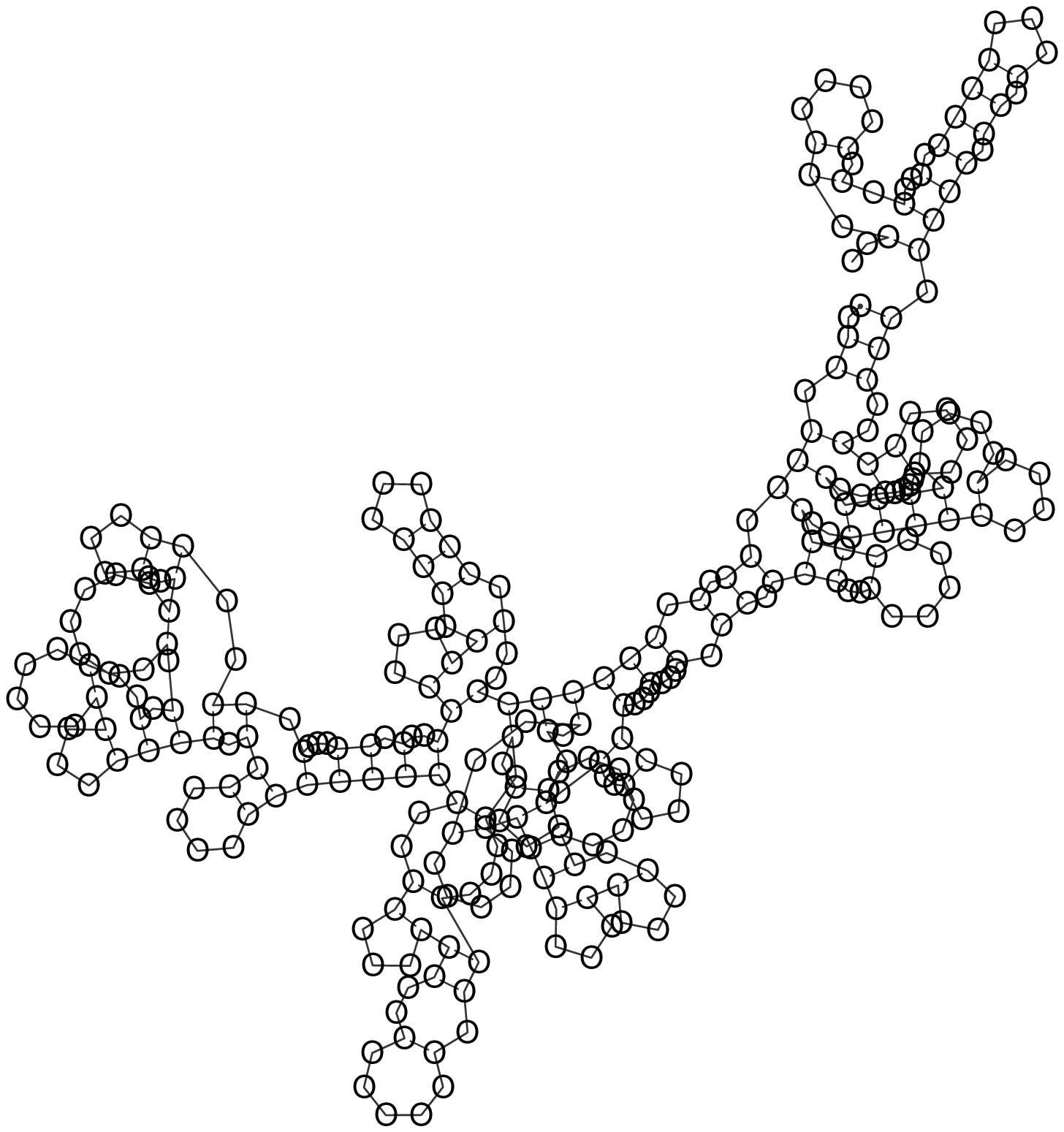}}
\Fig{\includegraphics[width=2cm]{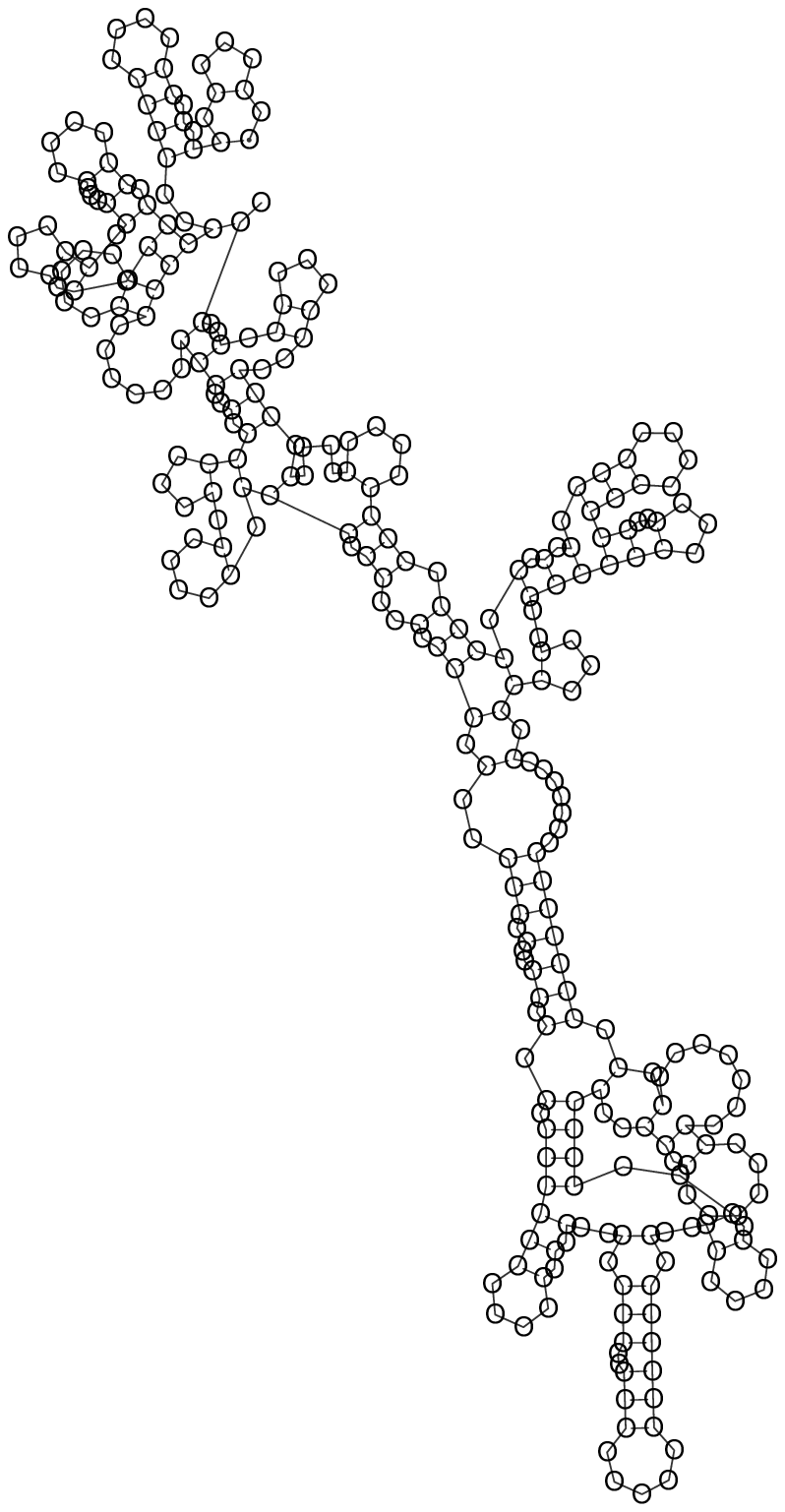}}
\Fig{\includegraphics[width=2cm]{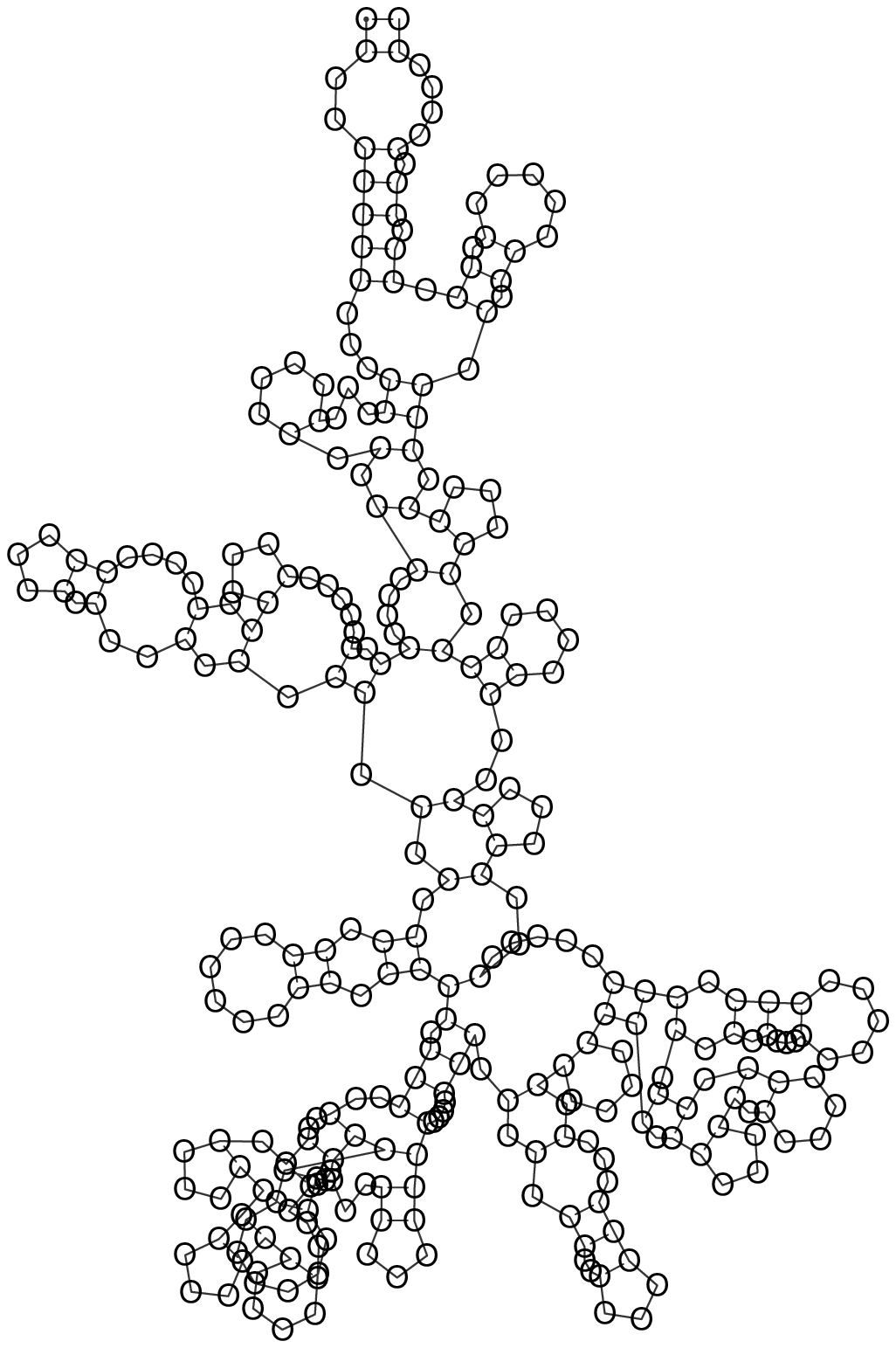}}
\Fig{\includegraphics[width=2cm]{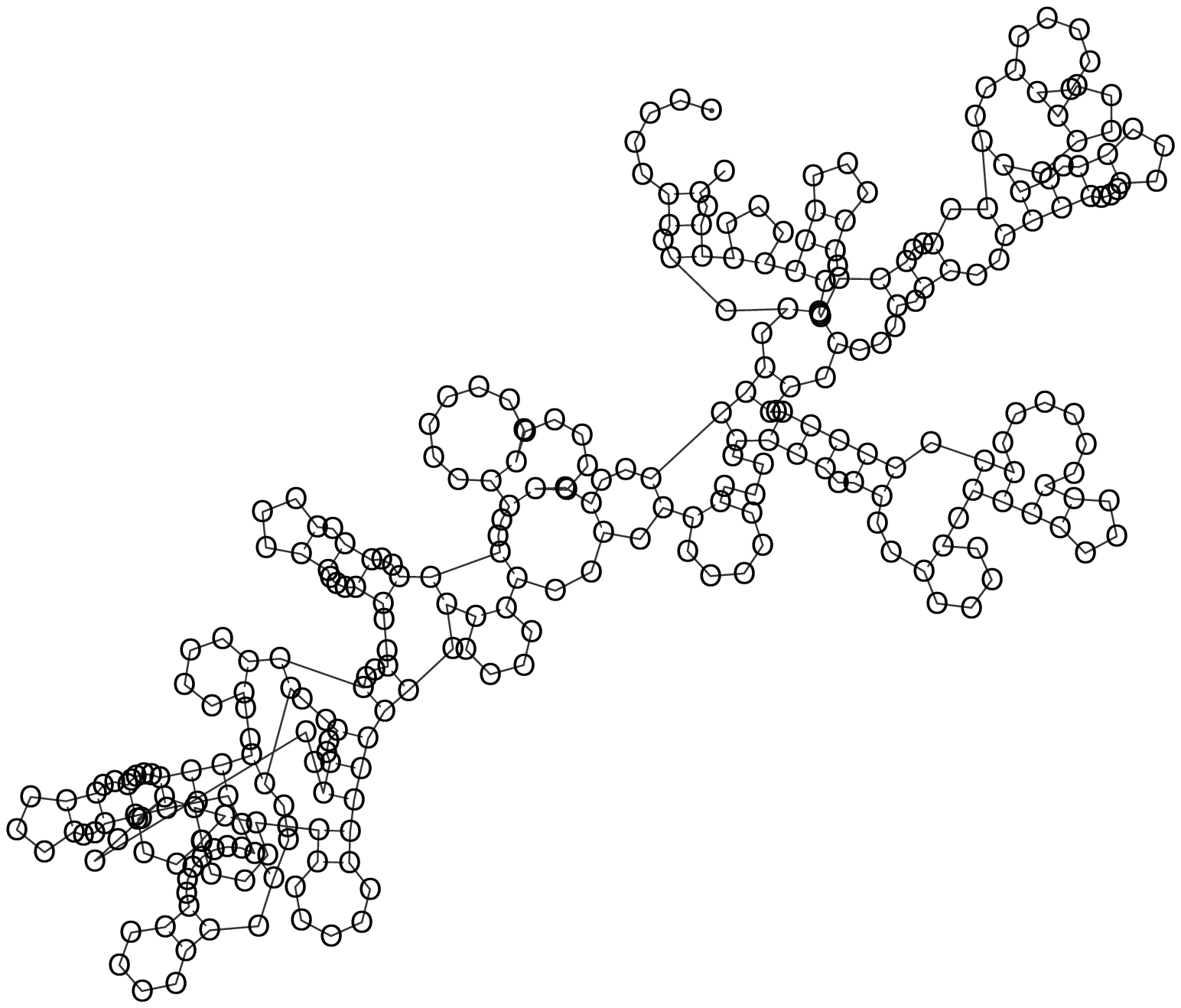}}
\Fig{\includegraphics[width=2cm]{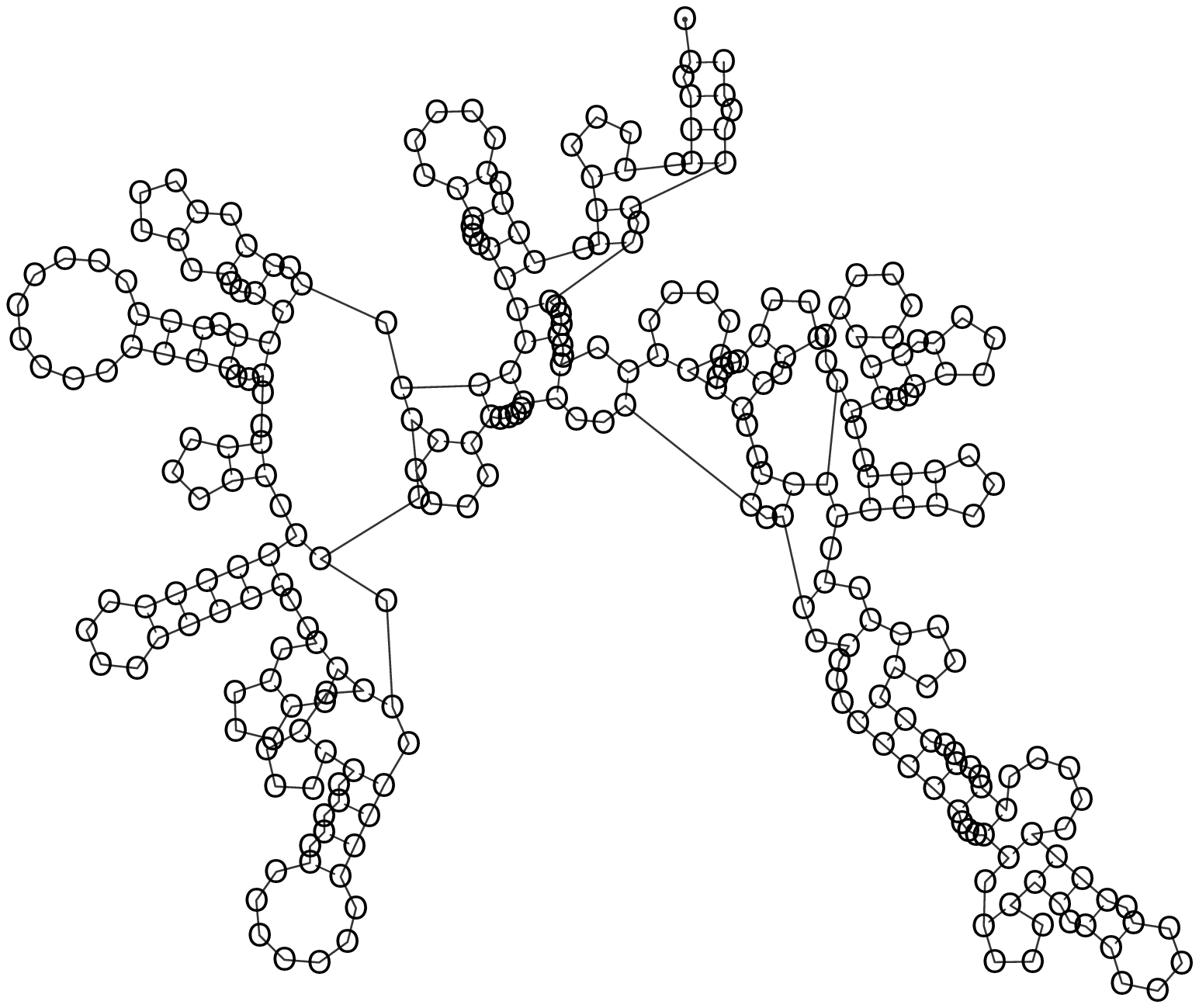}}\\
\Fig{\includegraphics[width=2cm]{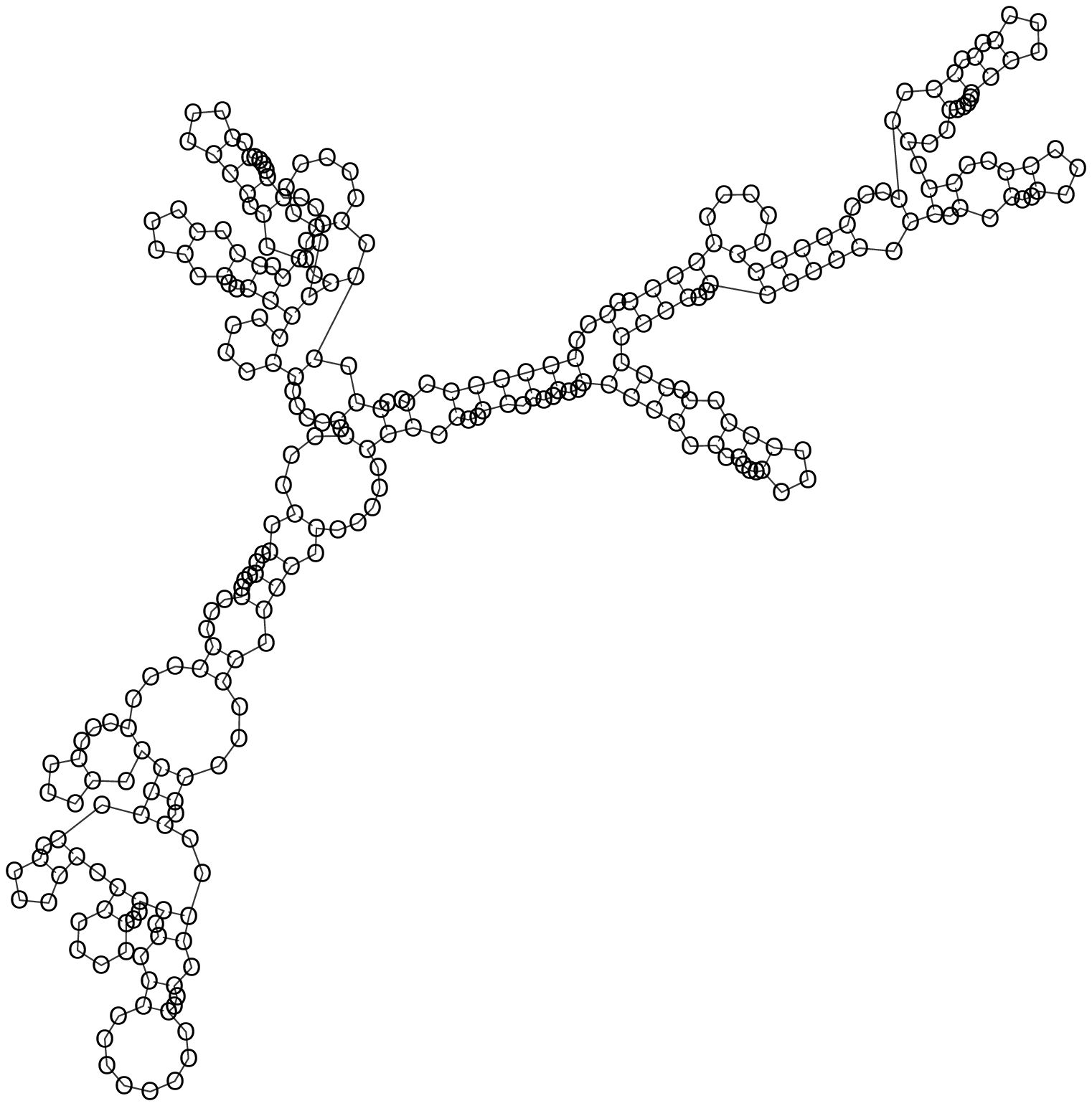}}
\Fig{\includegraphics[width=2cm]{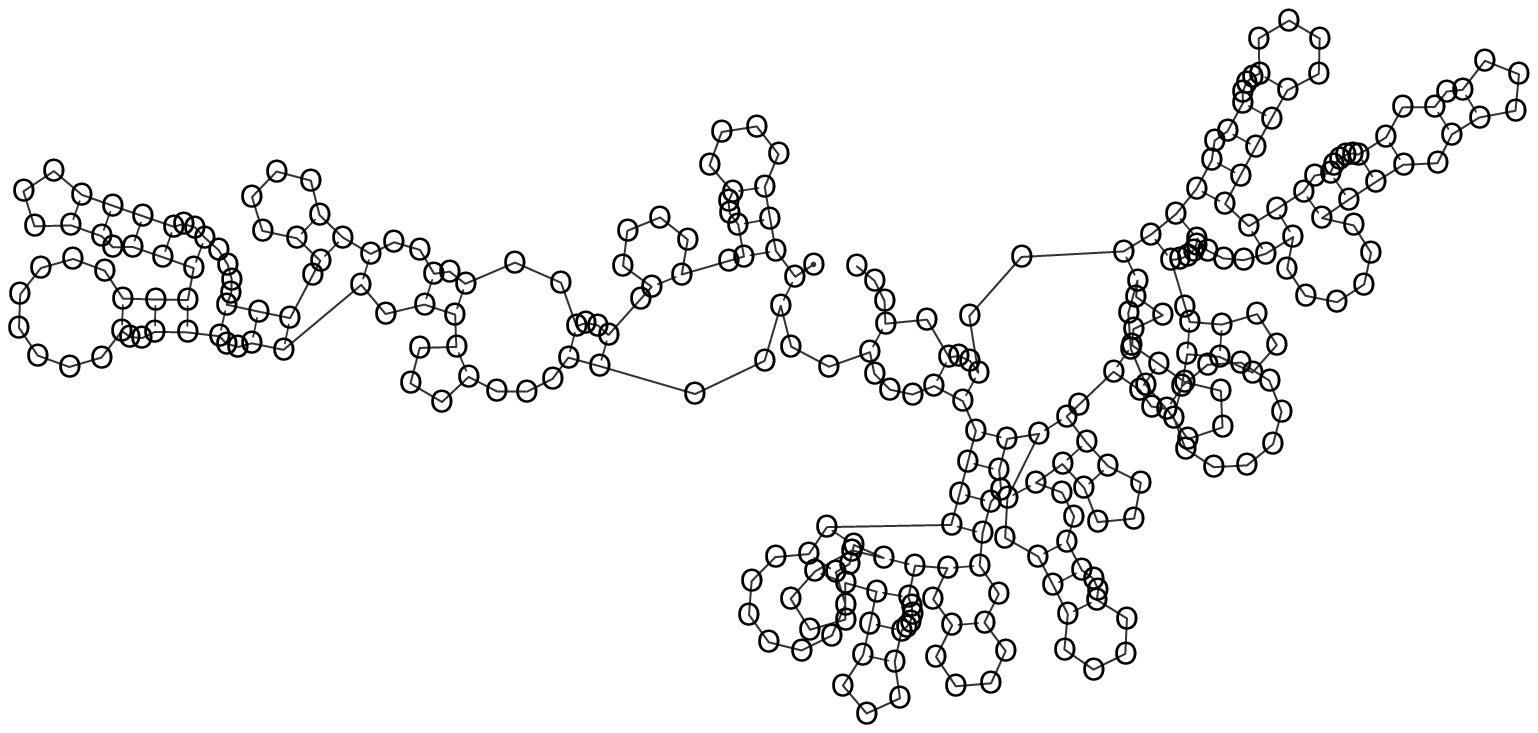}}
\Fig{\includegraphics[width=2cm]{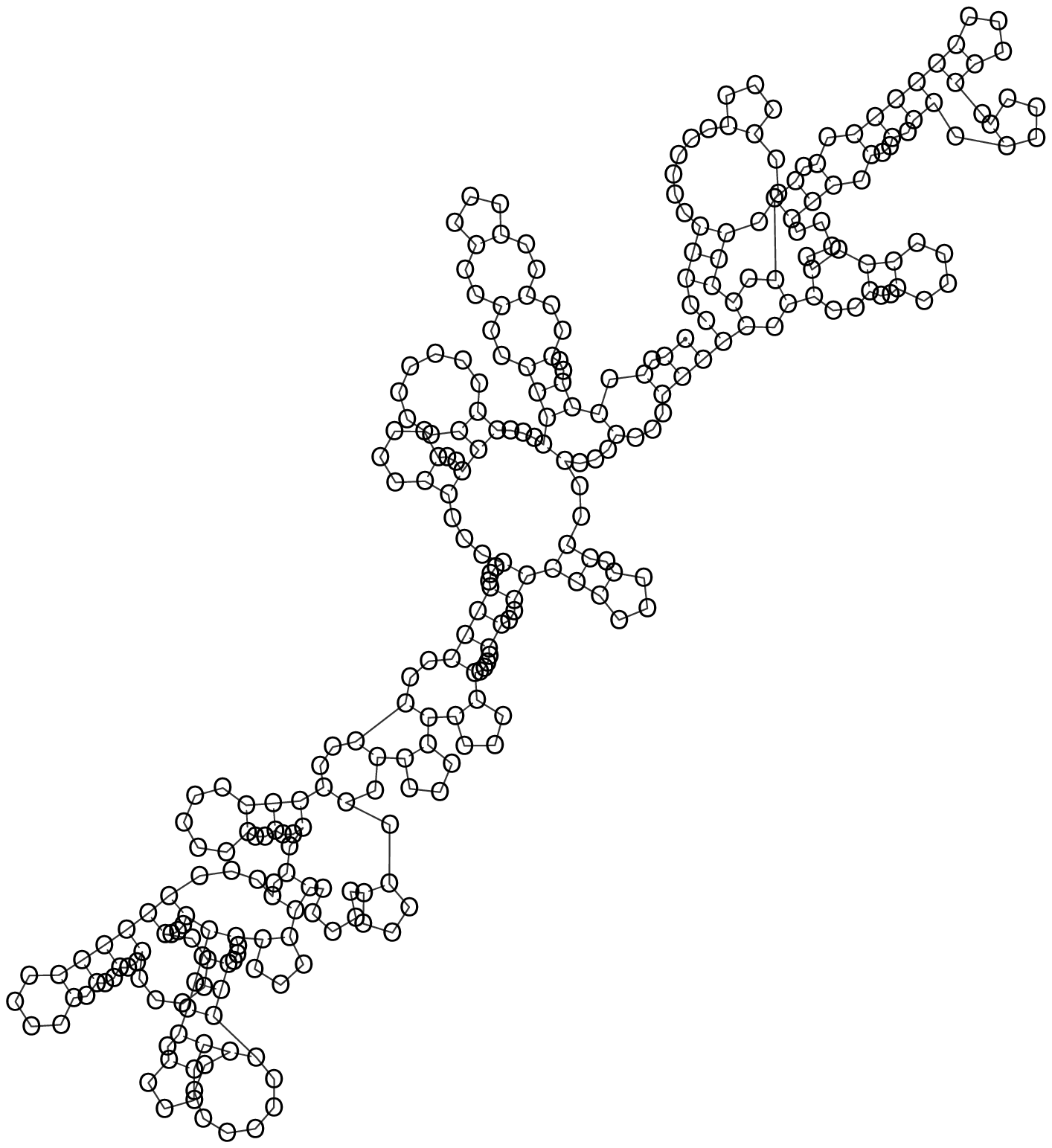}}
\Fig{\includegraphics[width=2cm]{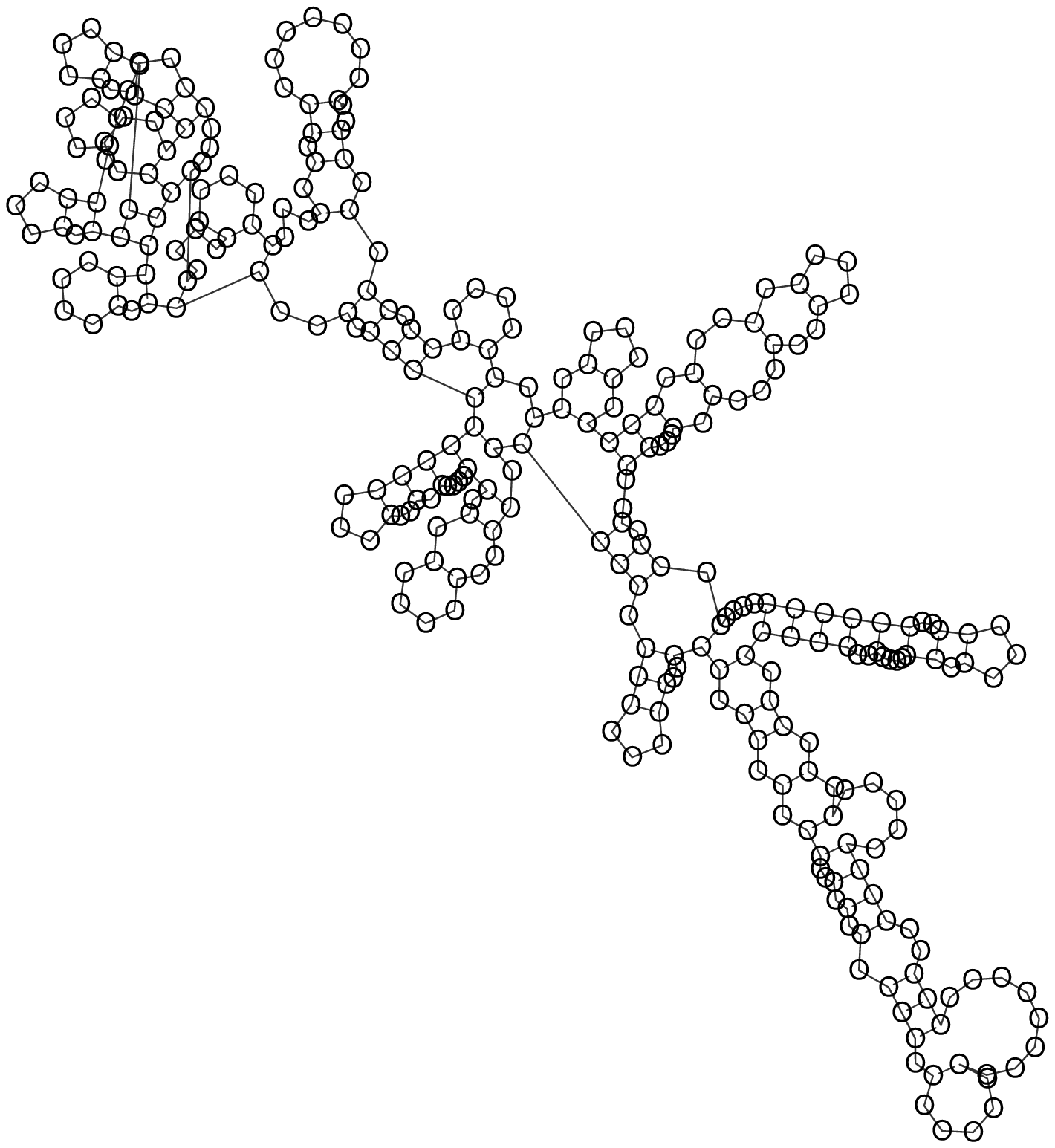}}
\Fig{\includegraphics[width=2cm]{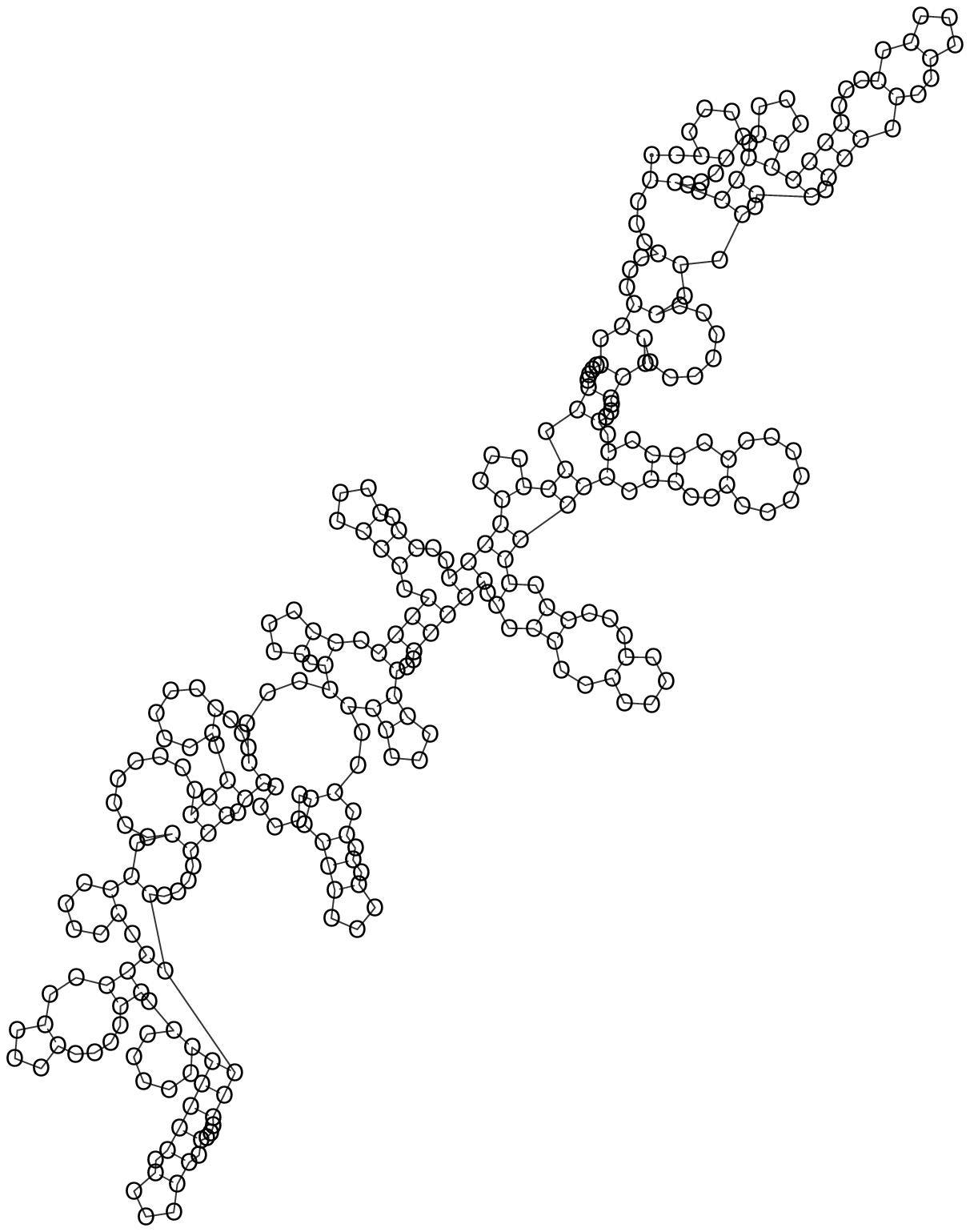}}
\Fig{\includegraphics[width=2cm]{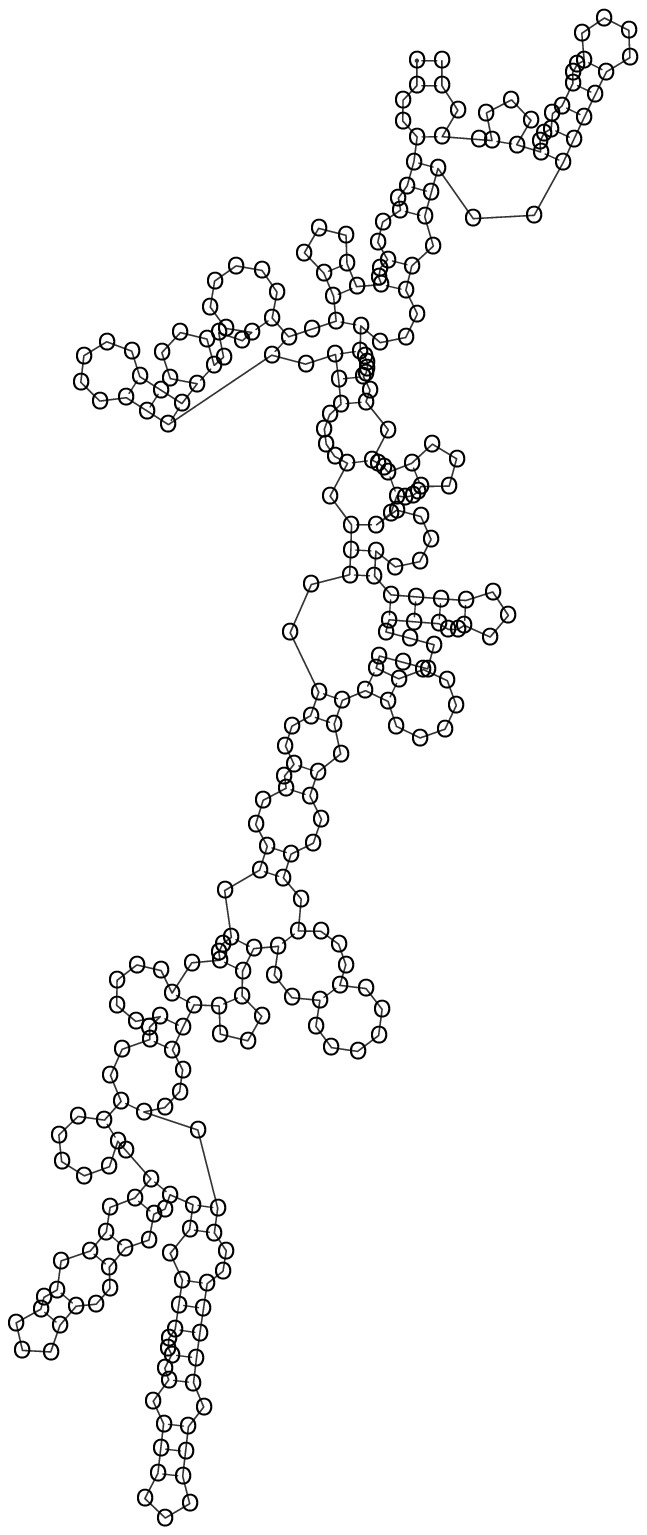}}
\end{tabular}}\\
{\bf Uniform model \UniformModel}
\\[0.5cm]
\fbox{
\begin{tabular}{c}
\Fig{\includegraphics[width=2cm]{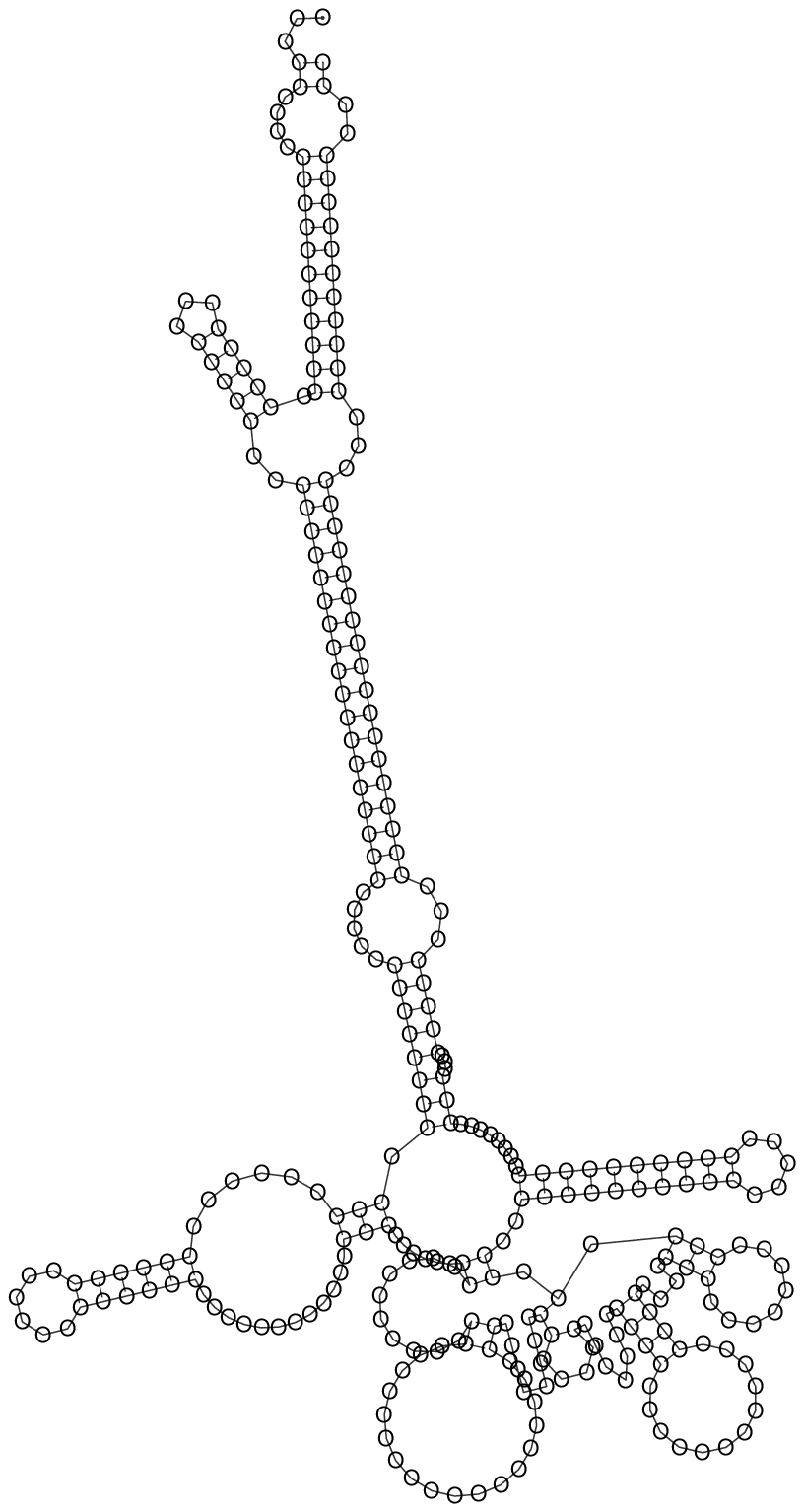}}
\Fig{\includegraphics[width=2cm]{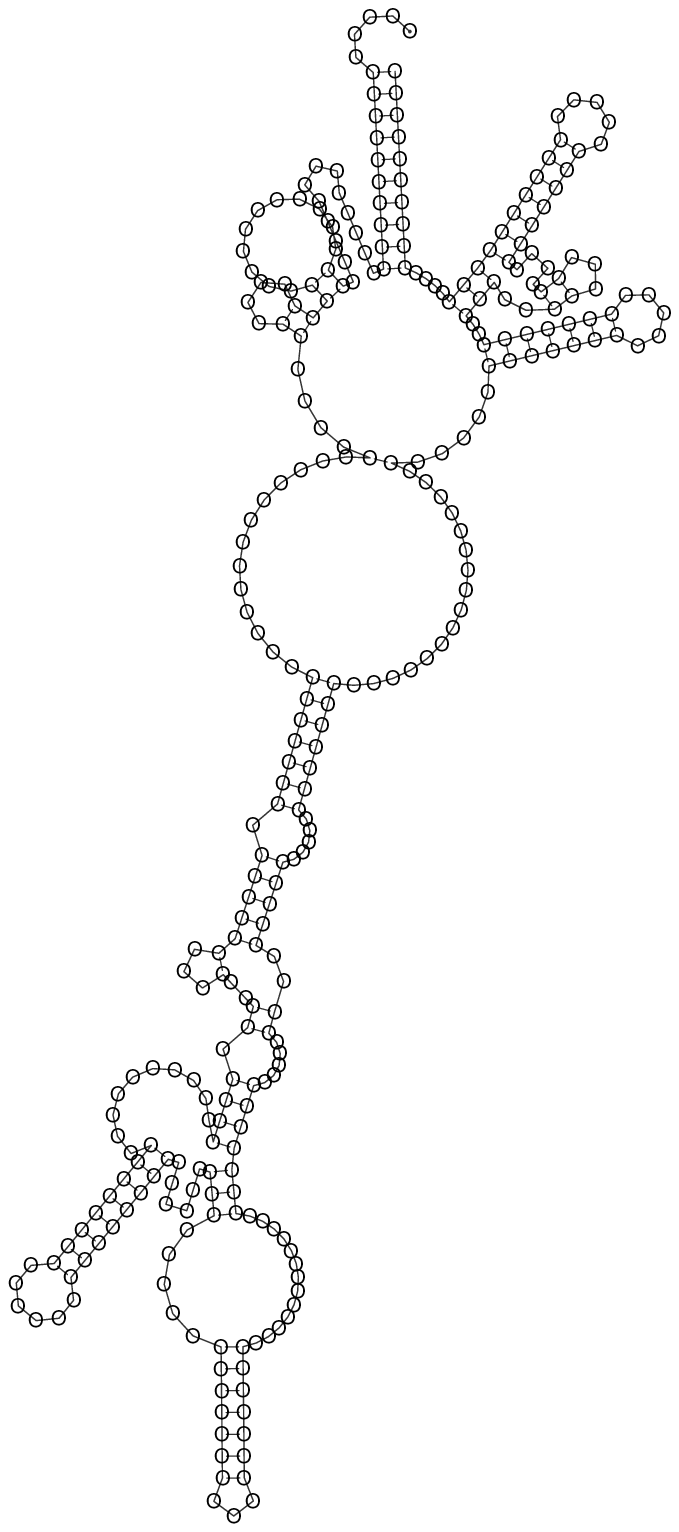}}
\Fig{\includegraphics[width=2cm]{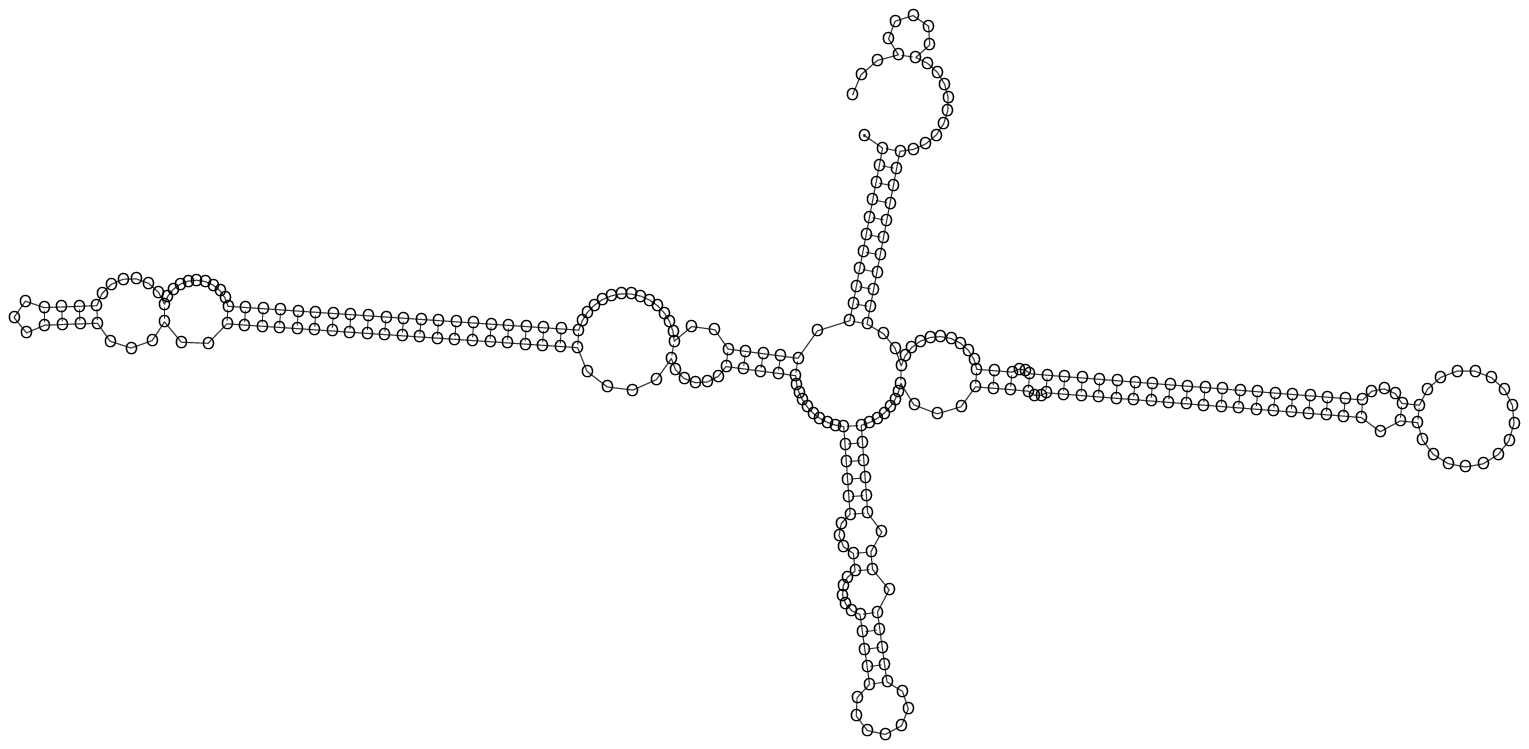}}
\Fig{\includegraphics[width=2cm]{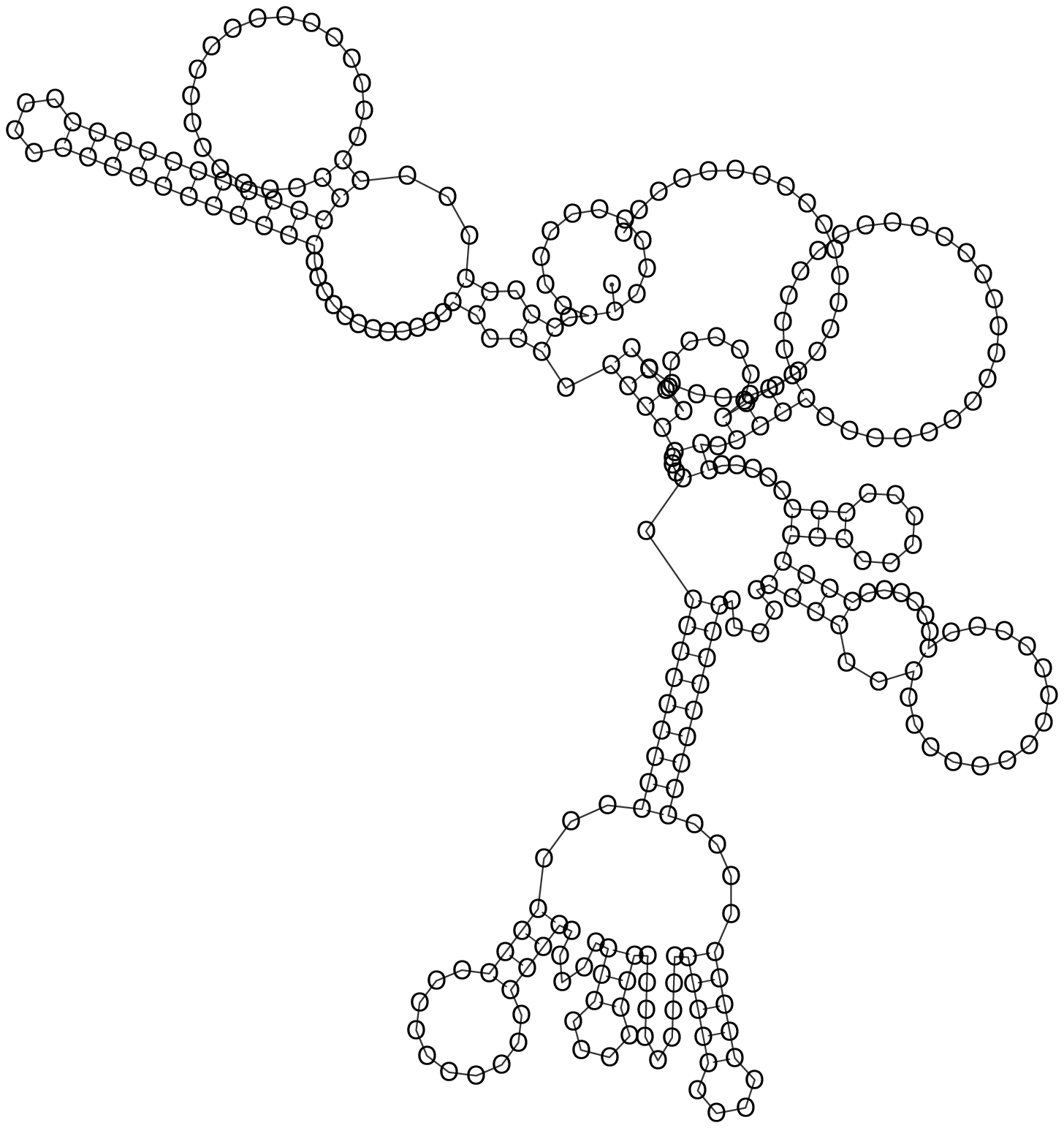}}
\Fig{\includegraphics[width=2cm]{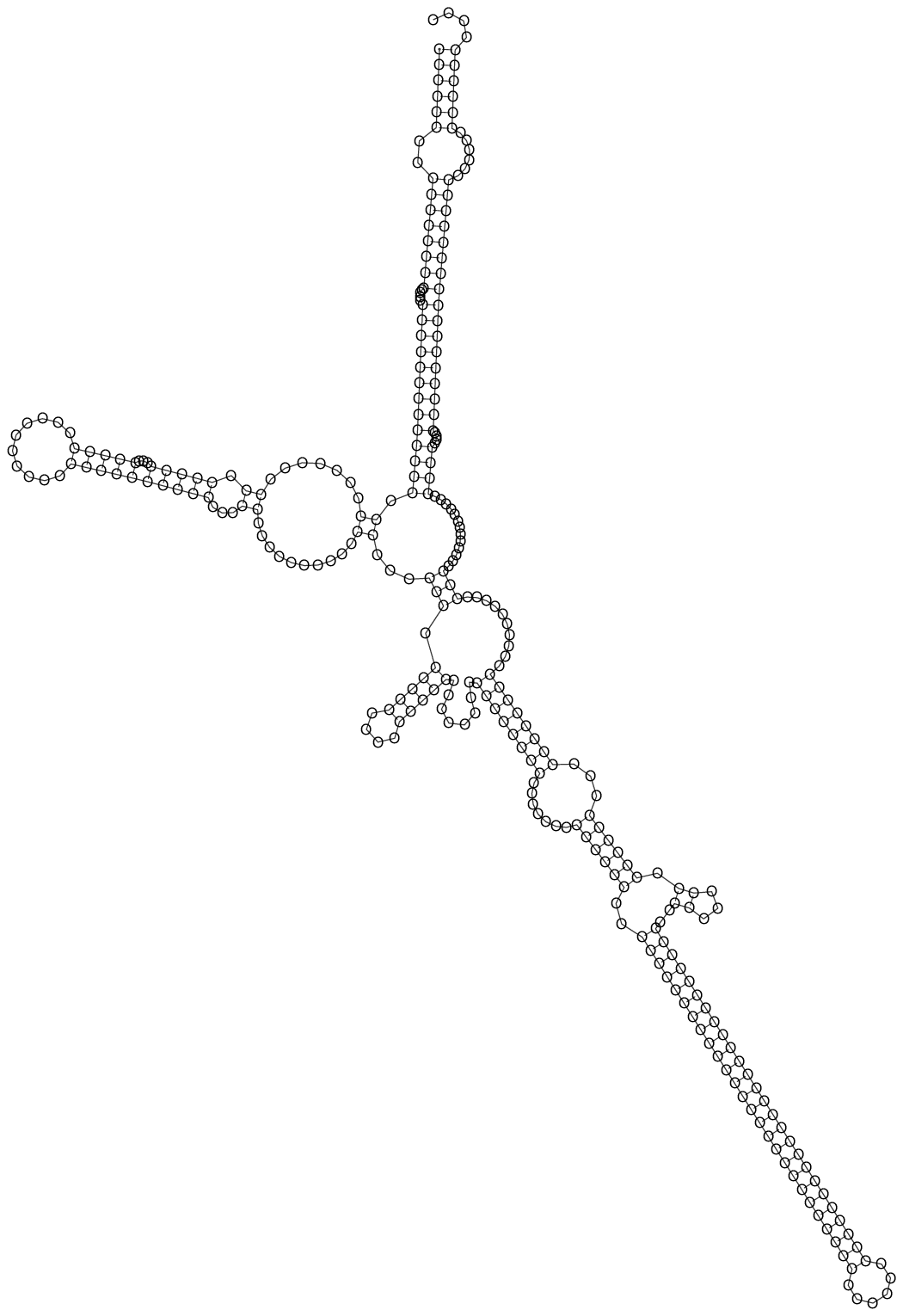}}
\Fig{\includegraphics[width=2cm]{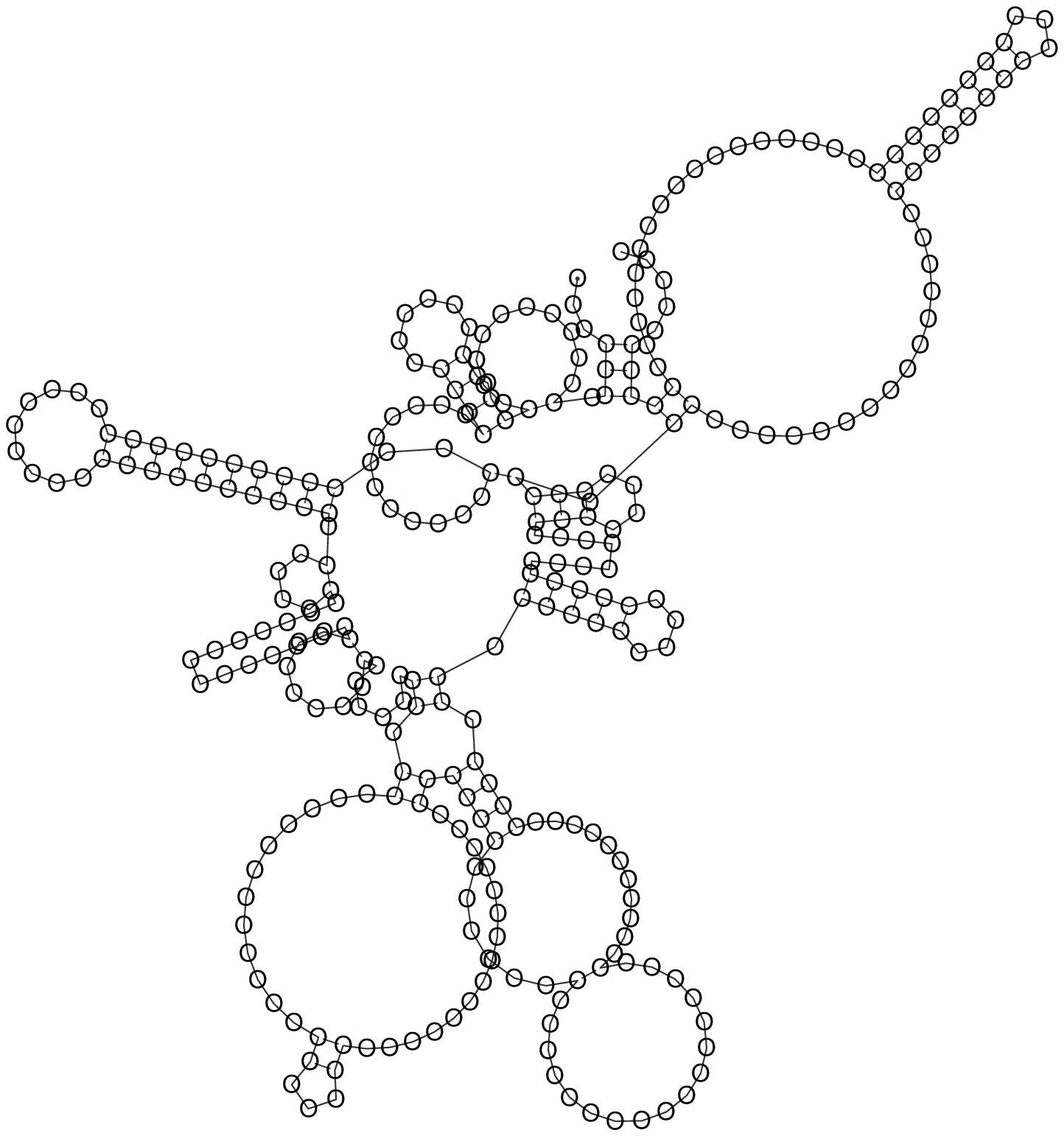}}\\
\Fig{\includegraphics[width=2cm]{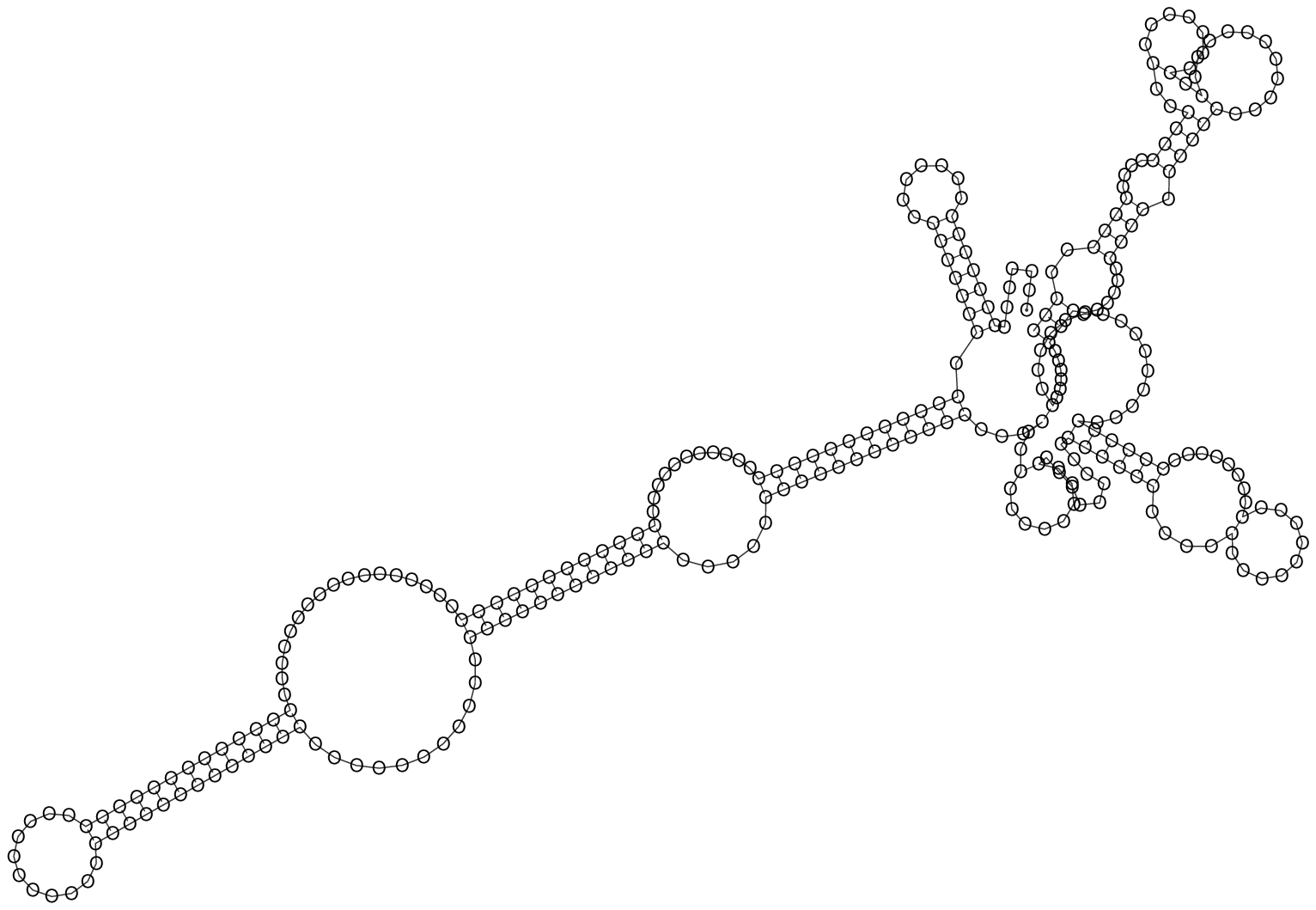}}
\Fig{\includegraphics[width=2cm]{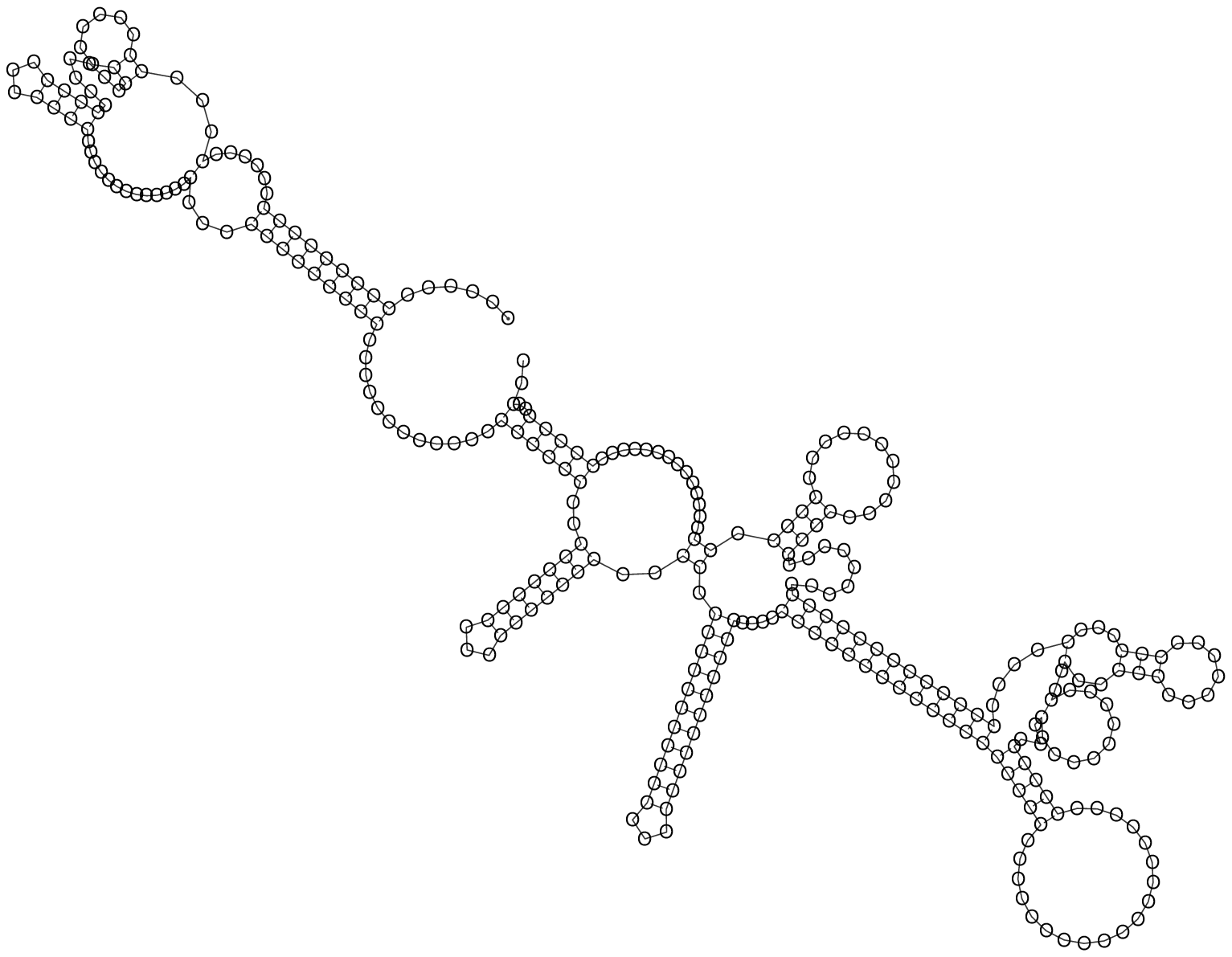}}
\Fig{\includegraphics[width=2cm]{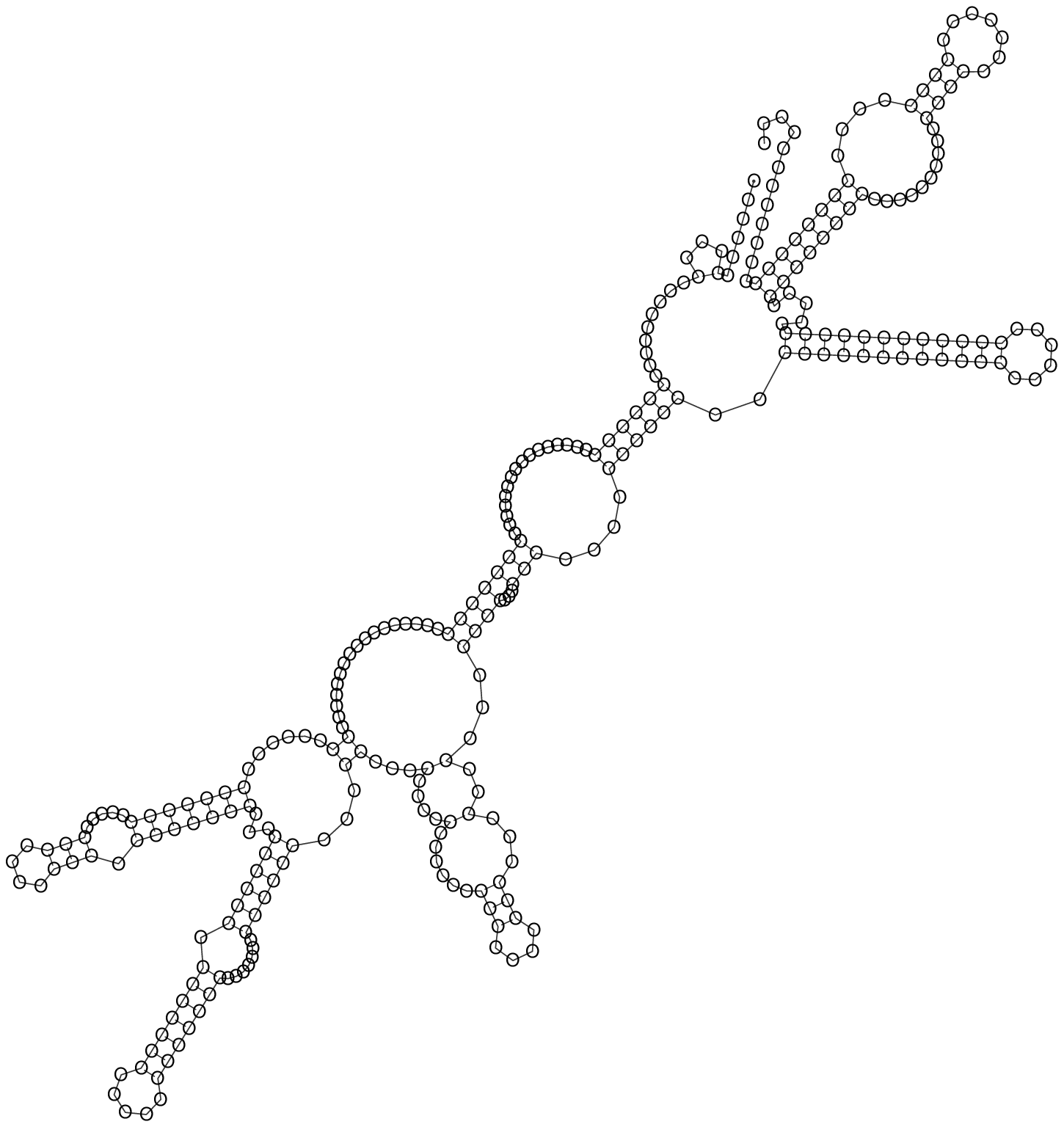}}
\Fig{\includegraphics[width=2cm]{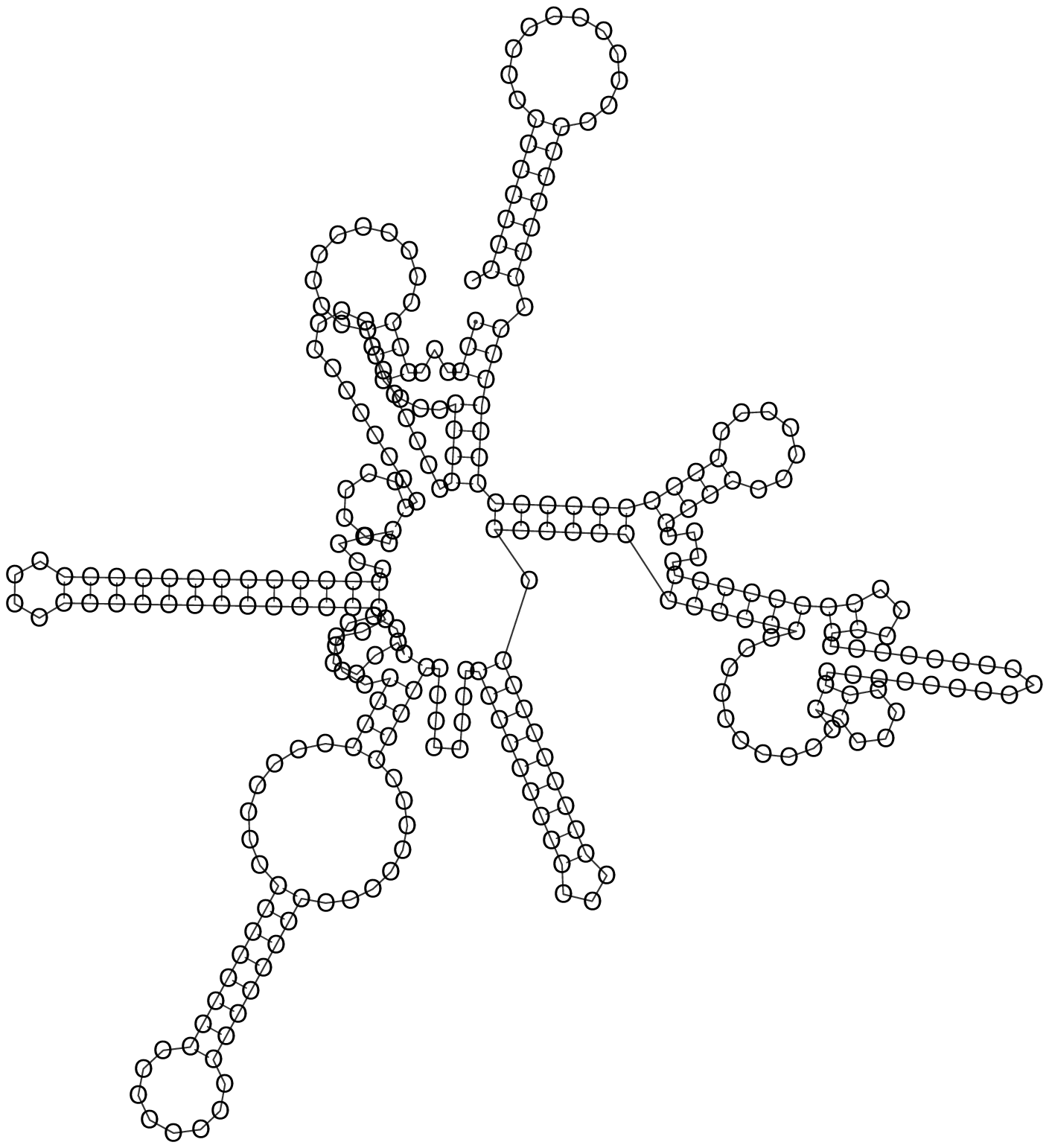}}
\Fig{\includegraphics[width=2cm]{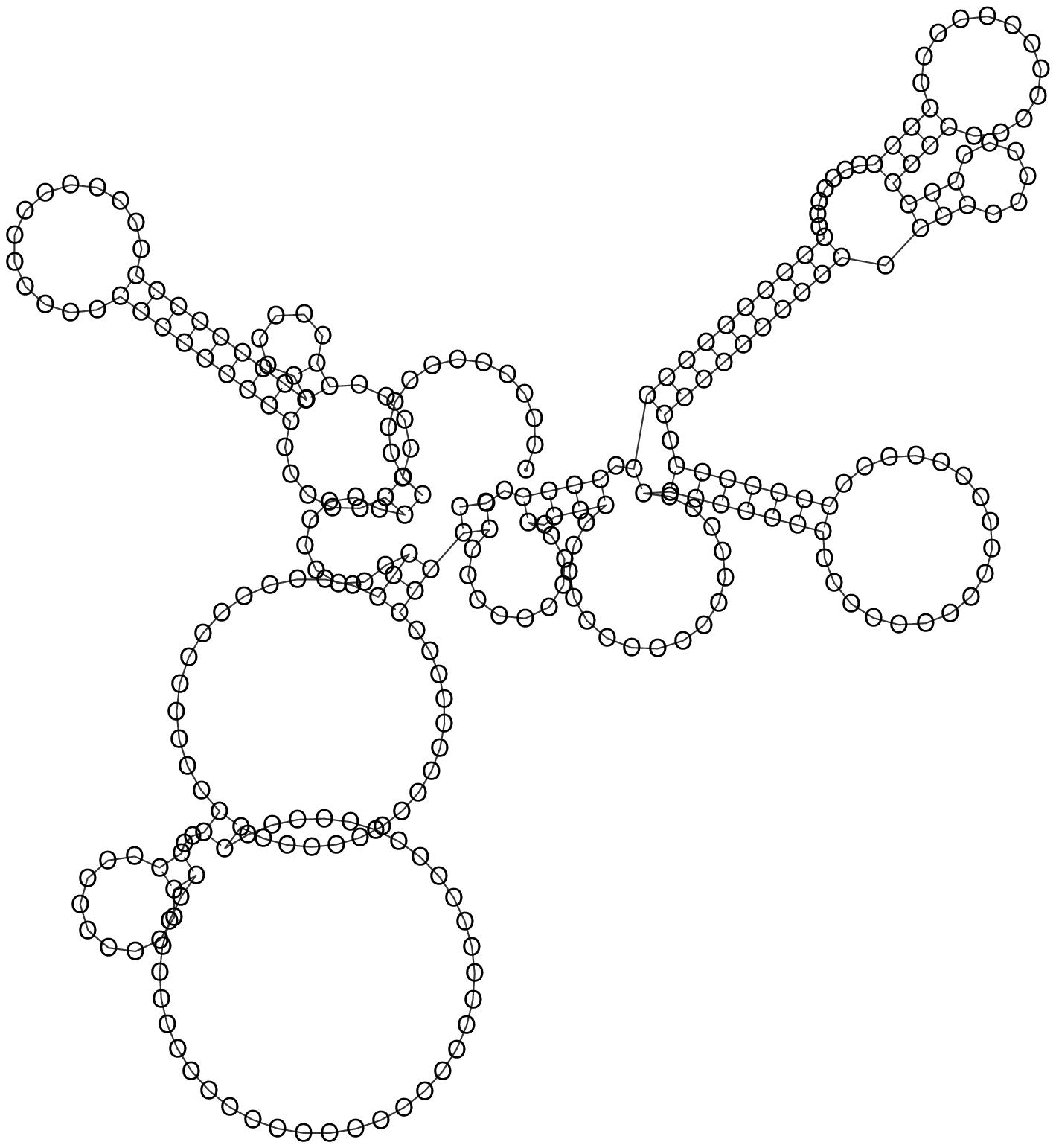}}
\Fig{\includegraphics[width=2cm]{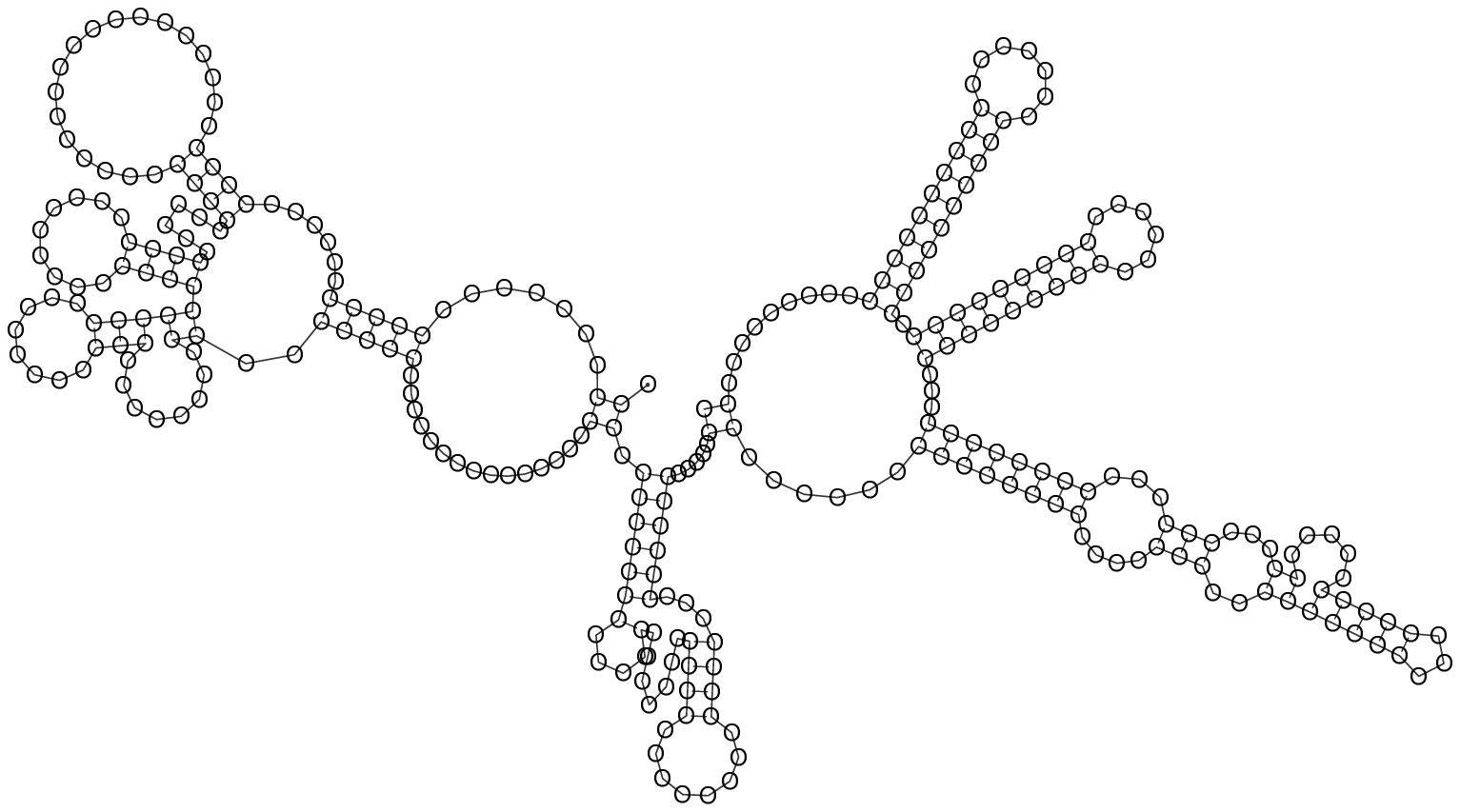}}
\end{tabular}}\\
{\bf Helices model \HelicesModel:} Constraints on expected number and length for hairpins.
\\[0.5cm]
\fbox{
\begin{tabular}{c}
\Fig{\includegraphics[width=2cm]{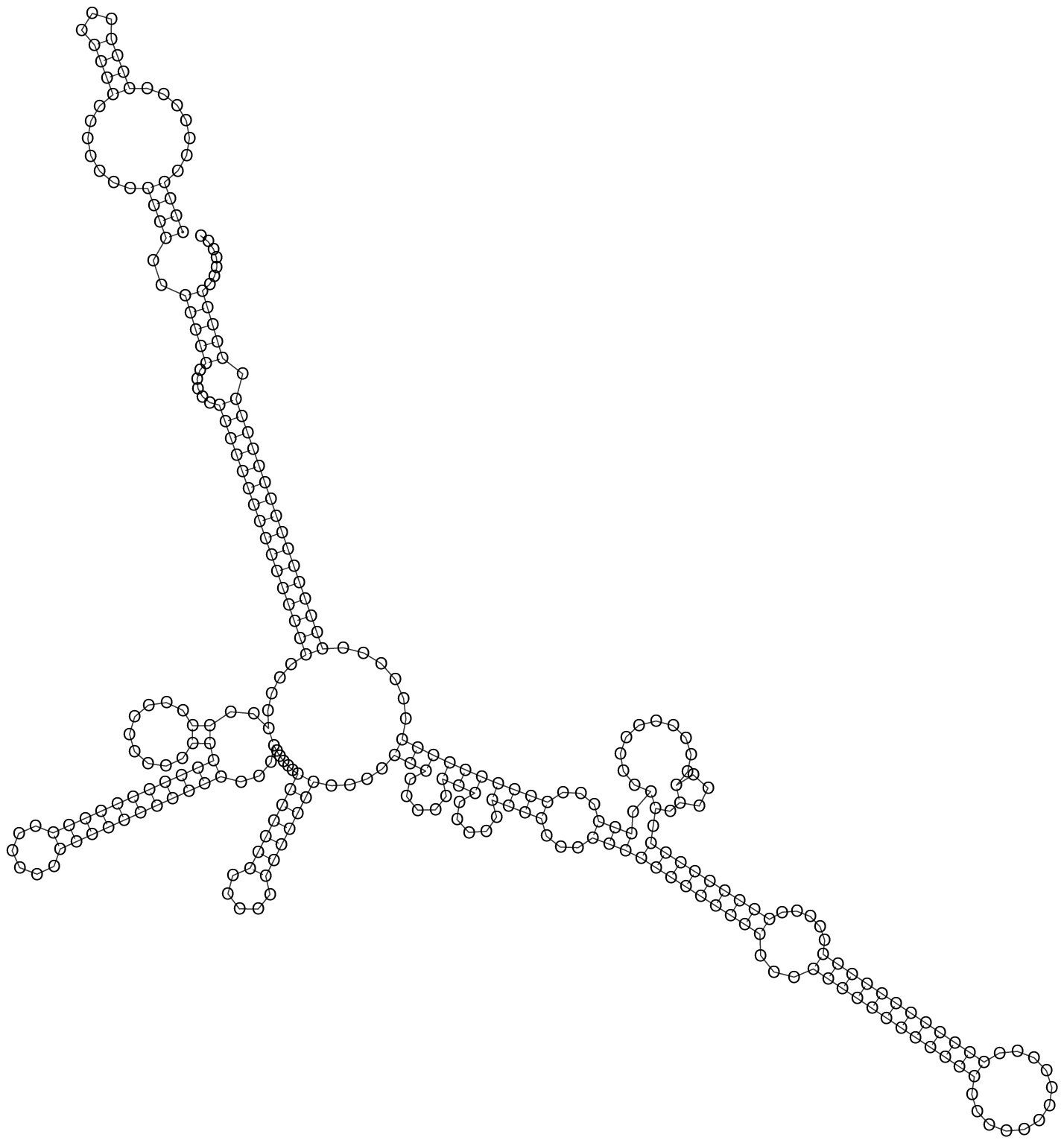}}
\Fig{\includegraphics[width=2cm]{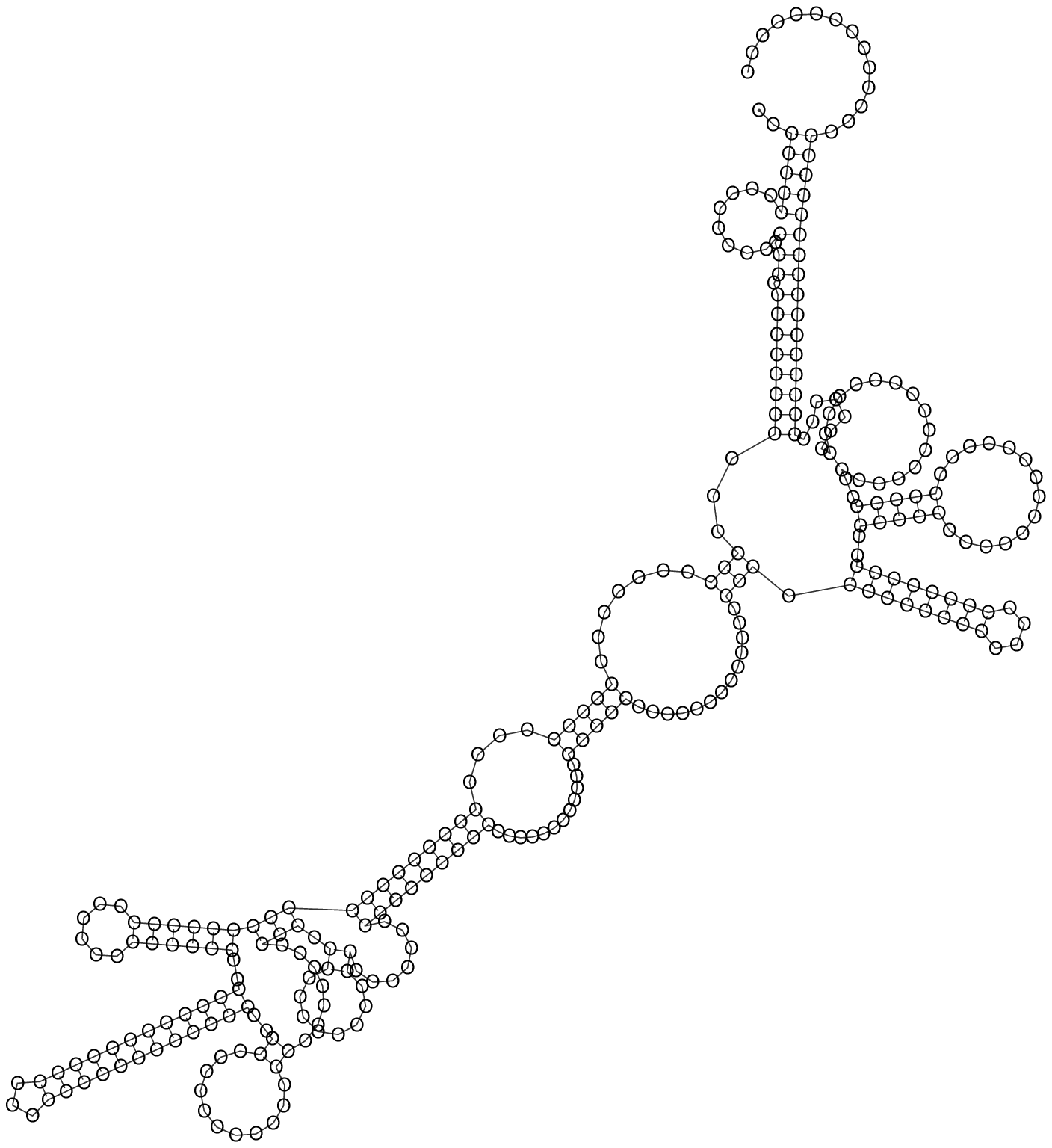}}
\Fig{\includegraphics[width=2cm]{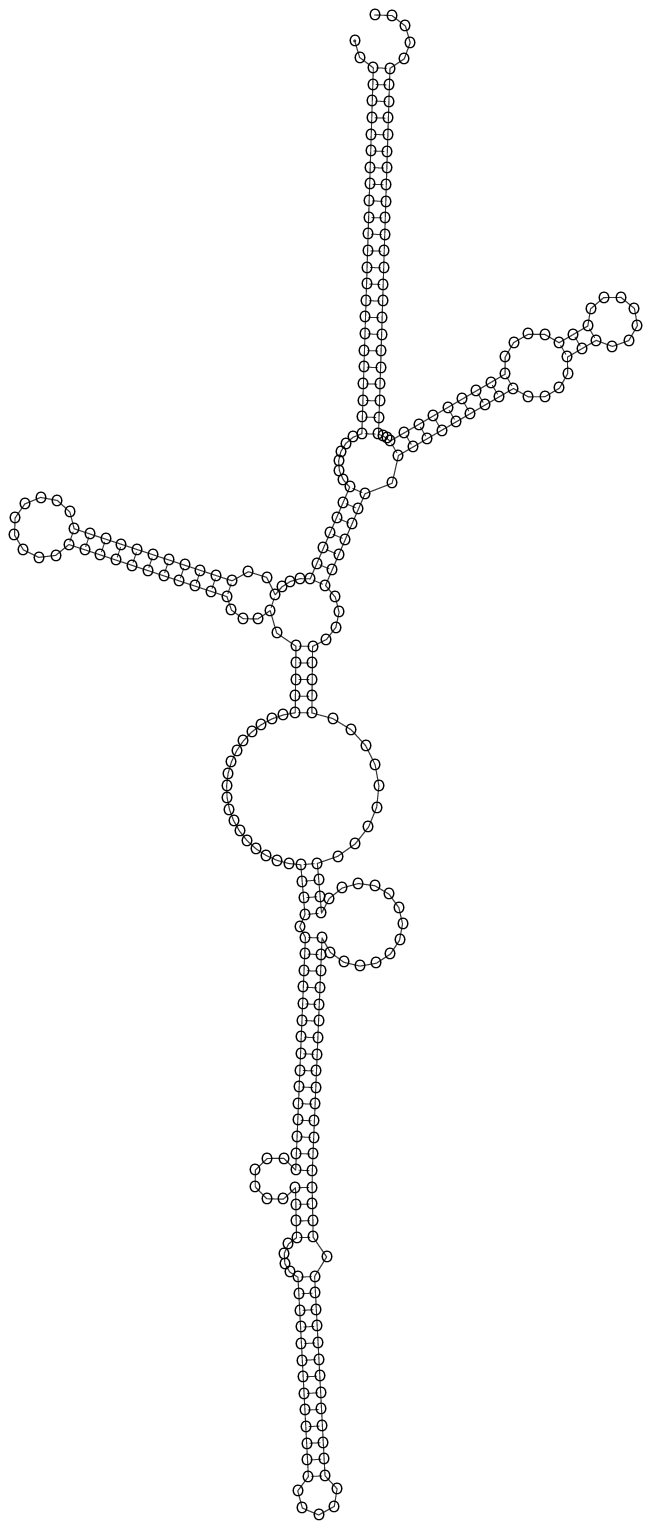}}
\Fig{\includegraphics[width=2cm]{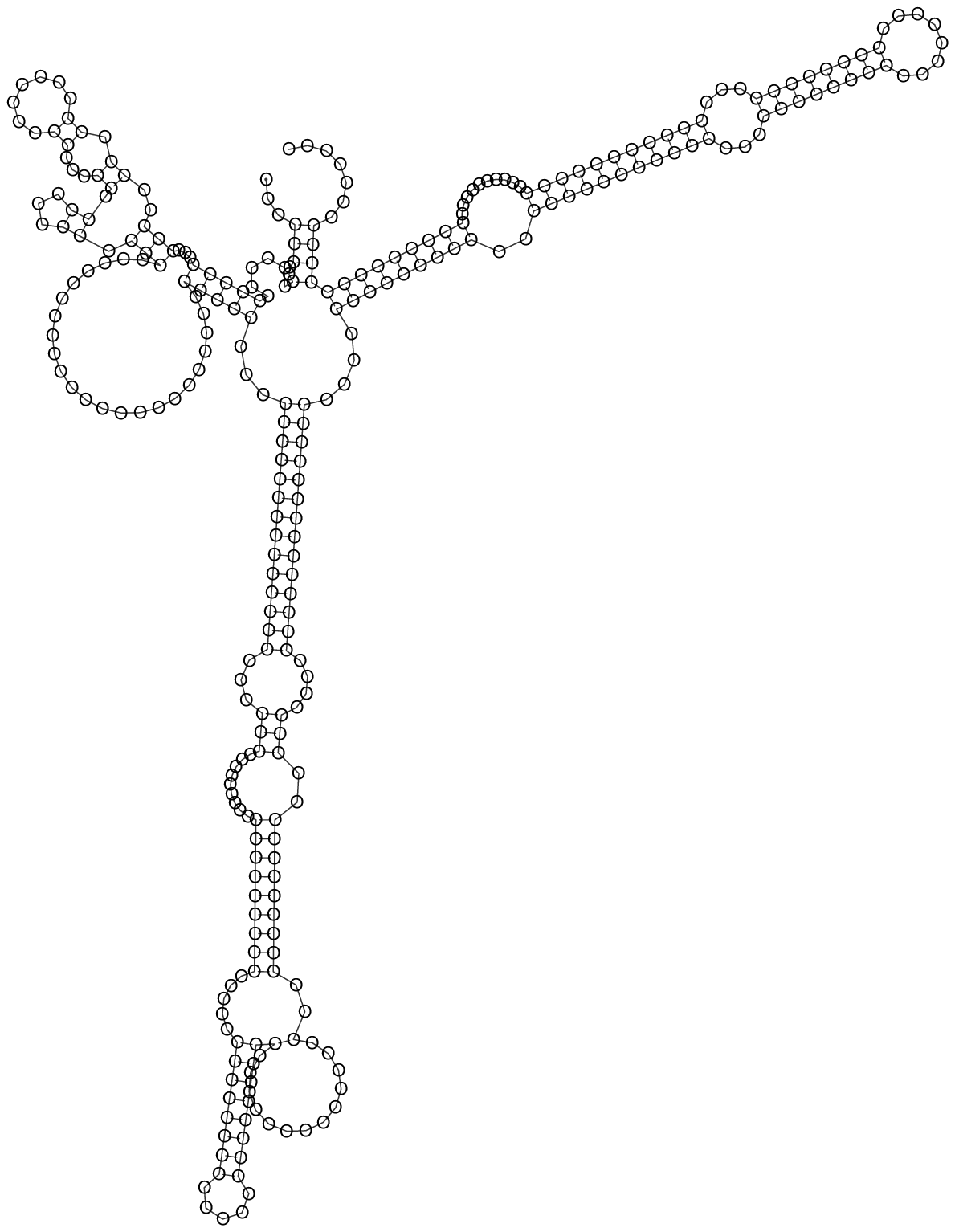}}
\Fig{\includegraphics[width=2cm]{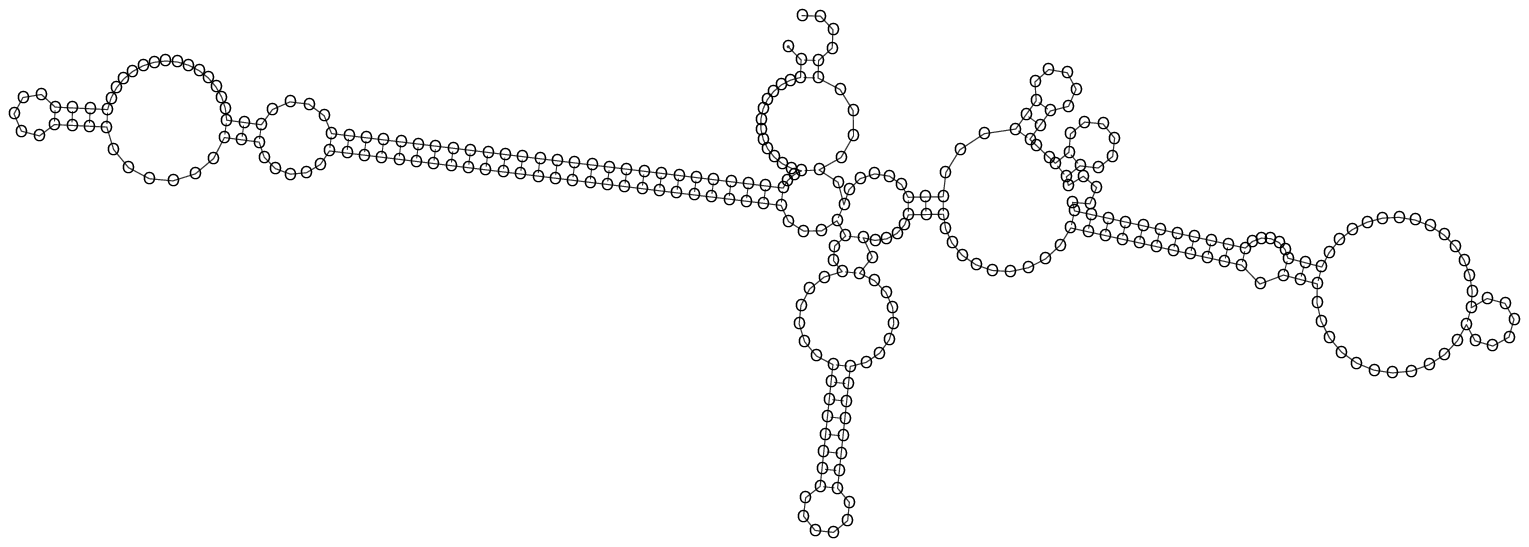}}
\Fig{\includegraphics[width=2cm]{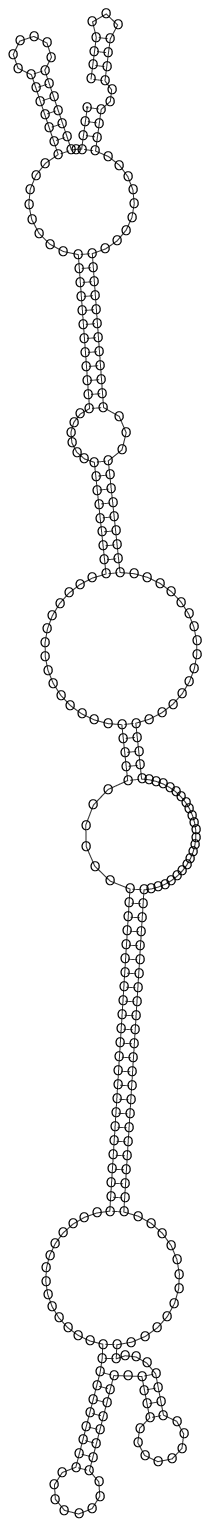}}\\
\Fig{\includegraphics[width=2cm]{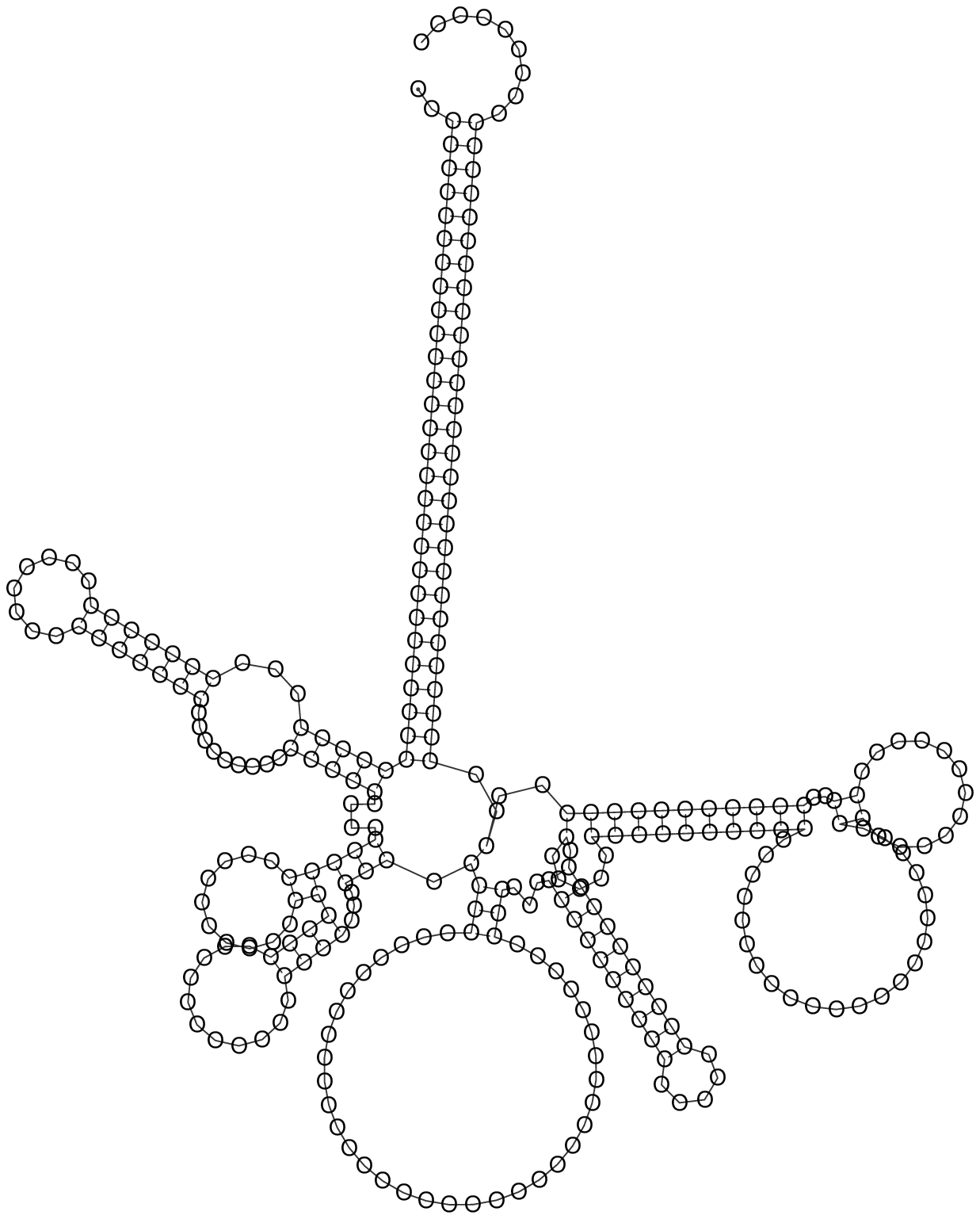}}
\Fig{\includegraphics[width=2cm]{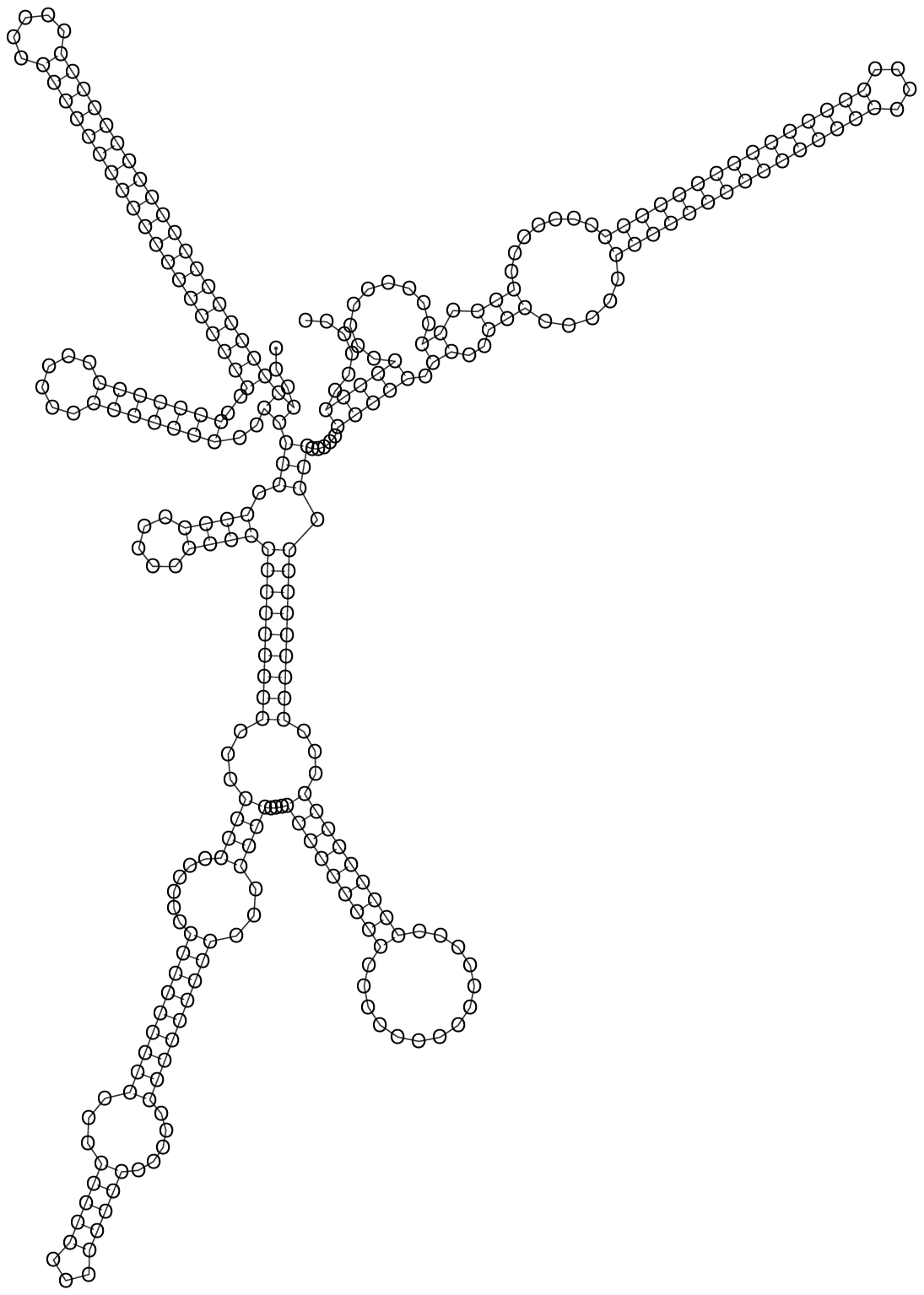}}
\Fig{\includegraphics[width=2cm]{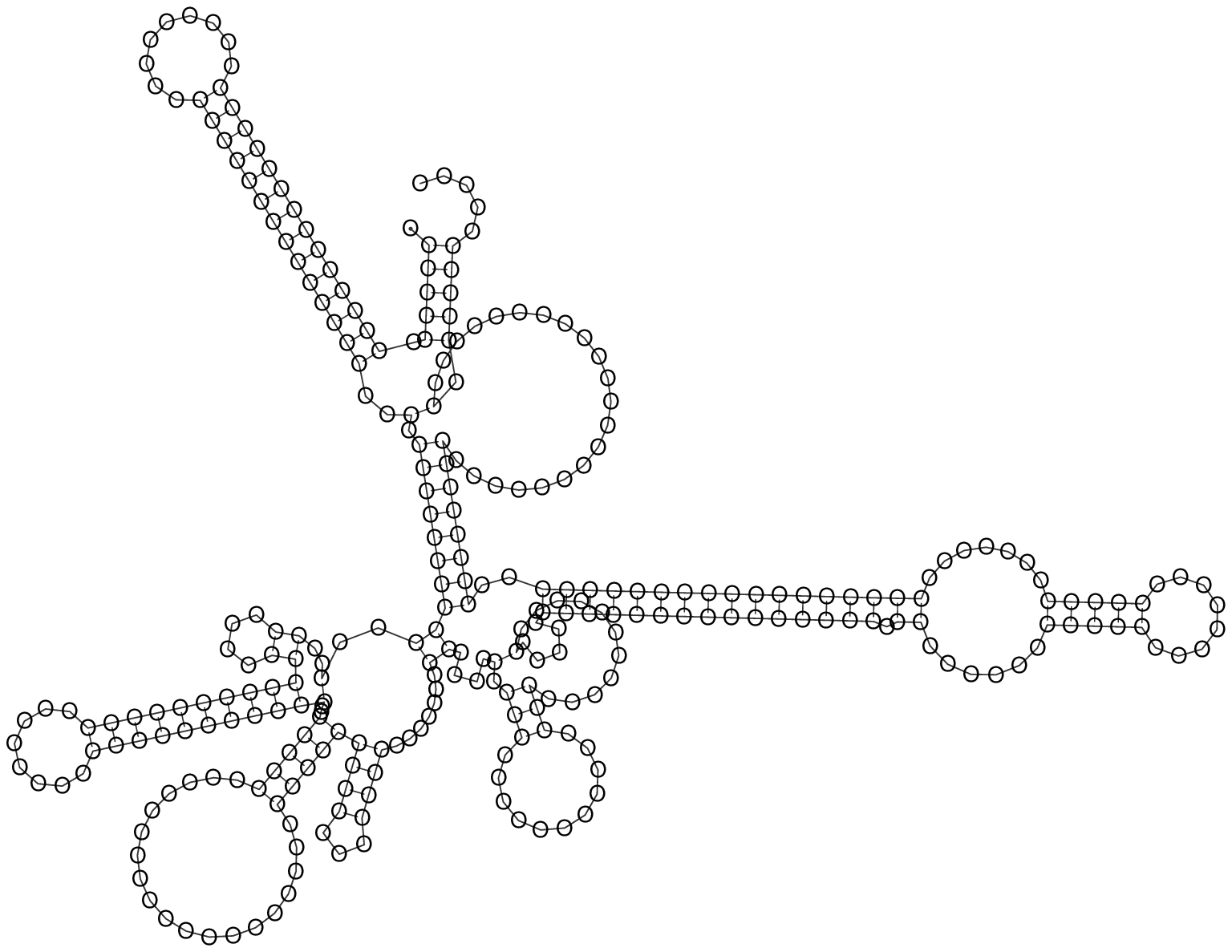}}
\Fig{\includegraphics[width=2cm]{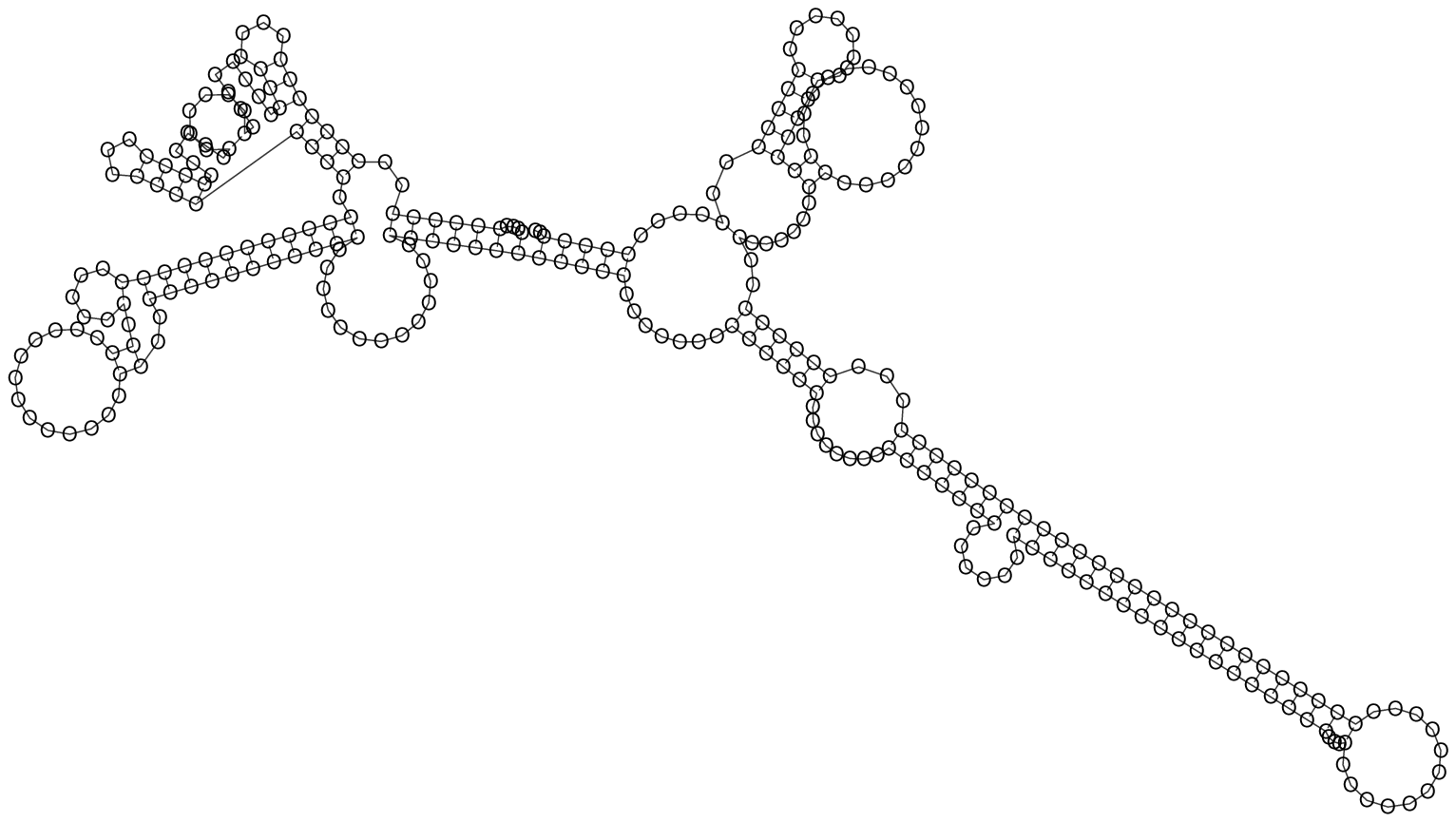}}
\Fig{\includegraphics[width=2cm]{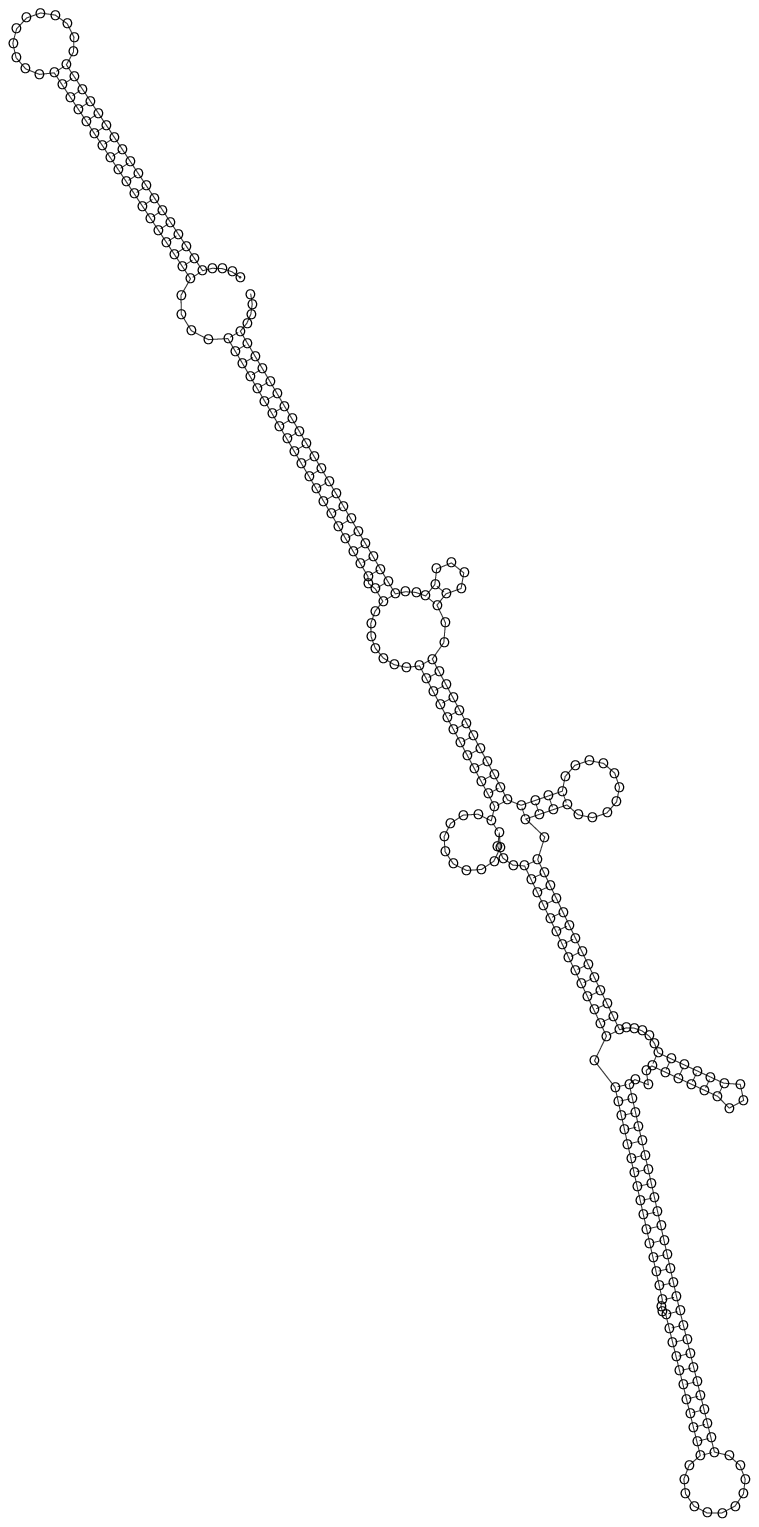}}
\Fig{\includegraphics[width=2cm]{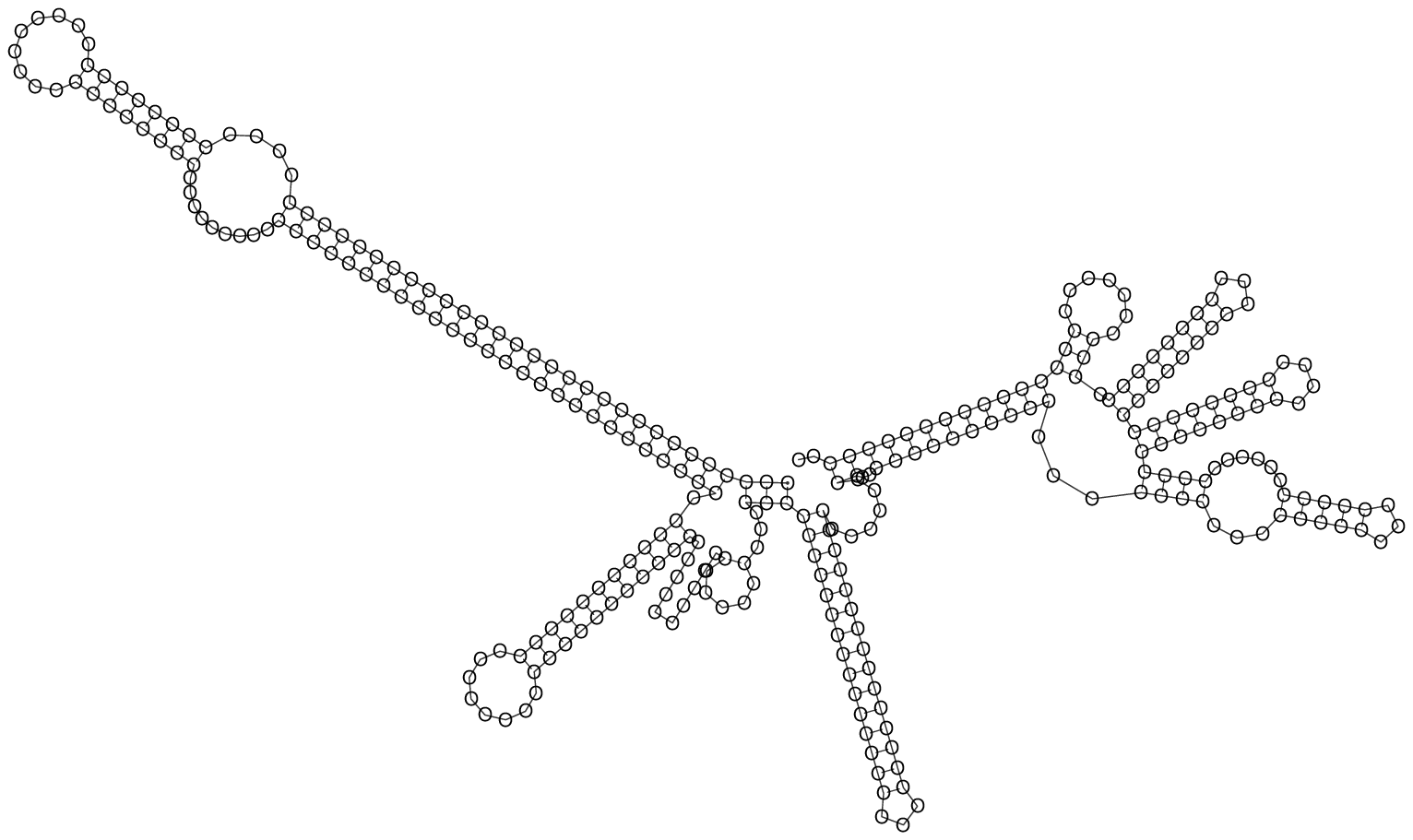}}
\end{tabular}}\\
{\bf Loops model \LoopsModel:} Constrained hairpins and multiple loops.
\\[0.5cm]
\fbox{
\begin{tabular}{c}
	\Fig{\includegraphics[width=2cm]{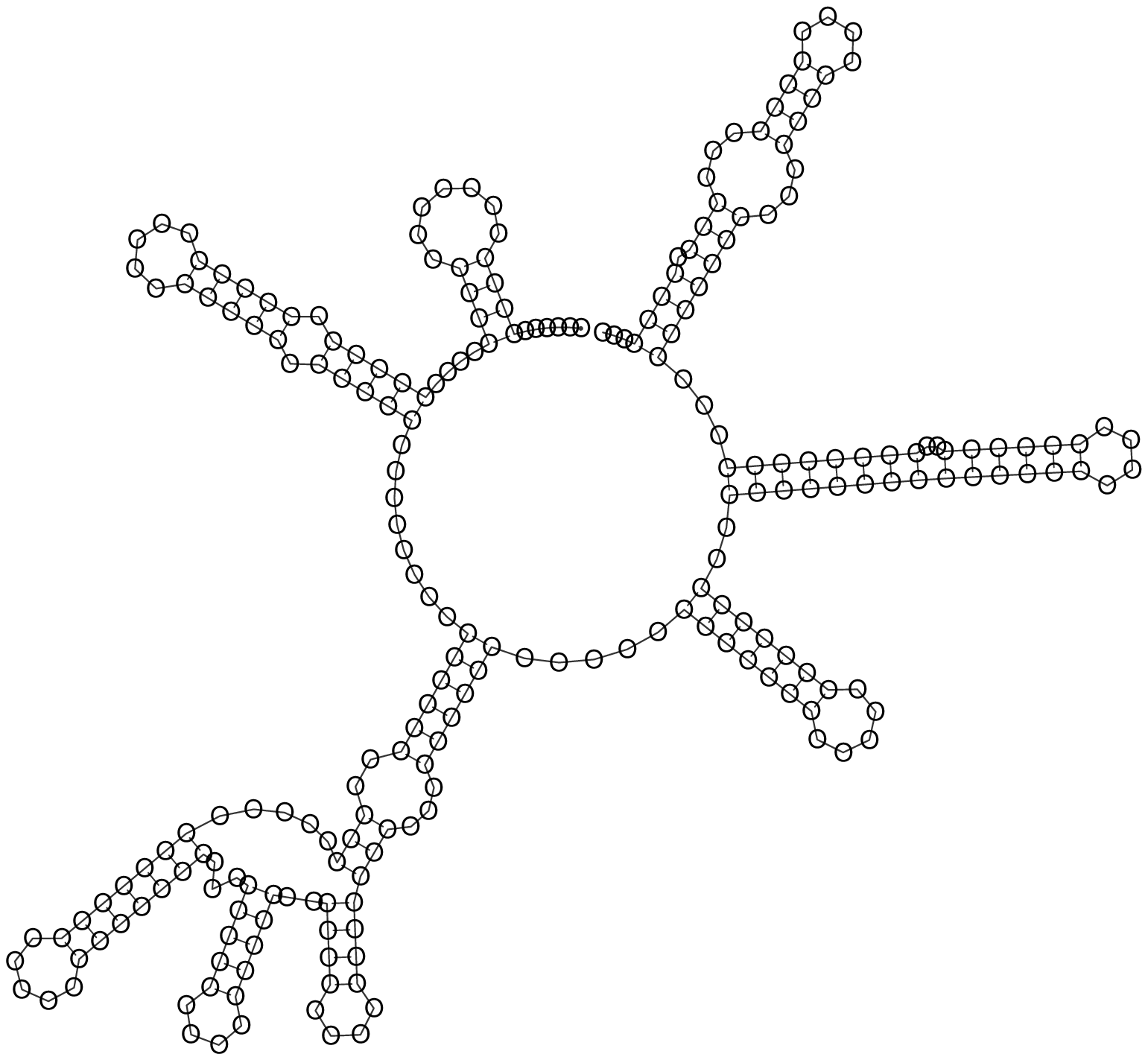}}
	\Fig{\includegraphics[width=2cm]{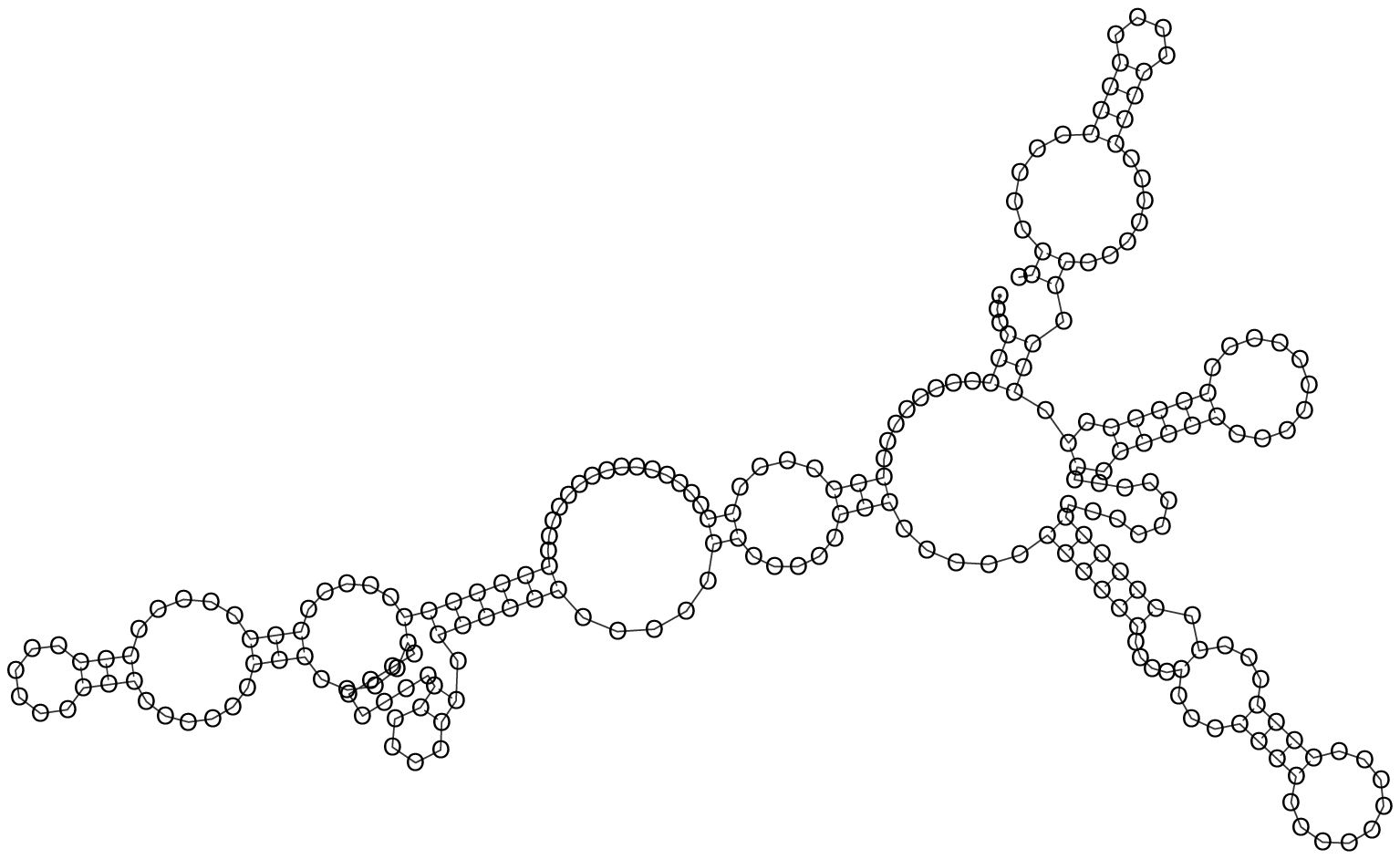}}
	\Fig{\includegraphics[width=2cm]{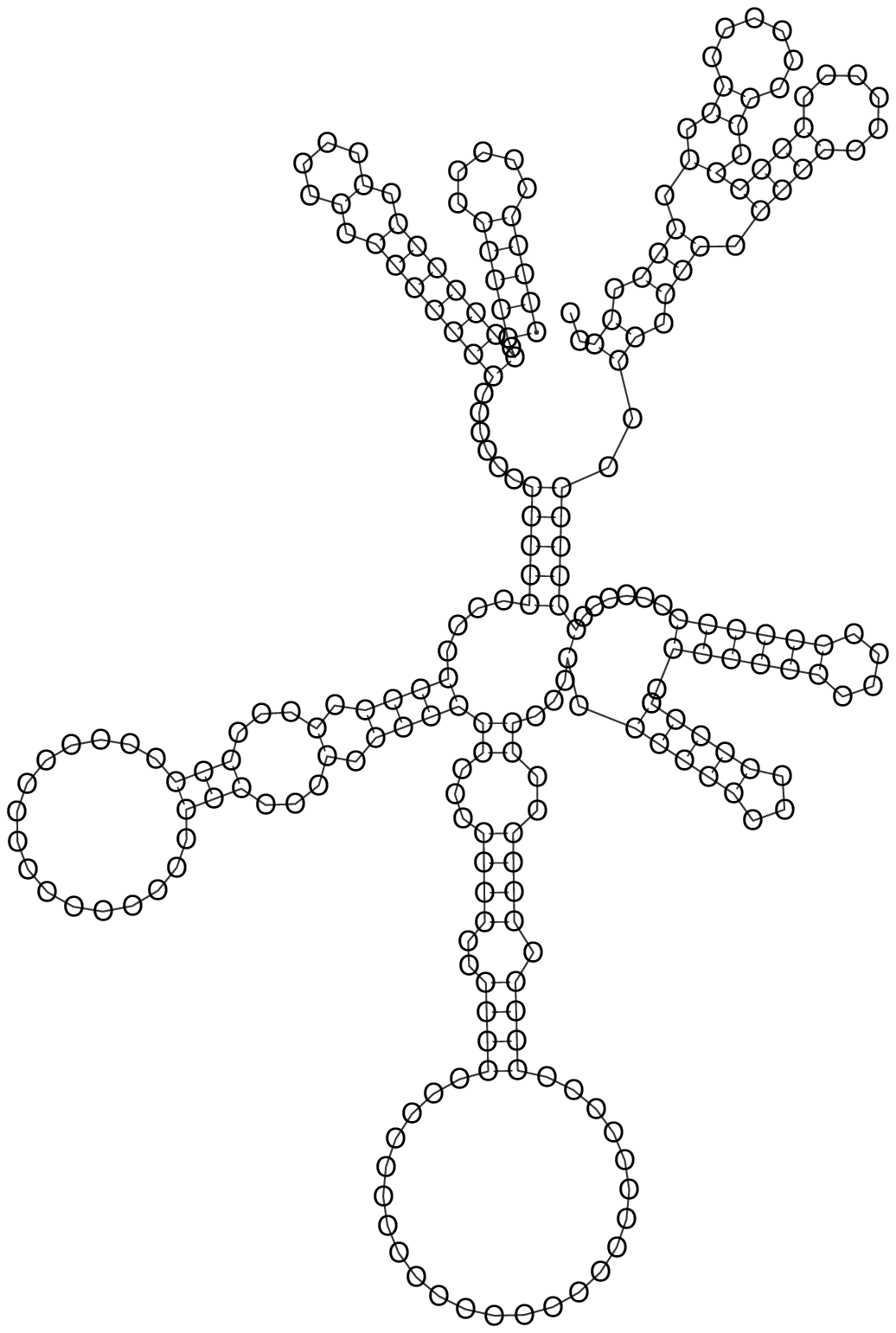}}
	\Fig{\includegraphics[width=2cm]{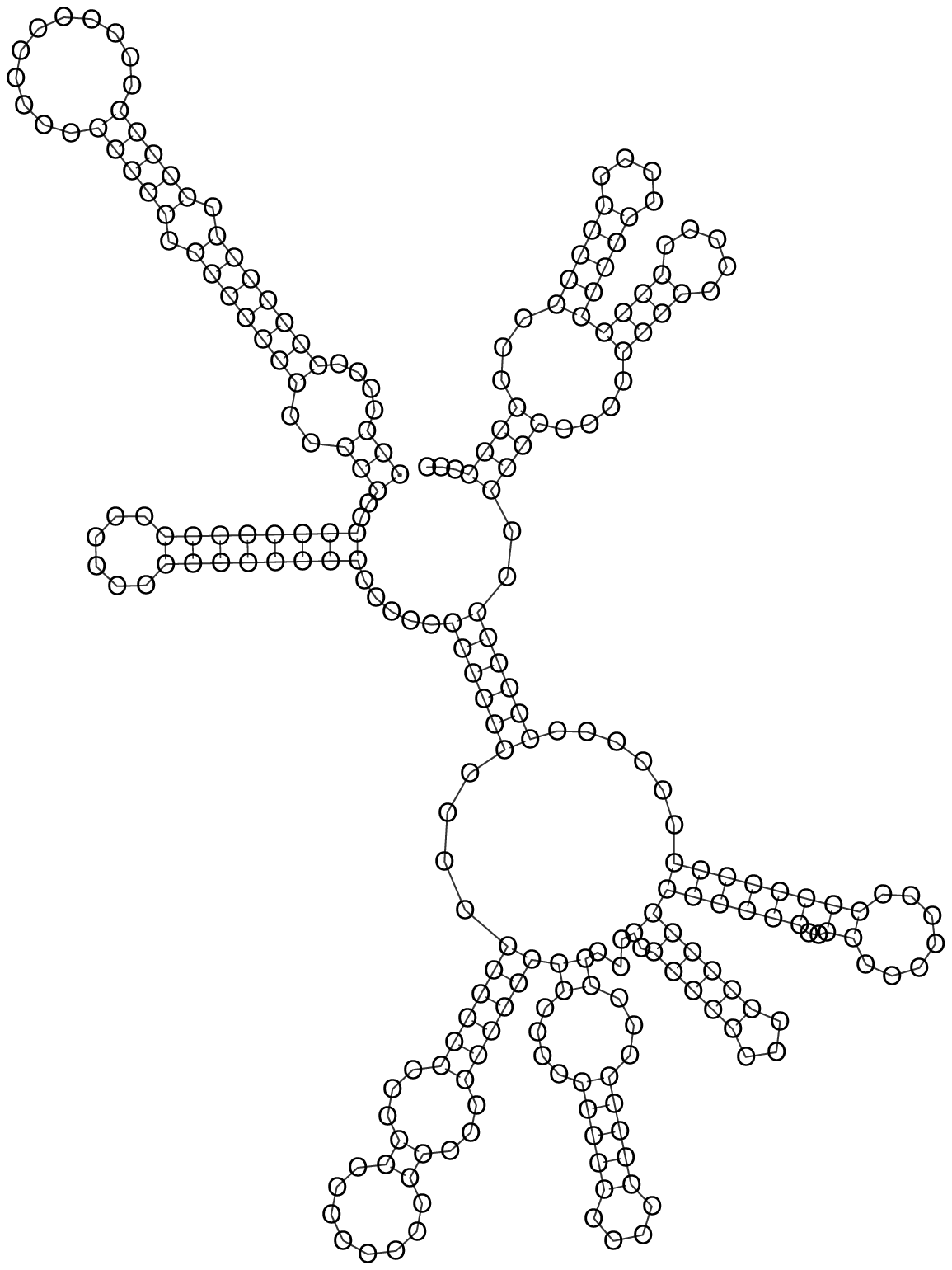}}
	\Fig{\includegraphics[width=2cm]{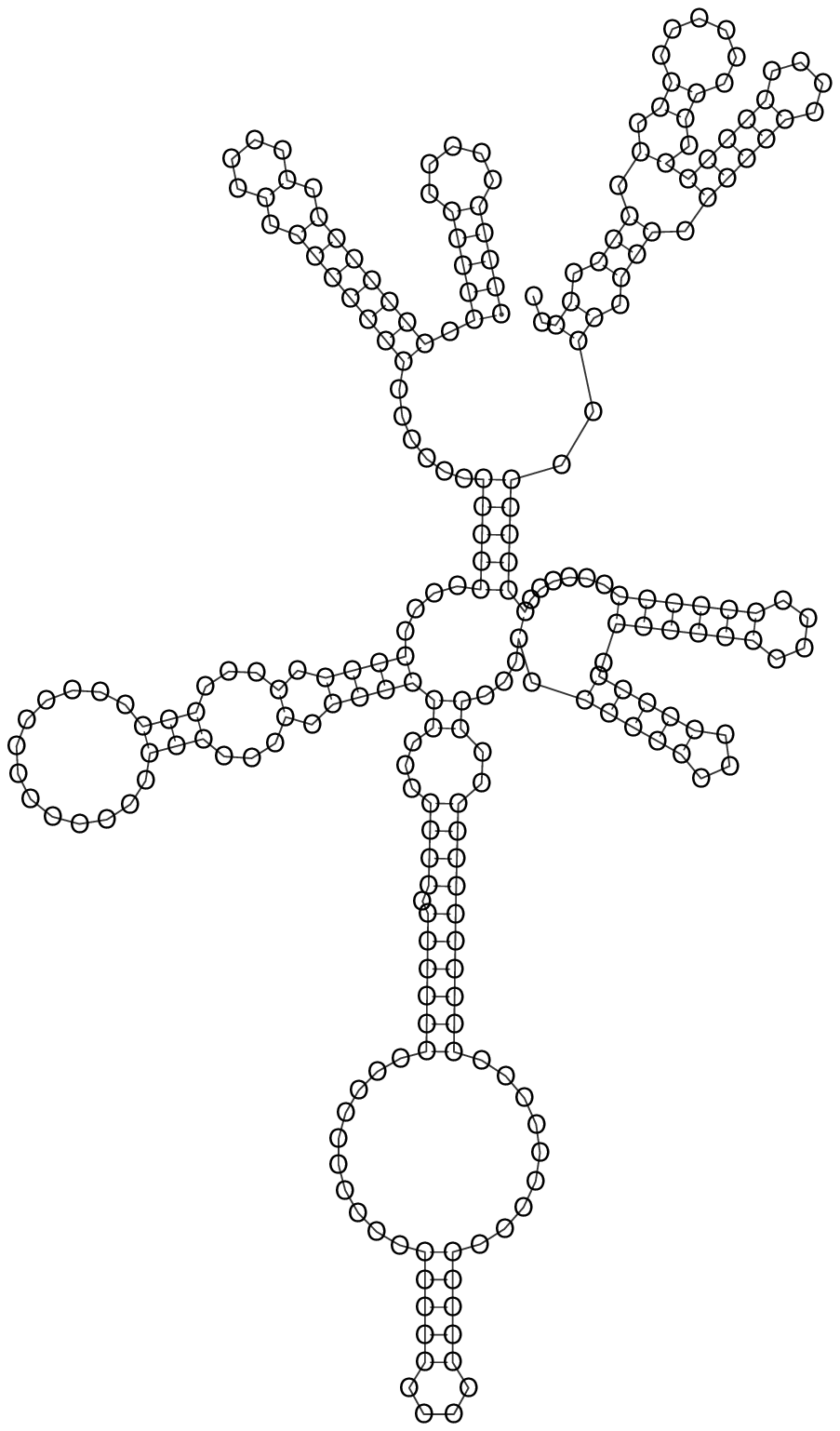}}
	\Fig{\includegraphics[width=2cm]{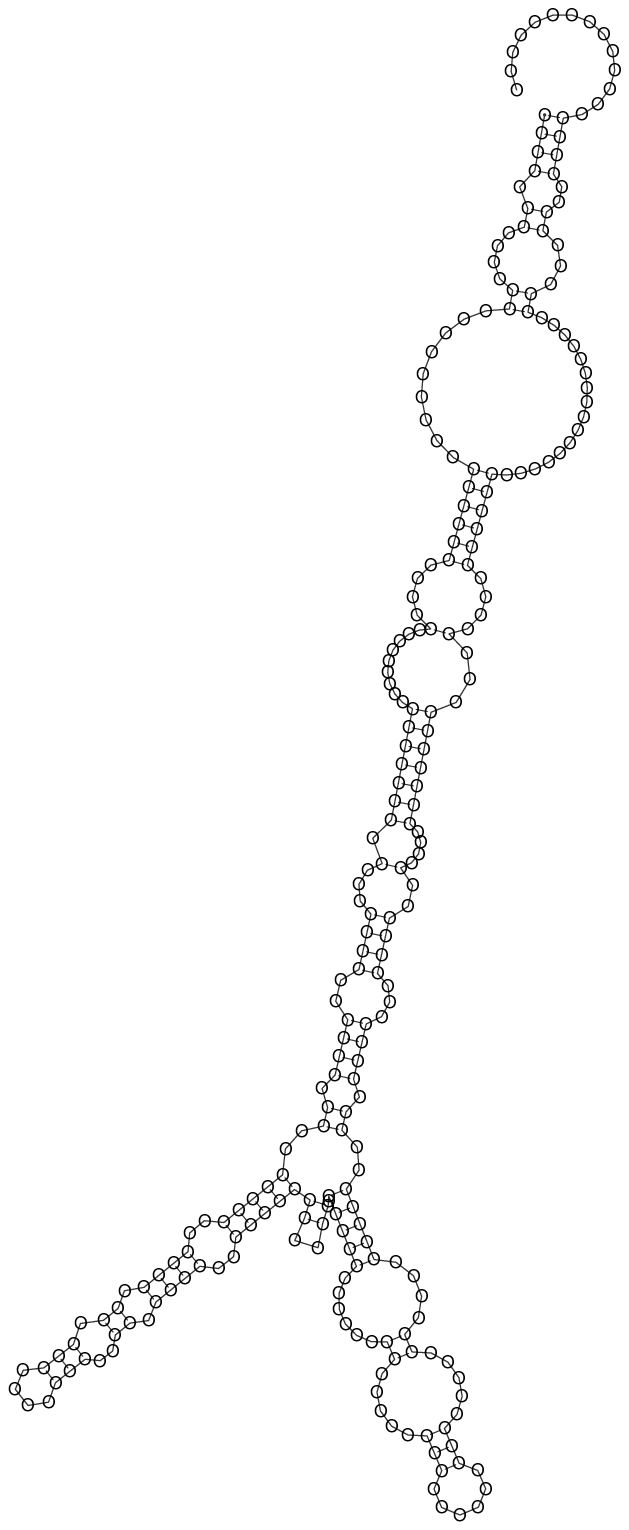}}\\
	\Fig{\includegraphics[width=2cm]{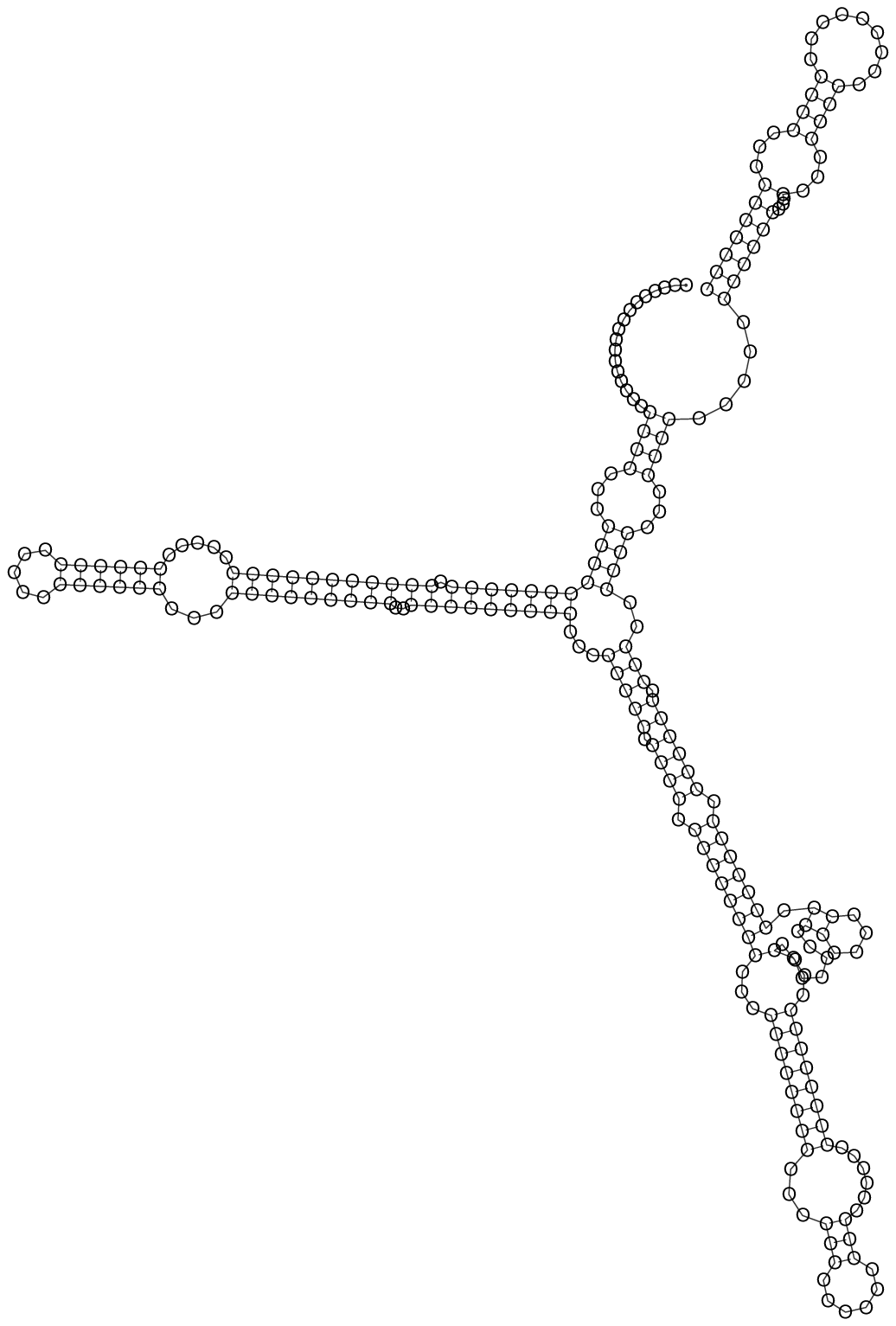}}
	\Fig{\includegraphics[width=2cm]{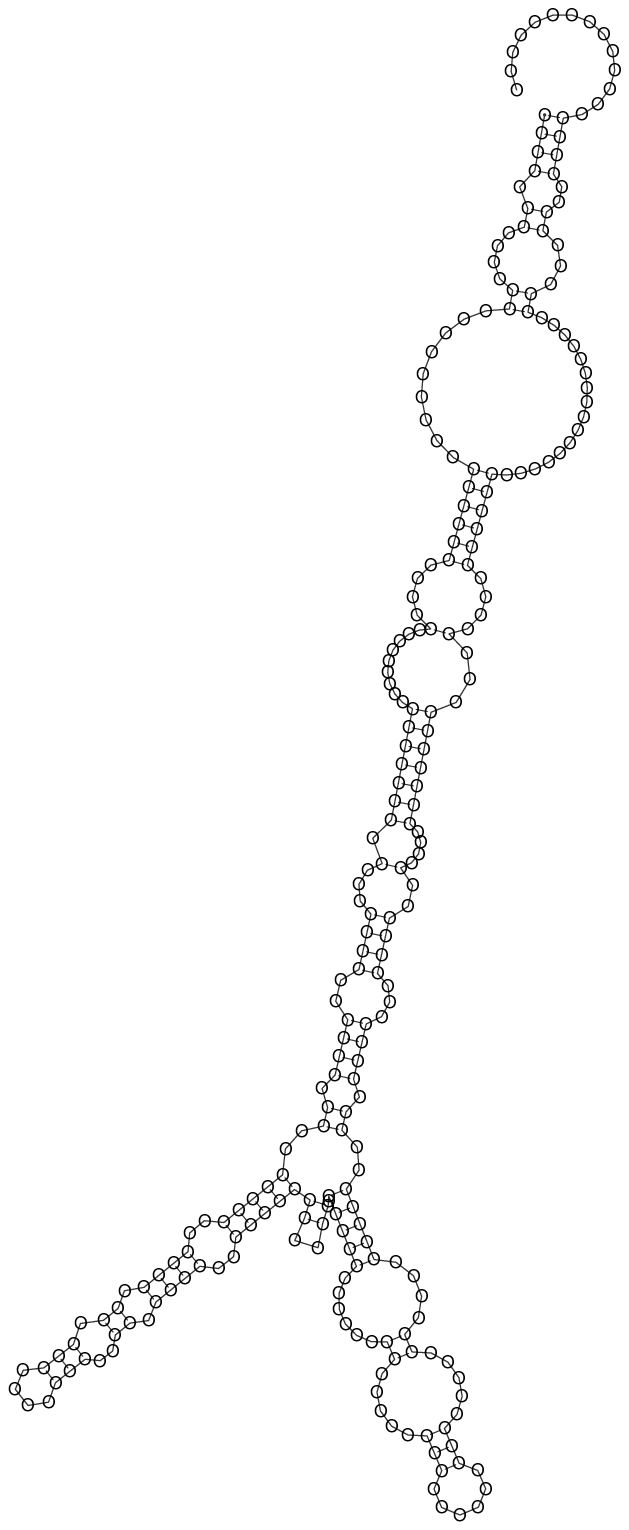}}
	\Fig{\includegraphics[width=2cm]{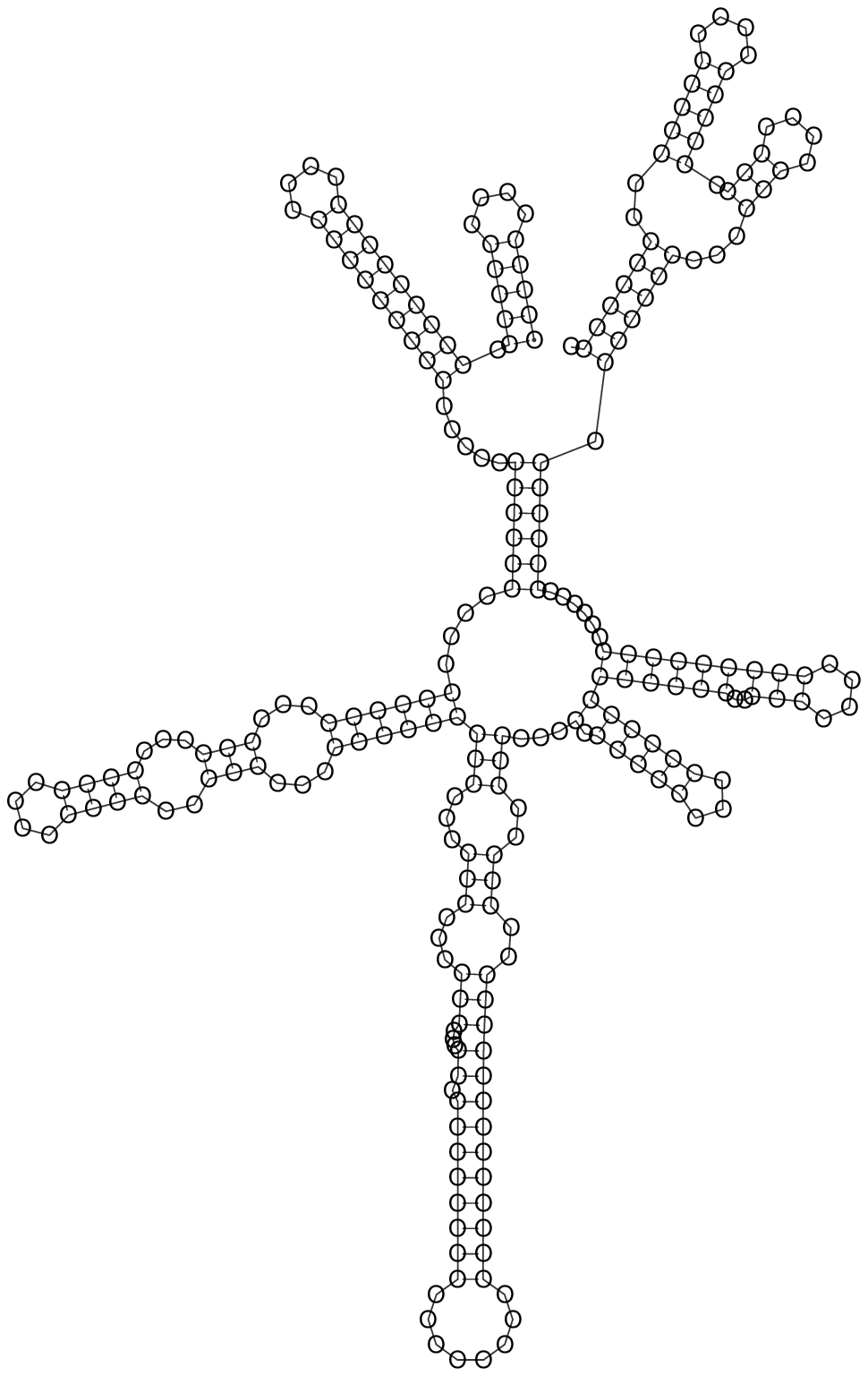}}
	\Fig{\includegraphics[width=2cm]{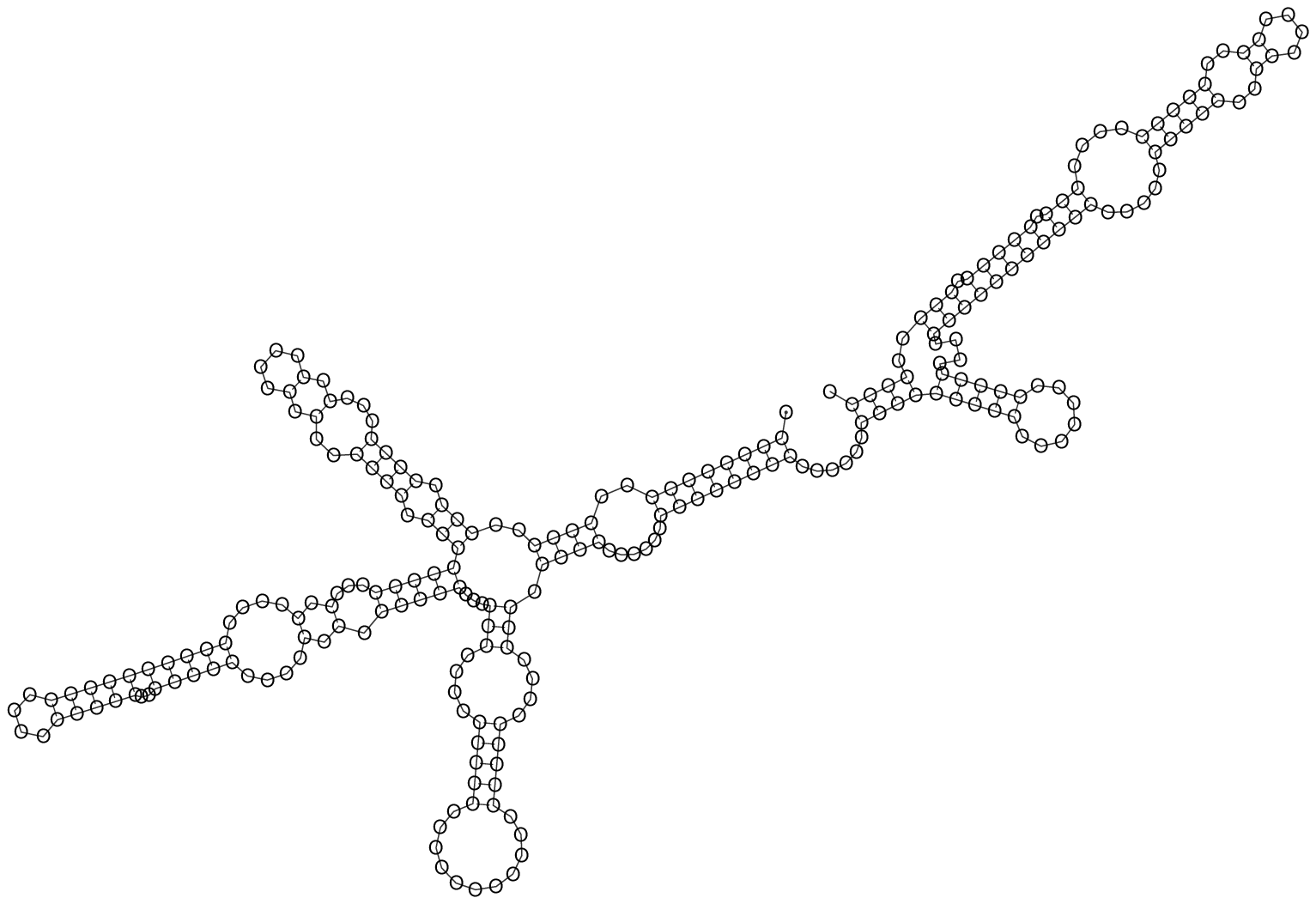}}
	\Fig{\includegraphics[width=2cm]{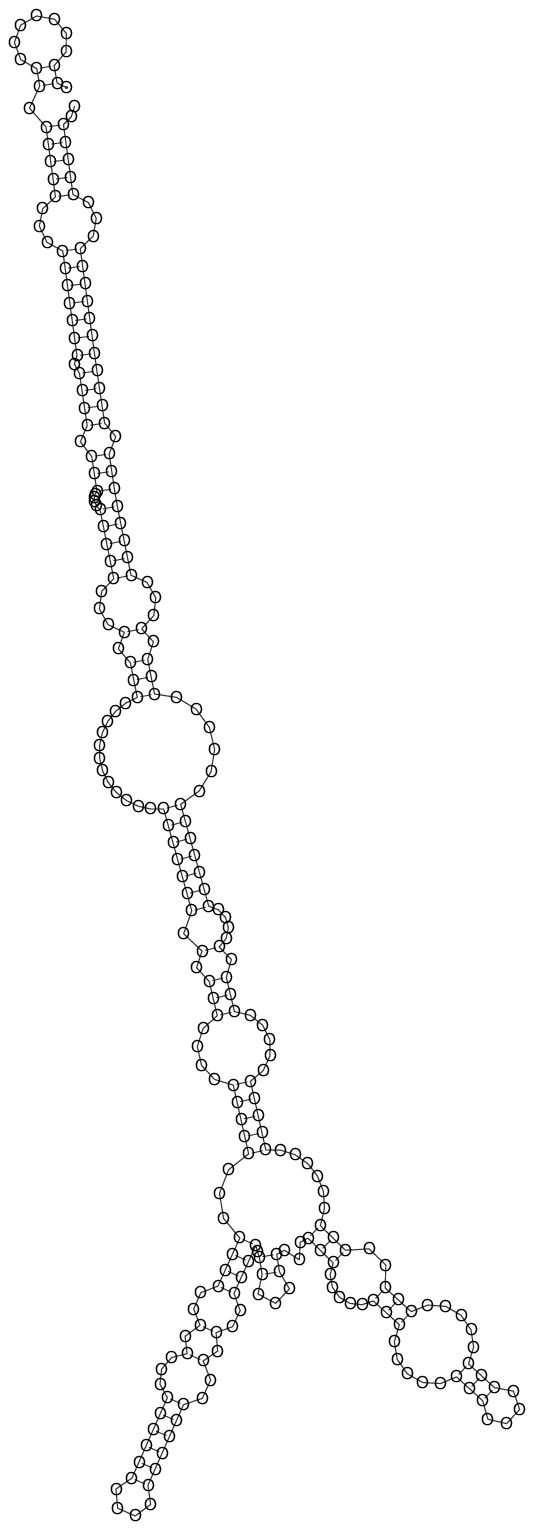}}
	\Fig{\includegraphics[width=2cm]{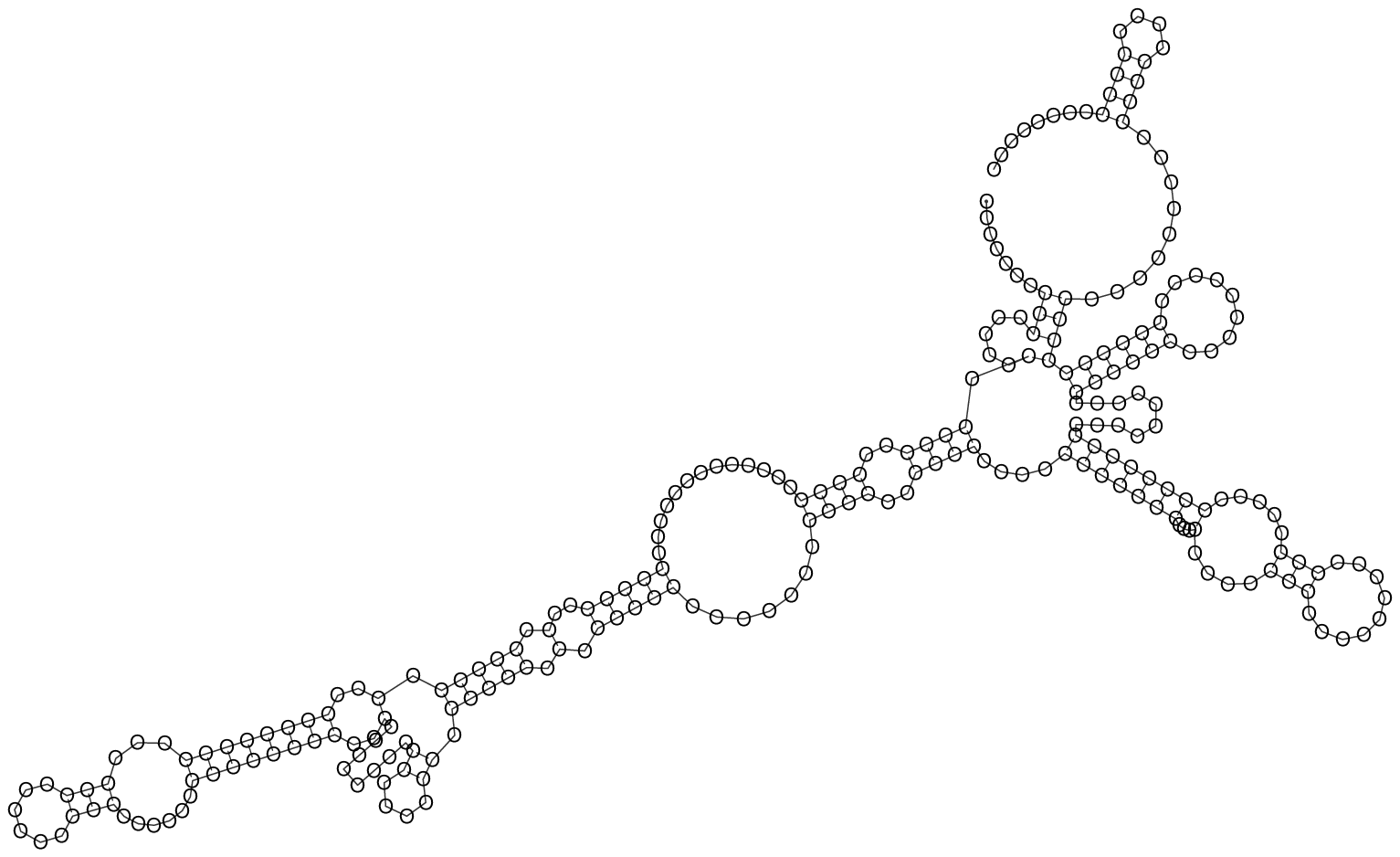}}
\end{tabular}}\\
{\bf Native structures:} Real structures of size $\pm$ 300 excerpted from~\cite{Mathews2004}.
\caption{Typical random structures of size $300$ in the three studied random models of increasing
fitness, and in real structures of similar size.}\label{fig:randomrnas}
\end{center}
\end{figure}
From these three models, it is possible to use our prototype to generate random structures of size 300, draw them using the
{\tt RNAPlot} tool from the \emph{Vienna} package~\cite{Hofacker-Fontana-1994} and compare them visually to the real ones. We observe
in Figure~\ref{fig:randomrnas} a clear progression from the messy \UniformModel\ to the more realistic \LoopsModel. This illustrates
the ability of our program to assist in the design of models for biological sequences and structures.

\section{Generation according to exact frequencies}
\label{secExact}

Here, given a targeted size $n$ and a $k$-tuple $(n_1, \ldots, n_k)$
of integers, our goal is to generate uniformly at random a structure
of $\CCS{C}{n}$ which contains exactly $n_i$ atoms $\At_i$ for all $1
\leq i \leq k$. Let $r$ be the number of occurrences of
undistinguished atoms in the structure: we have $r = n-\sum_{i=1}^k
n_i$. The principle of the method that we describe here is a natural
extension of the general outline given in Section~\ref{secUniform}.

A first general algorithm was given in~\cite{Denise00}
by two of the authors of this article. Here we present an
improvement of that algorithm.

\begin{Proposition}
  The generation of  $m$ structures of size $n=n_1+\cdots+n_k+r$ featuring exactly $n_i$
  occurrences of atom $\At_i$ can be performed in $\bigO{r^2\prod_{i=1}^k n_i^2 +m n k \log n}$
  arithmetic operations for general specifications, or in $\bigO{r\prod_{i=1}^k n_i +m n}$
  for regular specifications.
\end{Proposition}


\medskip
For any class $C$ given as a standard specification, we write $c_{j_1,
\ldots, j_k,r}$ for the number of structures of $C$ of size
$n=r+\sum_{i=1}^k j_i$, which contain $j_i$ atoms $\At_i$ for each $i\in[1,k]$, and $r$ other atoms. For
short, we can also write $c_{\bf j}$, where ${\bf j} = (j_1, \ldots, j_k, r)$.

Let us first outline the algorithm given in~\cite{Denise00}. The
preprocessing stage consists in computing a table of the $c_{j_1,
 \ldots, j_k, r}$ for  $\left\{0 \leq j_i
\leq n_i\right\}_{i\in[1,k]}$ and $0 \leq r \leq n-\sum_{i=0}^k n_i$. This requires computing a
table of $\Theta(r\prod_{i=1}^k n_i)$ entries, with the recurrences
stated in Table~\ref{tabR2}. Since $\Theta(r\prod_{i=1}^k n_i)$ arithmetic operations
are required to compute each entry, this preprocessing clearly takes time
$\Theta(r^2\prod_{i=1}^k j_i^2)$ for general specifications. For regular specifications, given using only
rules of the form $ C = T_i B$, $T_i = \At_i$ and $ C = 1$, only one of the entries associated with the $T_i$'s is
non-null, and the product rule can be evaluated in $\bigO{1}$ arithmetic operations, bringing the
preprocessing complexity down to $\Theta(r\prod_{i=1}^k n_i)$.
\begin{table}
\begin{equation*}
  \begin{array}{ccl}
  C =1 &\Rightarrow&
  c_{0,0,\ldots,0} = 1\;;\\
  C = \At_i
  &\Rightarrow&  c_{0,\ldots,0,1,0,\ldots,0} = 1 \hfill (j_i = 1)\;;
\\
  C = A + B &
  \Rightarrow&
  c_{\bf j} = a_{\bf j} + b_{\bf j}\;;
        \label{fix3}
\\
  C = A \ProdOp B &\Rightarrow& 
  c_{\bf j} =
      \sum_{\begin{array}{c}
             \scriptstyle j'_1+j''_1 = j_1 \\
             \scriptstyle \ldots \\
             \scriptstyle j'_k+j''_k = j_k \\
             \scriptstyle r'+r'' = r \\
          \end{array}}
    \hspace{-0.4cm} a_{j'_1,\ldots,j'_k,r'} b_{j''_1,\ldots,j''_k,r''}\;;

\\
  \Theta C = A \ProdOp B & \Rightarrow&
  c_{\bf j} = {\frac 1 n}
    \sum_{\begin{array}{c}
             \scriptstyle j'_1+j''_1 = j_1 \\
             \scriptstyle \ldots \\
            \scriptstyle j'_k+j''_k = j_k \\
             \scriptstyle r'+r'' = r \\
          \end{array}}
    \hspace{-0.4cm} a_{j'_1,\ldots,j'_k,r'} b_{j''_1,\ldots,j''_k,r''}\;;
\\
  C =\Theta A &\Rightarrow&
  c_{\bf j} = n a_{\bf j}.
\end{array}
\end{equation*}
\caption{Counting procedures for standard specifications in the case
  of the random generation according to exact frequencies.}
\label{tabR2}
\end{table}

Now, each step of the generation stage consists in choosing
a rewriting rule of the current class.
Suppose that, at a given step of generation of a structure having
distribution ${\bf j} = (j_1, \ldots, j_k, r)$, one has to choose a
rewriting rule for the class $C$. If $C = A + B$, one
generates a structure with distribution ${\bf j}$ deriving from $A$
with probability $a_{\bf j}/c_{\bf j}$, or deriving from $B$ with
probability $b_{\bf j}/c_{\bf j}$. If $C = A \ProdOp B$, one
chooses a vector ${\bf h}=(h_1, \ldots, h_k, s)$ with probability
$a_{\bf h}b_{{\bf j}-{\bf h}}/c_{\bf h}$. Then one generates
a structure deriving from $A$ having distribution ${\bf h}$ and
a structure from $B$ having distribution ${\bf j}-{\bf h}$.

This generation stage, which has a worst-case complexity in $\Theta(n\prod_{i=1}^k n_i)$,
can be improved drastically. Indeed, the bottleneck of the above procedure
is the $C = A \ProdOp B$ case, where there are $j_1 j_2 \ldots j_k r$ possible different choices.
Now, let $c^{(h_1,\ldots,h_i)}_{(j_1,\ldots,j_k,r)}$ be the number
of structures generated from $C$, having distribution $(j_1,\ldots,j_k,r)$ and
such that, for each $x\in[1,i]$, exactly $h_x$ of the targeted $j_x$
occurrences of atom $\At_x$ are generated from $A$. We have:
$$c^{(h_1,\ldots,h_i)}_{(j_1,\ldots,j_i,\ldots,j_k,r)} =
\sum_{h_{i+1}\le j_{i+1}} \ldots \sum_{h_k \le j_k} \sum_{r' \le r}
a_{h_1,\ldots,h_k,r'}
b_{j_1-h_1,\ldots,j_k-h_k,r-r'}.
$$
Now the probability of
counting $h_i$ atoms $\At_i$ in the structure from $A$, given that the
structure contains $h_1$ atoms $\At_1$,
\ldots, $h_{i-1}$ atoms $\At_{i-1}$ is:
$$
\Pr(h_i |
        h_1,\ldots,h_{i-1})=
             \frac{c^{(h_1,\ldots,h_i)}_{(j_1,\ldots,j_i,\ldots,j_k,r)}}
                  {c^{(h_1,\ldots,h_{i-1})}_{(j_1,\ldots,j_i,\ldots,j_k,r)}}
$$
and the probability of counting $h_1$ atoms $\At_1$
in the structure from $A$ is:
$$
\Pr(h_1 | \emptyset)=
             \frac{c^{(h_1)}_{(j_1,\ldots,j_k,r)}}
                  {c_{j_1,\ldots,j_k,r}}.
$$
This allows to choose the adequate decomposition $h_1,\ldots,h_k$ sequentially.
Since picking a suitable value for $h_i$ involves investigating at most $j_i$ alternatives, the
overhead compared to the classic generation is limited to a factor $\bigO{k}$.

Hence the whole algorithm is as follows:
\begin{enumerate}
\item {\em Preprocessing stage.} For any combinatorial class $C$ in
  the standard specification, compute a table of the
  $c^{(h_1,\ldots,h_i)}_{(j_1,\ldots,j_i,\ldots,j_k,r)}$ for $1 \le i
  \le k$, $\left\{0 \leq j_x \leq n_x\right\}_{x\in[1,k]}$ and $\left\{0 \leq h_x\leq j_x\right\}_{x\in[1,i]}$. This
  can be done with the same recurrences as for the previous approach.
  Indeed the  $c^{(h_1,\ldots,h_i)}_{(j_1,\ldots,j_k,r)}$ are
  in fact partial sums of the one involved in products, and can
  therefore be computed {\it on the fly} during the computation of coefficients
  $c_{j_1,\ldots,j_k,r}$. This gives a complexity in
  $\bigO{r^2\prod_{i=1}^k n_i^2}$
arithmetic operations, while requiring storage
  of $\Theta(k r\prod_{i=1}^k n_i)$  numbers.

  For regular specifications, the sums associated with product rules only have one
  non-null term, so we can add a specific counting procedure
    $$ C = T_i \ProdOp A\quad\Rightarrow\quad c_{j_1,\ldots,j_k,r} = c_{j_1,\ldots,j_i-1,\ldots,j_k,r}$$
  which lowers the time/space complexity to $\Theta(r\prod_{i=1}^k n_i)$.

  \item {\em Generation stage.} The $C \to 1$, $C \to \At_i$, and $C \to A + B$ rules
  are trivially borrowed from~\cite{Denise00}. In the case of product rules, a sequential choice of ${\bf h}$
  described above leads to an overall generation complexity in $\bigO{m n \log n}$ arithmetic operations through
  a Boustrophedon investigation (See~\cite{Flajolet94}) of eligible decompositions in each dimension.
\end{enumerate}
  \begin{Remark}[Multidimensional Boustrophedon]
  Let us discuss the improvement observed by adopting a Boustrophedon order of investigation in this multidimensional scheme.
  We remind that, during the generation stage for products ($\ProdOp$), the Boustrophedon search consists in investigating potential partitions of the
  targeted size \emph{from the edges toward the middle} ($(0,n)$,$(n,0)$,$(1,n-1)$,\ldots) instead of \emph{sequentially} ($(0,n)$, $(1,n)$, \ldots).
  In the unidimensional Boustrophedon generation~\cite{Flajolet94} the worst case complexity $f(n)$ of the generation follows
  \begin{equation}
    f(n) = \max_{a+b=n}(2\min(a,b)+f(a)+f(b))
  \end{equation}
  which has a $\bigO{n\log n}$ solution~\cite{GreKnu81}.
  In the multidimensional case, let $\mathbf{c}=(c_1,\ldots,c_k)$ be the targeted k-tuple of occurrences, then the
  worst case complexity of our algorithm is given by
  $$ g(\mathbf{c},r) = \max_{\substack{\mathbf{a},\mathbf{b},r',r'' \mbox{ \footnotesize s.t.}\\ a_i+b_i = c_i \\ r' +r''=r}}
  \left( 2\min(r',r'')+2\sum_{i=1}^k \min(a_i,b_i) + g(\mathbf{a},r')  + g(\mathbf{b},r'')\right)$$
  Let $|\mathbf{x}|=\sum_{i=1}^k x_i$, then one has
  $$2\min(r',r'')+\sum_{i=1}^k \min(a_i,b_i) \le \min(r'+|\mathbf{a}|,r''+|\mathbf{b}|).$$
  and a straightforward induction shows that
  $$g(\mathbf{c},r) \le f(|\mathbf{c}|+r) \in \bigO{n\log n}.$$
  In the
  case of regular specifications, only binary decisions appear and the generation can be performed in
  $\Theta(m n)$ operations\YannCom{Il faut vérifier pourquoi $k$ n'apparaît pas dans ces deux complexités...}.
  \end{Remark}

\section{Conclusion}
In this paper, we introduced and developed a new scheme for the non-uniform, yet controlled,
generation of combinatorial structures.
First we addressed the random generation according to expected frequencies, motivated both by bioinformatics
and computer science applications. We introduced the notion of weighted standard
specification, and derived a random generation algorithm based on the
so-called recursive approach taking $\bigO{m n \log n + n^{1+o(1)}}$ for the generation
of $m$ structures in the according to the weighted distribution.
We showed that computing asymptotic weights, i. e. weights
that are suitable for asymptotic targeted frequencies, can be reduced to solving an explicit algebraic system.
For fixed sizes, we gave two distinct algorithmic approaches for the opposite
problem, {\it i.e.} the computation of atom frequencies achieved by given weights,
without solving any functional algebraic system. The first works for every standard
specification and takes $\bigO{k\cdot n^4}$ arithmetic operations whereas the second works
for context-free languages and uses grammar transforms to compute all frequencies in
$\bigO{k\cdot n^2}$ arithmetic operations.
This allowed us to reformulate the problem of computing suitable weights as an
optimization problem, which we solved in a heuristic fashion.
Finally, we addressed the exact frequency generation and derived a recursive algorithm that generates $m$ words having a predefined atoms
distribution $(n_1,\ldots,n_k,r)$ in $\bigO{m n \log n + r^2\prod_{i=1}^k n_i^2}$ arithmetic operations.

\OnlyFinal{\section*{Acknowledgements}
We are very grateful to Philippe Flajolet for helpful discussions and
valuable suggestions. We also thank Olivier Roques and Fr\'ed\'eric
Sarron for their help at an early stage of the present work. This
research was supported in part by the French ACI IMPBio program, and by the
ANR projects BRASERO ANR-06-BLAN-0045 and GAMMA 07-2\_195422.}

\bibliographystyle{elsarticle-num}
\bibliography{nonunif}

\end{document}